\documentclass[%
 reprint,
 nofootinbib,
 amsmath,amssymb,
 aps,
]{revtex4-2}

\usepackage{graphicx}
\usepackage{dcolumn}
\usepackage{bm}
\usepackage{color}
\usepackage[normalem]{ulem}
\usepackage[version=4]{mhchem}
\usepackage{url}
\usepackage{amssymb}
\usepackage[bbgreekl]{mathbbol}

\newcommand{\stoiMatrix}{S}
\newcommand{\stoiVector}{s}

\newcommand{\flux}{j}

\newcommand{\Vc}[1]{\boldsymbol{#1}}
\newcommand{\dd}{\mathrm{d}}
\newcommand{\dt}{\dd t}
\newcommand{\defeq}{:=}

\newcommand{\kcoef}{k}
\newcommand{\Transpose}{T}
\newcommand{\X}{\mathcal{X}}

\newcommand{\cmMatrix}{\Gamma} 
\newcommand{\cmvector}{\gamma} 
\newcommand{\B}{B}

\newcommand{\Polytope}{\mathcal{P}}
\newcommand{\BD}{\mathcal{D}}
\newcommand{\KL}{\mathcal{D}_{KL}}

\newcommand{\tf}{f}

\newcommand{\frenecy}{\omega}

\newcommand{\Fspace}{\mathcal{F}}
\newcommand{\Jspace}{\mathcal{J}}
\newcommand{\EPR}{\dot{\Sigma}}
\newcommand{\Gibbs}{\varphi}

\newcommand{\Grad}{\mathrm{grad}}
\newcommand{\Div}{\mathrm{div}}

\newcommand{\FM}{G}
\newcommand{\diag}{\mathrm{diag}}
\newcommand{\identityM}{I}

\newcommand{\pEPR}{\dot{\Pi}}

\newcommand{\Real}{\mathbb{R}}
\newcommand{\Integer}{\mathbb{Z}}

\newcommand{\Dissp}{\Psi}

\newcommand{\Variety}{\mathcal{M}}

\newcommand{\Img}{\mathrm{Im}}
\newcommand{\Ker}{\mathrm{Ker}}

\newcommand{\IncMatrix}{B}

\newcommand{\eqnref}[1]{Eq. (\ref{#1})}

\newcommand{\fgref}[1]{Fig. \ref{#1}}

\newcommand{\Graph}{\mathbb{G}}
\newcommand{\node}{\mathbb{v}}
\newcommand{\edge}{\mathbb{e}}
\newcommand{\molX}{\mathbb{X}}

\usepackage{mathtools}

\begin{document}

\preprint{APS/123-QED}

\title{Geometry of Nonequilibrium Chemical Reaction Networks\\
and Generalized Entropy Production Decompositions}
\author{Tetsuya J. Kobayashi}
\email[E-mail me at:]{tetsuya@mail.crmind.net}
\altaffiliation[Also at ]{Universal Biology Institute, The University of Tokyo, 7-3-1, Hongo, Bunkyo-ku, 113-8654, Japan.}
\affiliation{Institute of Industrial Science, The University of Tokyo, 4-6-1, Komaba, Meguro-ku, Tokyo 153-8505 Japan}
\author{Dimitri Loutchko}
\affiliation{Institute of Industrial Science, The University of Tokyo, 4-6-1, Komaba, Meguro-ku, Tokyo 153-8505 Japan}
\author{Atsushi Kamimura}
\affiliation{Institute of Industrial Science, The University of Tokyo, 4-6-1, Komaba, Meguro-ku, Tokyo 153-8505 Japan}
\author{Yuki Sughiyama}
\affiliation{Institute of Industrial Science, The University of Tokyo, 4-6-1, Komaba, Meguro-ku, Tokyo 153-8505 Japan}
\date{\today}

\
\date{\today}

\begin{abstract}
We derive the Hessian geometric structure of nonequilibrium chemical reaction networks (CRN) on the flux and force spaces induced by the Legendre duality of convex dissipation functions and characterize their dynamics as a generalized flow.
With this structure, we can extend theories of nonequilibrium systems with quadratic
dissipation functions to more general ones with nonquadratic ones, which are pivotal for studying chemical reaction networks.
By applying generalized notions of orthogonality in Hessian geometry to chemical reaction networks, we obtain two generalized decompositions of the entropy production rate, each of which captures gradient-flow and minimum-dissipation aspects in nonequilibrium dynamics.
\end{abstract}

\maketitle


\section{Introduction}
Thermodynamics aims at establishing the general description of thermal systems.
Although such a description was obtained for equilibrium thermodynamics \cite{callen1985,sughiyama2021ArXiv211212403Cond-MatPhysicsphysics}, its extension to nonequilibrium systems has been limited to specific situations and models.
For near-equilibrium situations, Onsager and Machlup evaluated the entropy production rate using the linear approximation known as the force-flux relation \cite{onsager1931Phys.Rev.,onsager1931Phys.Rev.a,onsager1953Phys.Rev.,machlup1953Phys.Rev.}, which corresponds to a quadratic dissipation function.
With the recent development of macroscopic fluctuation theory and stochastic thermodynamics \cite{bertini2015Rev.Mod.Phys.,taniguchi2007JStatPhys,schmiedl2007J.Chem.Phys.}, this result was extended to far-from equilibrium situations in fluid dynamics and diffusion processes in a continuous space, but the dissipation functions are still quadratic even though they are generalized to be state dependent. 
Those systems are characterized geometrically via the inner product structure induced by the quadratic functions and associated formal Riemannian geometry.
However, the knowledge obtained from such situations and models is not directly applicable to other systems with a discrete state space or with nonlinearities in the governing equations, where the dissipation functions are no longer quadratic.
In such cases, the inner product structure is no longer an adequate mathematical basis.
Although Wasserstein geometry has recently been introduced into the thermodynamics of diffusion processes as a new geometric approach \cite{aurell2011Phys.Rev.Lett.,nakazato2021Phys.Rev.Research}, it also employs the formal Riemannian geometric structure associated with the Wasserstein distance via the Benamou-Brenier formula and thereby is not directly applicable to systems with nonquadratic dissipation functions \cite{jordan1997PhysicaD:NonlinearPhenomena,benamou2000Numer.Math.,ambrosio2006}.

In this work, we clarify that this problem can be resolved by using the Hessian (information) geometric structure of the flux and force spaces induced by their nonlinear Legendre duality via convex (not necessarily quadratic) dissipation functions.
For systems with quadratic convex functions, the force and the corresponding flux are related by a linear transformation induced by the Riemannian metric.
Hessian geometry or information geometry \cite{shima2007,amari2016,nielsen2020Entropy} enables us to relate them by a nonlinear Legendre transformation induced by the convex dissipation functions.
Because the Riemannian metric is regarded as the Legendre transformation with quadratic dissipation functions, Hessian geometry is a nonlinear extension of the inner product structure in Riemannian geometry.
Even with being a nonlinear extension, Hessian geometry still preserves many important aspects of the inner product structure in the form of generalized orthogonalities, Pythagorean theorem, among others \cite{amari2016,nielsen2020Entropy}.
Therefore, we can naturally and consistently extend various previous results for systems with quadratic dissipation functions to those with nonquadratic ones.

We derive and demonstrate the structure by focusing mainly on chemical reaction networks (CRNs) because they are representative thermodynamic systems with both discrete state space and nonlinearity in the governing rate equations \cite{hill2005,schnakenberg1976Rev.Mod.Phys.,ge2016ChemicalPhysics,rao2016Phys.Rev.X}. 
CRNs also include Markov jump processes (MJPs) on a graph, which is an important class of systems in stochastic thermodynamics with a discrete state space but with a linear governing equation\footnote{It should be noted that a linear governing equation does not necessarily mean that its dissipation functions are quadratic.}.
In addition, CRNs are also important in light of their biological and engineering applications \cite{beard2008,mikhailov2017,alon2019}. 
The dissipation functions of these systems are not always quadratic, but are more general convex functions.

The nonquadratic property makes it difficult to dissect the equilibrium-like aspects from their nonequilibrium ones by the notion of orthogonality from the usual inner product structure, as was done for overdamped diffusion processes \cite{dechant2022ArXiv210912817Cond-Mat}.
As a result, some entropy decomposition formulas have not yet been generalized to MJP and CNR.
By employing the generalized notions of orthogonality in Hessian geometry \cite{shima2007,amari2016,nielsen2020Entropy}, 
we demonstrate that different aspects of nonequilibrium CRNs can be dissected as generalized decompositions of the entropy production rate (EPR).
In particular, we derive a CRN version of the Maes-Neto\v{c}n\`{y} relation by generalizing the Helmholtz decomposition.
If the dissipation functions are quadratic, these results are automatically and formally reduced to the original results obtained for diffusion processes.
Thus, Hessian geometry provides a natural generalization for resolving the problem.
Finally, we also discuss how Hessian geometry can potentially fill the gap between quadratic and nonquadratic cases by geometrically capturing the nonlinear dual relation between force and flux,  unify the description of those systems, and thereby extend the applicability of nonequilibrium thermodynamics.
As such an example, we mention a relation of our results with the thermodynamic uncertainty relation, and describe how our formulation can contribute to network thermodynamics and variational characterizations of nonequilibrium systems.

This work is organized as follows. In sec. \ref{sec:model}, we define MJP and CRN. 
In sec. \ref{sec:Lugendre}, we introduce the Legendre duality between flux and force, and the associated notion of generalized flow.  
In sec. \ref{eq:Hessian}, we clarify the Hessian geometric structure in the flux-force space and generalized orthogonalities, which are one of the main contributions of this work.
In sec. \ref{sec:CRN}, we demonstrate how the generalized notions of orthogonality lead to different types of EPR decompositions and their geometric meaning.
In sec. \ref{sec:Simulation}, we verify the obtained decompositions using CRN that can have equilibrium, complex-balanced, and noncomplex balanced steady states. 
In sec. \ref{sec:Discussion}, We summarize our results and provide possible applications and contributions of the Hessian structure in the flux-force space to other thermodynamic problems.

\section{Models}\label{sec:model}
In this section, we define MJP and CRN and show how MJP can be regarded as a special case of CRN. 

\subsection{Markov Jump Processes}
A reversible Markov jump process describes random jumps of noninteracting particles on a graph $\Graph$ consisting of $N_{\node}$ vertices, $\{\node_{i}\}_{i\in[1,N_{\node}]}$, and $N_{\edge}$ oriented edges $\{\edge_{e}\}_{e\in[1,N_{\edge}]}$.
$\kcoef_{e}^{+}\ge 0$ is the forward jump rate from the head of the oriented edge $\edge_{e}$ to its tail.
$\kcoef_{e}^{-}\ge 0$ is the reverse jump rate from the tail to the head of $\edge_{e}$.
For infinitely many such particles, we consider $p_{i}(t)\in[0,1]$, the fraction of particles on vertex $\node_{i}$ at time $t$. 
Then, the forward and reverse one-way fluxes on the $e$th edge are 
\begin{align}
\flux^{+}_{e}(\Vc{p})&=\kcoef_{e}^{+}p_{\node^{+}(\edge_{e})}, & 
\flux^{-}_{e}(\Vc{p})&=\kcoef_{e}^{-}p_{\node^{-}(\edge_{e})}, \label{MJP_each_flux}
\end{align}
where $\node^{+}(\edge_{e})$ and $\node^{-}(\edge_{e})$ are the head and tail vertices of edge $\edge_{e}$\footnote{Here, we have abused the notation $\node^{+}(\edge_{e})$ to indicate the index of the vertex $\node^{+}(\edge_{e})$.}.
The total flux function is the difference of one-way flux functions as $\Vc{\flux}(\Vc{p})=\Vc{\flux}^{+}(\Vc{p})-\Vc{\flux}^{-}(\Vc{p})\in \Real^{N_{\edge}}$ where $\Vc{\flux}^{\pm}(\Vc{p})=(\flux^{\pm}_{1}(\Vc{p}),\cdots, \flux^{\pm}_{N_{\edge}}(\Vc{p}))^{\Transpose}$.
Then, the dynamics of the state vector $\Vc{p}(t)\defeq (p_{1}(t), \cdots, p_{N_{\node}}(t))^{\Transpose}\in \Real^{N_{\node}}_{\ge 0}$ is represented by the master equation:
\begin{align}
    \dot{\Vc{p}}=-\IncMatrix \Vc{\flux}(\Vc{p})=-\Div \Vc{\flux}(\Vc{p}),\label{eq:MJP_rateEq}
\end{align}
where $\IncMatrix \in \{0,\pm 1\}^{N_{\node}\times N_{\edge}}$ is the incidence matrix of graph $\Graph$ and $\Div\defeq\IncMatrix$.
More specifically, for $\IncMatrix=(b_{i,e})$, 
\begin{align}
b_{i,e}&\defeq+1 && \mbox{if $\node_{i}$ is the head of the edge $\edge_{e}$}, \notag \\
b_{i,e}&\defeq-1 && \mbox{if $\node_{i}$ is the tail of the edge $\edge_{e}$}, \notag \\
b_{i,e}&\defeq0 && \mbox{otherwise}. \notag
\end{align}
\eqnref{eq:MJP_rateEq} is the continuity equation for diffusion on a graph and $\IncMatrix$ can be regarded as the discrete divergence operator on a graph \cite{grady2010}.
We also define the head and tail incidence matrices, respectively, as $\IncMatrix^{+}\defeq\max[\IncMatrix,0]$ and $\IncMatrix^{-}\defeq\max[-\IncMatrix,0]$. Thus, $\IncMatrix=\IncMatrix^{+}-\IncMatrix^{-}$.

Then, the flux functions are compactly described in a vector form as 
\begin{align}
    \Vc{\flux}^{\pm}(\Vc{p})=\Vc{\kcoef}^{\pm}\circ (\IncMatrix^{\pm})^{\Transpose}\Vc{p},\label{eq:MJP_rate}
\end{align}
where $\circ$ is the component-wise product of two vectors.
In this work, we assume that all edges describe reversible transitions, i.e., $\kcoef_{e}^{\pm}>0$ for all $e$.
While the representation of the master equation by \eqnref{eq:MJP_rateEq} is different from that used conventionally in stochastic thermodynamics, they are equivalent and this representation suits our purpose of unveiling the relationship between MJP and CRN as well as the underlying geometric structure.

\begin{figure}[t]
\includegraphics[width=0.48\textwidth, bb=0 0 500 200]{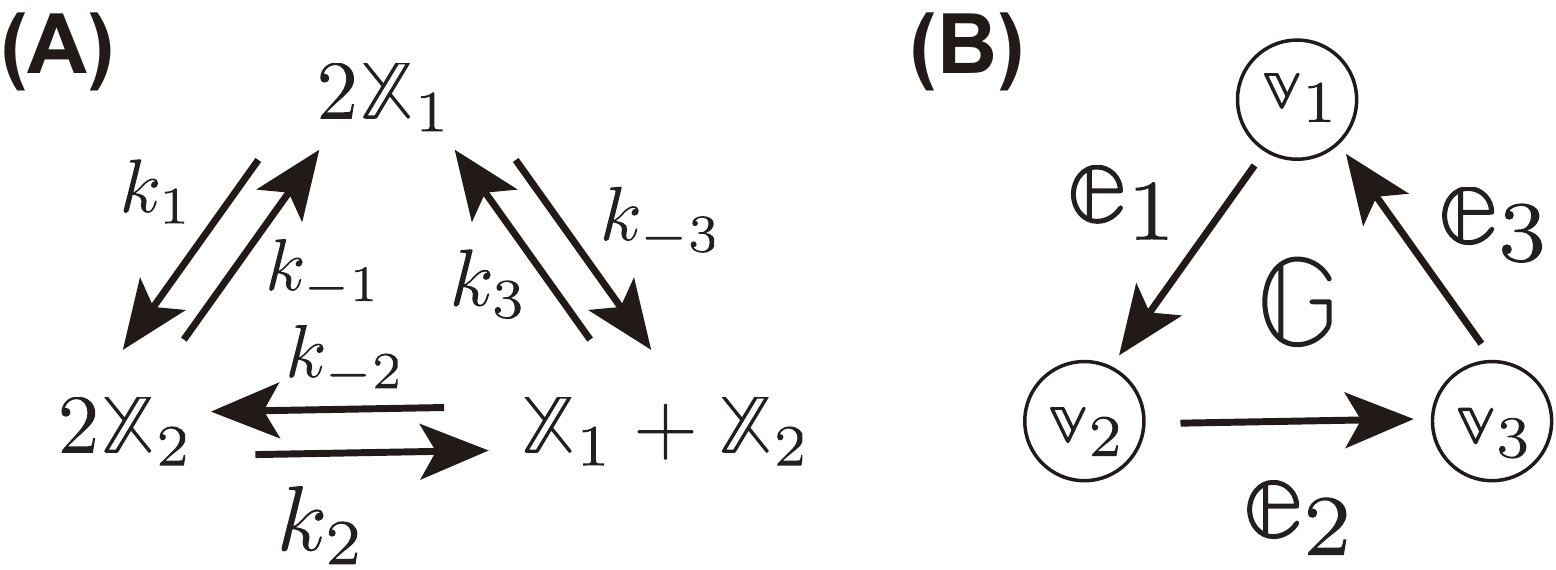}
\caption{Diagrammatic illustration of CRN (A) Diagrammatic representation of chemical reaction equations with reaction rate constants $\{\kcoef_{i}\}_{i=1,2,3}$. (B) The graph-theoretic structure of the CRN in (A). $\node_{1}$, $\node_{2}$, and $\node_{3}$ correspond to complex $2\molX_{1}$, $2\molX_{2}$, and $\molX_{1}+\molX_{2}$, respectively. Each directed edge represents a pair of forward and reverse reactions, and the direction of the edge indicates the direction of the forward reaction.}
\label{fig:CRN}
\end{figure}

\subsection{Chemical Reaction Networks}
Deterministic chemical reaction networks (CRN) are an important class of macroscopic thermodynamic models in light of their historical role played in thermodynamics since Gibbs \cite{gibbs1878Am.J.Sci.} and of their wide range of applications in engineering and biology \cite{beard2008,mikhailov2017,feinberg2019,alon2019}.
A reversible CRN is composed of a collection of forward and reverse reaction pairs(\fgref{fig:CRN} (A)), the $e$th forward reaction of which is described by the following chemical reaction equation \cite{beard2008}:
\begin{align}
\alpha_{1,e}\molX_{1}+\cdots+\alpha_{N_{\molX},e}\molX_{N_{\molX}}\rightarrow \beta_{1,e}\molX_{1}+\cdots+\beta_{N_{\molX},e}\molX_{N_{\molX}},\label{eq:mthReaction}
\end{align}
where $\molX_{i}$ is the $i$th molecular species, $N_{\molX}$ is the number of different kinds of molecular species, and $\alpha_{i,e}\in \mathbb{N}_{\ge 0}$ and $\beta_{i,e}\in \mathbb{N}_{\ge 0}$ are the numbers of molecule $\molX_{i}$ involved as the reactants and products of the $e$th reaction, respectively.
The stoichiometric vector of the $e$th forward reaction is defined as $\Vc{\stoiVector}_{e}\defeq(\beta_{1,e}-\alpha_{1,e},\cdots, \beta_{N,e}-\alpha_{N,e})^{\Transpose}$.
The stoichiometric matrix is $\stoiMatrix\defeq (\Vc{\stoiVector}_{1},\cdots, \Vc{\stoiVector}_{N_{\edge}})$.
The stoichiometric vector of the $e$th reverse reaction is obtained by just changing the sign of the forward one: $-\Vc{\stoiVector}_{m}$.
Thus, the stoichiometric matrix $\stoiMatrix$ defined only for the forward reactions is sufficient to characterize a reversible CRN.
Let $\flux_{e}^{+}(\Vc{x})$ and $\flux_{e}^{-}(\Vc{x})$ be the one-way fluxes of the $e$th forward and reverse reactions in which $\Vc{x}=(x_{1},\cdots,x_{N_{\molX}})^{\Transpose}\in \X \defeq \Real_{>0}^{N_{\molX}}$ represents the concentration of molecules.
The total flux is $\Vc{\flux}(\Vc{x})=\Vc{\flux}^{+}(\Vc{x})-\Vc{\flux}^{-}(\Vc{x})$ where $\Vc{\flux}^{\pm}(\Vc{x})=(\flux^{\pm}_{1}(\Vc{x}), \cdot, \flux^{\pm}_{N_{\edge}}(\Vc{x}))^{\Transpose}$.
Then, we have the deterministic chemical rate equation (CRE) \cite{beard2008,rao2016Phys.Rev.X,feinberg2019} as 
\begin{align}
    \dot{\Vc{x}}=\stoiMatrix \Vc{\flux}(\Vc{x}) =-\Div \Vc{\flux}(\Vc{x}).\label{eq:CRN_rateEq}
\end{align}
\eqnref{eq:CRN_rateEq} is the continuity equation for the CRN with the divergence operator $\Div=-\stoiMatrix$. 

The intrinsic graph structure of CRN can be manifested by considering the sets of reactants and products as vertices of the graph $\Graph$ connected by the reaction edges (\fgref{fig:CRN} (B)).
Specifically, the set of reactants $\node_{e}^{+}\defeq (\alpha_{1,e}\molX_{1}+\cdots+\alpha_{1,e}\molX_{N_{\node}})$ and the set of products $\node_{e}^{-}\defeq(\beta_{1,e}\molX_{1}+\cdots+\beta_{1,e}\molX_{N_{\node}})$ are regarded as the head and tail vertices of the $e$th reaction edge, respectively.
Such sets (vertices) are called complexes in the CRN theory \cite{feinberg2019}.
Because each complex is a set of molecular species, a CRN is a special kind of hypergraph in which multiple molecular species are connected by oriented edges (reactions) \cite{klamt2009PLOSComputationalBiology}. 
Reflecting this hypergraph nature, $\stoiMatrix$ can be decomposed as 
\begin{align}
    \stoiMatrix=-\cmMatrix \IncMatrix=-\cmMatrix (\IncMatrix^{+}-\IncMatrix^{-}),\label{eq:CRN_stoiMatrix}
\end{align}
where $N_{\molX}$ is the number of different vertices (complexes), $\IncMatrix$ is the incidence matrix of the complex graph $\Graph$, and $\cmMatrix=(\Vc{\cmvector}_{1}, \cdots, \Vc{\cmvector}_{N_{\node}})\in\Integer^{N_{\molX}\times N_{\node}}$ where $\Vc{\cmvector}_{i}$ specifies the molecular species involved in the $i$th vertex (complex) as $\node_{i}\defeq (\cmvector_{1,i}\molX_{1}+\cdots+\cmvector_{N_{\molX},i}\molX_{N_{\molX}})$.

If we adopt the law of mass action kinetics, the $e$th forward and reverse reaction fluxes can be represented as
\begin{align}
    \flux^{\pm}_{e}(\Vc{x})=\kcoef^{\pm}_{e} \sum_{i=1}^{N_{\node}}b_{i,e}^{\pm}\prod_{j=1}^{N_{\molX}}x_{j}^{\gamma_{j,i}},\label{CNR_each_flux}
\end{align}
where $\kcoef_{e}^{+}\in \Real_{\ge 0}$ and $\kcoef_{e}^{-}\in \Real_{\ge 0}$ are the reaction rate constants of the $e$th forward and reverse reactions, respectively.
In vector form, we can compactly represent it as 
\begin{align}
    \Vc{\flux}^{\pm}(\Vc{x})=\Vc{\kcoef}^{\pm}\circ (\IncMatrix^{\pm})^{\Transpose}\Vc{x}^{\cmMatrix^{\Transpose}},\label{eq:CRN_rate}
\end{align}
where $\Vc{x}^{\Vc{\cmvector}}\defeq \prod_{j=1}^{N_{\molX}} x_{j}^{\cmvector_{j}}\in \Real_{>0}$ and $\Vc{x}^{\cmMatrix^{\Transpose}}\defeq (\Vc{x}^{\Vc{\cmvector}_{1}}, \cdots, \Vc{x}^{\Vc{\cmvector}_{N_{\node}}})^{\Transpose}$.
We should note an important relation, $(\IncMatrix^{\pm})^{\Transpose}\Vc{x}^{\cmMatrix^{\Transpose}}=\Vc{x}^{(\cmMatrix\IncMatrix^{\pm})^{\Transpose}}$, which holds because every column vector of $\IncMatrix^{\pm}$ contains only one $+1$ and the others are $0$ (see also Appendix for notation).

By comparing \eqnref{eq:MJP_rateEq} and \eqnref{eq:MJP_rate} with \eqnref{eq:CRN_rateEq} and \eqnref{eq:CRN_rate}, we can see that CRNs contain MJPs as a special case.
Specifically, if we identify $\Vc{x}$ with $\Vc{p}$ and set $\cmMatrix=\identityM$, where $\identityM$ is the identity matrix,
\eqnref{eq:CRN_rateEq} and \eqnref{eq:CRN_rate} are reduced to \eqnref{eq:MJP_rateEq} and \eqnref{eq:MJP_rate}, respectively. 
In other words, a MJP is a CRN, each complex of which contains only one molecular species.
Such a CRN is called a monomolecular reaction network and its thermodynamic nature has been investigated in the context of equilibrium and nonequilibrium chemical thermodynamics, especially by Hill \cite{hill2005}.
Because CRNs include MJPs, we work only on CRNs in the following sections\footnote{The state of MJP is the probability vector $\Vc{p}(t)$. Thus, the conservation of probability $\sum_{i}p_{i}(t)=1$ should be satisfied. To regard MJP as a CNR, we here do not impose such a conservation in advance because the conservation of probability over time, i.e., $\sum_{i}p_{i}(t)=\mathrm{const.}$ is automatically satisfied by the property of the incidence matrix $\IncMatrix$ such that $\Vc{1}^{\Transpose}\IncMatrix=\Vc{0}$. Thus, by restricting the dynamics of the system with the initial condition satisfying $\sum_{i}x_{i}(0)=1$, the constraint $\sum_{i}p_{i}(t)=1$ is automatically obtained. }.

\section{Legendre Duality of flux-force}\label{sec:Lugendre}
Next, we introduce the Legendre duality between flux and force for CRNs\footnote{We abbreviate the thermodynamic force as force here.}, and summarize their relation to entropy production.
The specific type of convex function introduced here was recently derived in the Macroscopic Fluctuation Theorem (MFT) \cite{maes2017Phys.Rev.Lett.,kaiser2018JStatPhys,renger2018Entropy,renger2021DiscreteContin.Dyn.Syst.-S,patterson2021ArXiv210314384Math-Ph}.

For a given pair of one-way fluxes $\Vc{\flux}^{\pm}$, the total flux and force are obtained as 
\begin{align}
    \Vc{\flux}\defeq\Vc{\flux}^{+}-\Vc{\flux}^{-}\in \Jspace,\qquad \Vc{\tf}\defeq\frac{1}{2}\ln\frac{\Vc{\flux}^{+}}{\Vc{\flux}^{-}}\in \Fspace \label{eq:flux_force},
\end{align}
where $\Jspace=\mathbb{V}$ is an $N_{\edge}$-dimensional vector space and $\Fspace=\mathbb{V}^{*}$ is its linear-algebraic dual space\footnote{The definition of the force here includes $1/2$, which does not appear in the conventional definition of thermodynamic force. This is because we adopt the derivation of this form of force from the large deviation theory\cite{patterson2021ArXiv210314384Math-Ph}. We can remove it by including $1/2$ in the definition of the dissipation functions.}.
The force of this form comes from the local detailed balance condition \cite{maes2021SciPostPhys.Lect.Notes}, and is also consistent with macroscopic chemical thermodynamics with the mass action kinetics \cite{schnakenberg1976Rev.Mod.Phys.,rao2016Phys.Rev.X,sughiyama2021ArXiv211212403Cond-MatPhysicsphysics}.
If $\Vc{\flux}$ and $\Vc{\tf}$ are defined as in \eqnref{eq:flux_force}, the entropy production rate (EPR) is obtained as 
\begin{align}
\EPR\defeq 2 \langle \Vc{\flux},\Vc{\tf}\rangle=(\Vc{\flux}^{+}-\Vc{\flux}^{-})^{\Transpose}\ln\frac{\Vc{\flux}^{+}}{\Vc{\flux}^{-}}\ge 0, \label{eq:EPR_def}
\end{align}
where $\langle\Vc{\flux}, \Vc{\tf}\rangle\defeq \sum_{e=1}^{N_{\edge}}\flux_{e}\tf_{e}$ is the usual bilinear pairing on $\mathbb{V} \times \mathbb{V}^{*}$.

As \eqnref{eq:flux_force} implies, the pair of flux and force has a nonlinear relationship. 
To show that the relation is a Legendre duality, we introduce the frenetic activity \cite{maes2017}\footnote{The definition of the activity here includes $2$, which does not appear in the definition of the activity in \cite{maes2017}. This is because we adopt the derivation of this form of activity from the large deviation theory\cite{patterson2021ArXiv210314384Math-Ph}. We can remove it by including $2$ in the definition of the dissipation functions.}: \begin{align}
    \Vc{\frenecy}\defeq 2 \sqrt{\Vc{\flux}^{+}\circ\Vc{\flux}^{-}}\in \mathbb{R}_{> 0}^{N_{\edge}}. \label{eq:frenecy}
\end{align}
Then, we have $\Vc{\flux}=\frac{\Vc{\frenecy}}{2}\circ \left[e^{\Vc{\tf}}-e^{\Vc{-\tf}}\right]$. 
Thus, if $\Vc{\frenecy}$ is given, the force $\Vc{\tf}$ can be converted to the corresponding flux $\Vc{\flux}$ that satisfies \eqnref{eq:flux_force}. 
This relation between the pair $(\Vc{\flux},\Vc{\tf})$ is a one-to-one Legendre duality (LD) induced by the following strictly convex smooth functions: 
\begin{align}
    \Dissp^{*}_{\Vc{\frenecy}}(\Vc{\tf})&\defeq \Vc{\frenecy}^{\Transpose} \left[\cosh(\Vc{\tf})-\Vc{1}\right],\notag \\
    \Dissp_{\Vc{\frenecy}}(\Vc{\flux})&\defeq \Vc{\flux}^{\Transpose} \sinh^{-1}\left(\frac{\Vc{\flux}}{\Vc{\frenecy}}\right) - \Vc{\frenecy}^{\Transpose} \left[\sqrt{\Vc{1}+\left(\frac{\Vc{\flux}}{\Vc{\frenecy}}\right)^{2}}-\Vc{1}\right],\label{eq:DissipationFunctions}
\end{align}
which lead to the Legendre transformations: 
\begin{align}
    \Vc{\flux}&=\partial_{\Vc{\tf}} \Dissp^{*}_{\Vc{\frenecy}}(\Vc{\tf})=\Vc{\frenecy}\circ \sinh(\Vc{\tf}),\label{eq:LegendreTJ}\\
    \Vc{\tf}&=\partial_{\Vc{\flux}} \Dissp_{\Vc{\frenecy}}(\Vc{\flux})=\sinh^{-1}\left(\frac{\Vc{\flux}}{\Vc{\frenecy}}\right)\label{eq:LegendreTF}.
\end{align}
We can easily verify that \eqnref{eq:LegendreTJ} and \eqnref{eq:LegendreTF} are equivalent to \eqnref{eq:flux_force} by direct computation (see also Appendix ).
In the following, we abbreviate $\partial_{\Vc{\tf}} \Dissp^{*}$ and $\partial_{\Vc{\flux}}\Dissp_{\Vc{\frenecy}}$ with $\partial \Dissp^{*}$ and $\partial \Dissp_{\Vc{\frenecy}}$ because we do not use differentiation of $\Dissp^{*}$ and $\Dissp$ with respect to other variables in this work.
Moreover, the pair $(\Vc{\flux},\Vc{\tf})$ satisfies the Legendre identity:
\begin{align}
\Dissp^{*}_{\Vc{\frenecy}}(\Vc{\tf})+ \Dissp_{\Vc{\frenecy}}(\Vc{\flux})-\langle \Vc{\flux},\Vc{\tf}\rangle=0. \label{eq:LegendreIdentity}
\end{align}
The potential functions $\Dissp^{*}_{\Vc{\frenecy}}(\Vc{\tf})$ and $\Dissp_{\Vc{\frenecy}}(\Vc{\flux})$ of the form in \eqnref{eq:DissipationFunctions} were recently derived via large deviation functions of the corresponding microscopic models in MFT \cite{maes2017Phys.Rev.Lett.,renger2018Entropy,renger2021DiscreteContin.Dyn.Syst.-S,patterson2021ArXiv210314384Math-Ph}, where they are called dissipation functions.
Both dissipation functions are nonnegative and symmetric:
\begin{align}
    \Dissp_{\Vc{\frenecy}}(\Vc{\flux})=\Dissp_{\Vc{\frenecy}}(-\Vc{\flux})\ge 0,\qquad \Dissp^{*}_{\Vc{\frenecy}}(\Vc{\tf})=\Dissp^{*}_{\Vc{\frenecy}}(-\Vc{\tf})\ge 0,
\end{align}
and satisfy $
\min_{\Vc{\tf}}\Dissp^{*}_{\Vc{\frenecy}}(\Vc{\tf})=\Dissp^{*}_{\Vc{\frenecy}}(\Vc{0})=0$ and $\min_{\Vc{\flux}}\Dissp_{\Vc{\frenecy}}(\Vc{\flux})=\Dissp_{\Vc{\frenecy}}(\Vc{0})=0$.
From these properties, we can verify that $\Vc{\tf}=\Vc{0}$ and $\Vc{\flux}=\Vc{0}$ are Legendre dual: $\Vc{0}=\partial\Dissp^{*}_{\Vc{\frenecy}}(\Vc{0})$ and $\Vc{0}=\partial\Dissp_{\Vc{\frenecy}}(\Vc{0})$,
This is consistent with the thermodynamic requirement that if the force is zero, the corresponding flux becomes zero, and vice versa.

In addition, from the Legendre identity (\eqnref{eq:LegendreIdentity}), the nonnegativity of the EPR can also be attributed to the nonnegativity of the dissipation functions:
\begin{align}
    \EPR/2= \langle \Vc{\flux},\Vc{\tf}\rangle = \Dissp^{*}_{\Vc{\frenecy}}(\Vc{\tf})+ \Dissp_{\Vc{\frenecy}}(\Vc{\flux})\ge 0.  
\end{align}
It should be noted that the dissipation functions naturally provide a decomposition of the EPR into nonnegative terms.

In the following, a pair of flux and force with the same decoration, e.g., $(\Vc{\flux},\Vc{\tf})$, $(\Vc{\flux}',\Vc{\tf}')$, or $(\Vc{\flux}(\Vc{x}),\Vc{\tf}(\Vc{x}))$, represents a Legendre dual pair linked by \eqnref{eq:LegendreTJ} and \eqnref{eq:LegendreTF}.
Because of the one-to-one LD, the CRE (\eqnref{eq:CRN_rateEq}) can be represented as :
\begin{align}
    \dot{\Vc{x}}=\stoiMatrix \Vc{\flux}(\Vc{x})=&\stoiMatrix \partial \Dissp^{*}_{\Vc{\frenecy}(\Vc{x})}[\Vc{\tf}(\Vc{x})]\notag \\
    &=-\Div\partial \Dissp^{*}_{\Vc{\frenecy}(\Vc{x})}[\Vc{\tf}(\Vc{x})].\label{eq:CRN}
\end{align}
The equation of this form is a generalized flow driven by the force $\Vc{\tf}(\Vc{x})$ \cite{ambrosio2006,renger2018Entropy,patterson2021ArXiv210314384Math-Ph}.
The representation of the dynamics in this form is not specific to CRN or MJP. Thus, it can cover other systems such as overdamped diffusion by appropriately defining $\Div$, $\Dissp(\Vc{x})$, $\Dissp^{*}(\Vc{x})$, and $\Vc{\tf}(\Vc{x})$.

It should be noted that the derivation of Legendre duality and associated quantities and relations is not dependent on the specific functional form of the one-way fluxes, i.e., \eqnref{eq:CRN_rate} or \eqnref{eq:MJP_rate}, the former of which is from the kinetic law of mass action assumption.
Thus, the result here may be applied to a wider class of kinetic laws.

\section{Hessian Geometry and generalized orthogonalities}\label{eq:Hessian}
\begin{figure}[ht]
\includegraphics[width=0.49\textwidth, bb=0 0 600 400]{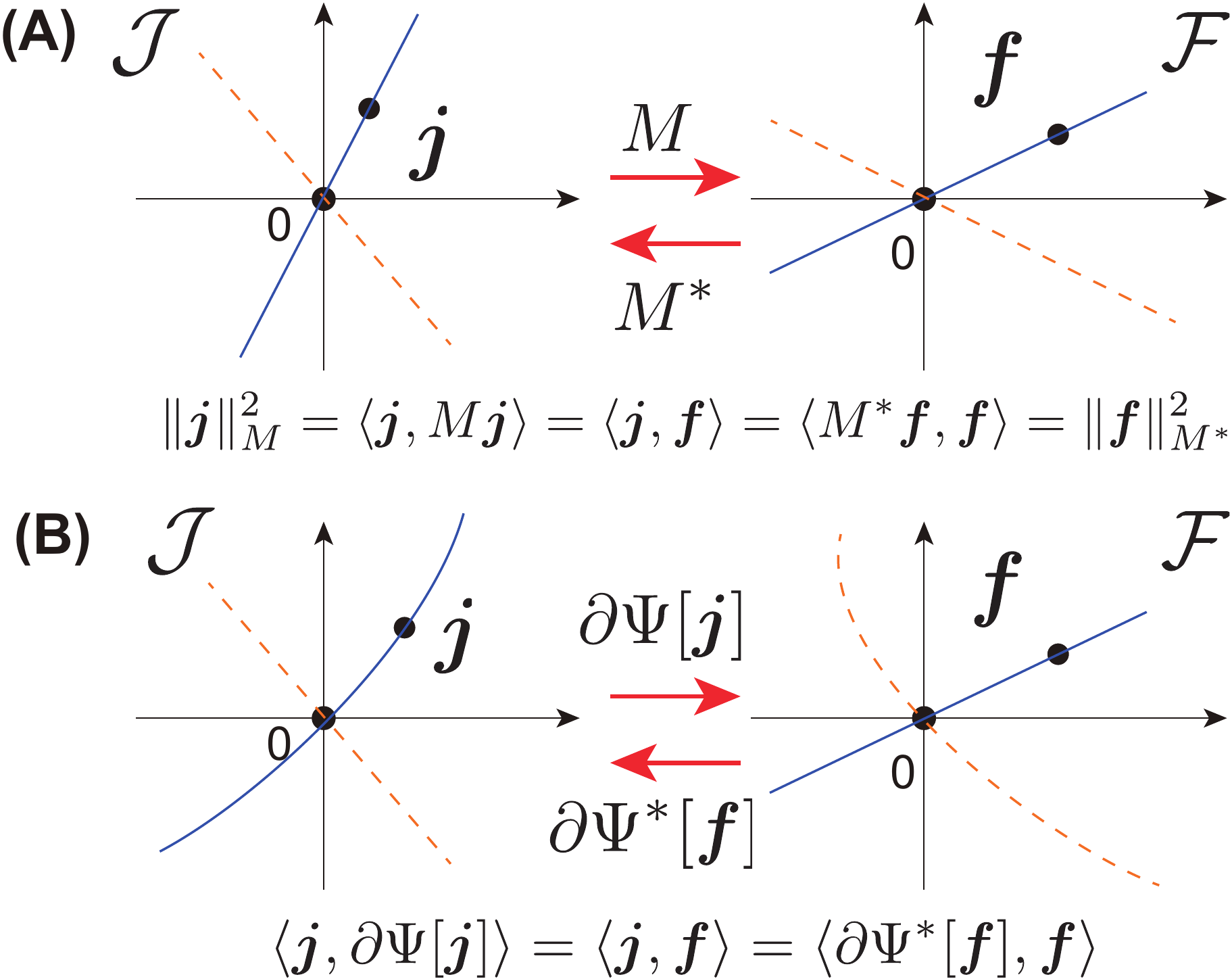}
\caption{(A) Linear-algebraic dual spaces $(\Jspace, \Fspace)$ with the inner product structure defined by a metric via $M$ and $M^{*}$. Linear subspaces in $\Jspace$, e.g., blue solid and red dashed lines, are mapped to linear subspaces in $\Fspace$. 
(B) Linear-algebraic dual spaces $(\Jspace, \Fspace)$ with the Hessian geometric structure defined by Legendre transformations $\partial \Dissp$ and $\partial^{*} \Dissp$. A linear subspace in $\Jspace$, e.g., red dashed line, is mapped to a curved subspace in $\Fspace$ whereas a linear subspace in $\Fspace$, e.g., blue solid line, is mapped to a curved subspace in $\Jspace$.}
\label{fig:Rieman_Hesse}
\end{figure}

Because of the nonlinearity of Legendre duality with the nonquadratic dissipation functions, we can no longer employ the inner product structure between flux and force and the associated formal Riemannian geometric notions (\fgref{fig:Rieman_Hesse}).
We clarify that this problem is resolved by employing Hessian geometry and its geometric notions\footnote{It should be noted that, in differential geometry, Hessian geometry is a class of Riemannian geometry with a metric given by a Hessian matrix, which additionally induces the Legendre dual structure. }.
The elucidation of the fundamental roles of Hessian geometry in nonequilibrium dynamics of MJP and CRN is one of the main contributions of this work.

\subsection{Hessian geometry}
Hessian geometry is the geometry induced by convex potential functions $\Dissp(\Vc{\flux})$ and $\Dissp^{*}(\Vc{\tf})$ to a pair of dual spaces $\Jspace$ and $\Fspace$ \cite{shima2007,amari2016}\footnote{Hessian geometry is generally defined on affine manifolds. However, because we relate the Hessian geometric structure with inner product one, we here restrict $\Jspace$ and $\Fspace$ to be the dual vector spaces for simplicity of presentation.}.
In this section, $\Dissp(\Vc{\flux})$ and $\Dissp^{*}(\Vc{\tf})$ are not restricted to the specific form of \eqnref{eq:DissipationFunctions} to obtain general results.
We consider the case for each fixed $\Vc{\frenecy}$\footnote{This also means fixed $\Vc{x}$ if $\Vc{\frenecy}$ is a function of $\Vc{x}$. This is similar to the inner product structure of the tangent and cotangent spaces at each point $\Vc{x}$ on a Reimannian manifold.}. 
Thus, the dependence of $\Dissp(\Vc{\flux})$ and $\Dissp^{*}(\Vc{\tf})$ on $\Vc{\frenecy}$ is omitted for the sake of notational simplicity.
They are assumed more generally to be just smooth and strictly convex functions satisfying the Legendre duality:
\begin{align}
    \Dissp^{*}(\Vc{\tf}) &= \max_{\Vc{\flux}}\left[\langle\Vc{\flux},\Vc{\tf} \rangle-\Dissp(\Vc{\flux}) \right],\\
    \Dissp(\Vc{\flux}) &= \max_{\Vc{\tf}}\left[\langle\Vc{\flux},\Vc{\tf} \rangle-\Dissp^{*}(\Vc{\tf}) \right].
\end{align}
Instead of the association of $\Vc{\flux}$ and $\Vc{\tf}$ by a linear transformation $\Vc{\flux}=M^{*}\Vc{\tf}$ on the inner product space (\fgref{fig:Rieman_Hesse} A), $\Vc{\flux}$ and $\Vc{\tf}$ are associated by the Legendre transformation(\fgref{fig:Rieman_Hesse} B):
\begin{align}
    \Vc{\flux}&=\partial_{\Vc{\tf}}\Dissp^{*}(\Vc{\tf}), & \Vc{\tf}&=\partial_{\Vc{\flux}}\Dissp(\Vc{\flux}).\label{eq:LT2}
\end{align}
Hessian geometry is fundamental for capturing nonlinear geometry induced by the convex functions. 
As an important application, it has played the essential role in describing the geometry of statistical models in information geometry \cite{amari2016,nielsen2020Entropy} and that of equilibrium thermodynamics\cite{kobayashi2021ArXiv211214910Phys.,sughiyama2021ArXiv211212403Cond-MatPhysicsphysics,sughiyama2022ArXiv220109417Cond-MatPhysicsphysics}.

To regard the Hessian geometric structure as a nonlinear generalization of the inner product structure\footnote{It should be noted that the Hessian geometric structure is not necessarily restricted to this interpretation as an extension of the inner product structure.}, we additionally assume that $\Dissp(\Vc{\flux})$ satisfies the symmetry condition $\Dissp(\Vc{\flux})=\Dissp(-\Vc{\flux})$.
From this symmetry condition for $\Dissp(\Vc{\flux})$, we also obtain the symmetry of $\Dissp^{*}(\Vc{\tf})=\Dissp^{*}(-\Vc{\tf})$ via the Legendre duality.
From the symmetries, the minimum of $\Dissp(\Vc{\flux})$ and $\Dissp^{*}(\Vc{\tf})$ is attained at $\Vc{0}$: $
\arg \min_{\Vc{\flux}}\Dissp(\Vc{\flux})=\Vc{0}$ and $\arg \min_{\Vc{\tf}}\Dissp^{*}(\Vc{\tf})=\Vc{0}$.
Because the Legendre transformation \eqnref{eq:LT2} is independent of a constant in $\Dissp(\Vc{\flux})$ and $\Dissp^{*}(\Vc{\tf})$, without losing generality, we can assume that $\min_{\Vc{\flux}}\Dissp(\Vc{\flux})=\Dissp(\Vc{0})=\min_{\Vc{\flux}}\Dissp(\Vc{\flux})=\Dissp(\Vc{0})=0$.
Thus, 
\begin{align}
    \Dissp(\Vc{\flux})=\Dissp(-\Vc{\flux})\ge 0,\qquad \Dissp^{*}(\Vc{\tf})=\Dissp^{*}(-\Vc{\tf})\ge 0. \label{eq:Dissp_sym}
\end{align}
In the following, we call $\Dissp(\Vc{\flux})$ and $\Dissp^{*}(\Vc{\tf})$ with this property dissipation functions.

To demonstrate that the Hessian geometric structure includes the inner product structure, suppose that $\Dissp(\Vc{\flux})$ is a quadratic function as 
\begin{align}
\Dissp(\Vc{\flux})=\frac{1}{2}\langle\Vc{\flux},M\Vc{\flux} \rangle:=\frac{1}{2}\|\Vc{\flux}\|^{2}_{M}, \label{eq:qDisNormP}
\end{align}
where $M\in \Real^{N_{\edge}\times N_{\edge}}$ is a positive definite matrix.
Then, its Legendre dual becomes
\begin{align}
\Dissp^{*}(\Vc{\tf})=\frac{1}{2}\langle M^{-1}\Vc{\tf},\Vc{\tf} \rangle=\frac{1}{2}\langle M^{*}\Vc{\tf},\Vc{\tf} \rangle:=\frac{1}{2}\|\Vc{\tf}\|^{2}_{{M}^{*}},\label{eq:qDisNormD}
\end{align}
where ${M}^{*}\defeq {M}^{-1}$.
Thus, $\Dissp(\Vc{\flux})$ and $\Dissp^{*}(\Vc{\tf})$ are reduced to the squared norms associated with the metric ${M}$ and ${M}^{*}$.
The Legendre transformations become 
\begin{align}
\Vc{\flux} & = {M}\Vc{\tf} = ({M}^{*})^{-1}\Vc{\tf}, & \Vc{\tf} & = {M}^{*}\Vc{\flux} = {M}^{-1}\Vc{\flux}.
\end{align}
These are the linear pairing of $\Vc{\flux}$ and $\Vc{\tf}$ via the metric ${M}$.
Furthermore, for the paired $\Vc{\flux}$ and $\Vc{\tf}$ by ${M}$, 
$\Dissp(\Vc{\flux})$ and $\Dissp^{*}(\Vc{\tf})$ are essentially identical to the entropy production as 
\begin{align}
    \frac{1}{2} \langle\Vc{\flux},\Vc{\tf} \rangle=\Dissp(\Vc{\flux})=\Dissp^{*}(\Vc{\tf})=\EPR/4. \label{eq:Dissps_EPR_quadratic}
\end{align}
However, this identity of $\Dissp(\Vc{\flux})$ and $\Dissp^{*}(\Vc{\tf})$ no longer holds for a nonquadratic $\Dissp(\Vc{\flux})$ and $\Dissp^{*}(\Vc{\tf})$ pair.
In addition, for quadratic cases, the generalized flow (\eqnref{eq:CRN}) reduces to the flow on a Riemannian manifold:
\begin{align}
    \dot{\Vc{x}}=-\Div M^{*}_{\Vc{x}}[\Vc{\tf}(\Vc{x})].
\end{align}
As a result, various notions of geometry in the inner product space are generalized in Hessian geometry.

\subsection{Generalized distance and Bregman divergence}
Owing to the nonlinearity, the notion of distance, i.e., is generalized at least into two versions.

The first version is to define a generalized distance via the dissipation functions as
\begin{align}
    \mathcal{D}_{H}(\Vc{\flux},\Vc{\flux}')&\defeq \Dissp(\Vc{\flux}-\Vc{\flux}'),\\ 
    \mathcal{D}^{*}_{H}(\Vc{\tf},\Vc{\tf}')&\defeq \Dissp^{*}(\Vc{\tf}-\Vc{\tf}').
\end{align}
From \eqnref{eq:Dissp_sym}, $\mathcal{D}_{H}(\Vc{\flux},\Vc{\flux}')$ is nonnegative, symmetric $\mathcal{D}_{H}(\Vc{\flux},\Vc{\flux}')=\mathcal{D}_{H}(\Vc{\flux}',\Vc{\flux})$ and satisfies $\mathcal{D}_{H}(\Vc{\flux},\Vc{\flux}')=0$ if and only if $\Vc{\flux}=\Vc{\flux}'$.
The same is true for $\mathcal{D}^{*}_{H}(\Vc{\tf},\Vc{\tf}')$.
In addition, they are reduced to the usual metric induced by the squared norm for quadratic cases, as shown in \eqnref{eq:qDisNormP} and \eqnref{eq:qDisNormD}.

The second version is the Bregman divergence, which is defined as 
\begin{align*}
    \BD[\Vc{\flux}\|\Vc{\flux}']&\defeq\Dissp(\Vc{\flux})-\Dissp(\Vc{\flux}')-\langle\Vc{\flux}-\Vc{\flux}', \partial_{\Vc{\flux}}\Dissp(\Vc{\flux}')\rangle\\ 
    \BD^{*}[\Vc{\tf}'\|\Vc{\tf}]&\defeq\Dissp^{*}(\Vc{\tf})-\Dissp^{*}(\Vc{\tf}')-\langle\partial_{\Vc{\tf}}\Dissp^{*}(\Vc{\tf}'), \Vc{\tf}-\Vc{\tf}'\rangle.
\end{align*}
The Bregman divergence $\BD[\Vc{\flux}\|\Vc{\flux}']$ is nonnegative and strictly convex for $\Vc{\flux}$\footnote{Not necessarily for $\Vc{\flux}'$.}  and also attains the minimum $0$ if and only if $\Vc{\flux}=\Vc{\flux}'$\footnote{It should be noted that the symmetry property of the dissipation function \eqnref{eq:Dissp_sym} is not required to derive these properties.}.
However, it is generally asymmetric $\BD[\Vc{\flux};\Vc{\flux}']\neq \BD[\Vc{\flux}';\Vc{\flux}]$.
If $\Dissp(\Vc{\flux})$ is quadratic (\eqnref{eq:qDisNormP}),  $\BD[\Vc{\flux}\|\Vc{\flux}']$ is reduced to $\BD[\Vc{\flux}\|\Vc{\flux}']=\frac{1}{2}\|\Vc{\flux}-\Vc{\flux}'\|^{2}_{{M}}$ and $\BD^{*}[\Vc{\tf}\|\Vc{\tf}']$ is reduced to $\BD^{*}[\Vc{\tf}\|\Vc{\tf}']=\frac{1}{2}\|\Vc{\tf}-\Vc{\tf}'\|^{2}_{{M}^{*}}$.
It should be noted that for any flux-force pairs $(\Vc{\flux},\Vc{\tf})$ and $(\Vc{\flux}',\Vc{\tf}')$, the divergences $\BD[\Vc{\flux}\|\Vc{\flux}']$ and $\BD^{*}[\Vc{\tf}'\|\Vc{\tf}]$ are the same:
\begin{align}
    \BD[\Vc{\flux}\|\Vc{\flux}']=\BD^{*}[\Vc{\tf}'\|\Vc{\tf}]. \label{eq:Bregman_same}
\end{align}
Thus, $\BD[\Vc{\flux}\|\Vc{\flux}']$ and $\BD^{*}[\Vc{\tf}'\|\Vc{\tf}]$ are different representations of the same quantity in the flux space $\Jspace$ and the force space $\Fspace$, respectively.
As a result, $\BD[\Vc{\flux}\|\Vc{\flux}']$ and $\BD^{*}[\Vc{\tf}'\|\Vc{\tf}]$ are also the same as the following mixed representation:
\begin{align}
    \BD[\Vc{\flux};\Vc{\tf}']:=\Dissp(\Vc{\flux})+\Dissp^{*}(\Vc{\tf}')-\langle\Vc{\flux}, \Vc{\tf}'\rangle, \label{eq:BD}
\end{align}
i.e., $\BD[\Vc{\flux}\|\Vc{\flux}'] = \BD[\Vc{\flux};\Vc{\tf}'] =\BD^{*}[\Vc{\tf}'\|\Vc{\tf}]$ (see also Appendix ).

\subsection{Generalized Hilbert orthogonality}
The two generalized distances naturally lead to two generalized notions of orthogonality.

For any two force vectors $\Vc{\tf}_{S}, \Vc{\tf}_{A}\in \Fspace$, the first orthogonality is defined to satisfy
\begin{align}
    \Dissp^{*}(\Vc{\tf}_{S}+\Vc{\tf}_{A})=\Dissp^{*}(\Vc{\tf}_{S}-\Vc{\tf}_{A}).\label{eq:ort1st}
\end{align}
This definition comes from the fact that, if $\Vc{\tf}_{S}$ and $\Vc{\tf}_{A}$ are orthogonal in an inner product space, then $\|\Vc{\tf}_{S}+\Vc{\tf}_{A}\|_{{M}^{*}}=\|\Vc{\tf}_{S}-\Vc{\tf}_{A}\|_{{M}^{*}}$ holds.
We call this orthogonality a generalized Hilbert orthogonality or simply Hilbert orthogonality \cite{renger2021ArXiv211112164Cond-MatPhysicsmath-Ph}.
Then an orthogonal decomposition of a given force $\Vc{\tf}$ into $\Vc{\tf}=\Vc{\tf}_{S}+\Vc{\tf}_{A}$ is obtained by finding an iso-dissipation force $\Vc{\tf}_{iso}$ satisfying 
$\Dissp^{*}(\Vc{\tf})=\Dissp^{*}(\Vc{\tf}_{iso})$ and by computing $\Vc{\tf}_{S}$ and $\Vc{\tf}_{A}$ as
\begin{align}
\Vc{\tf}_{S}&=\frac{\Vc{\tf}+\Vc{\tf}_{iso}}{2}, & \Vc{\tf}_{A}=\frac{\Vc{\tf}-\Vc{\tf}_{iso}}{2}.
\end{align}
Note that $\Vc{\tf}_{iso}=\Vc{\tf}_{S}-\Vc{\tf}_{A}$ and also that we can have different decompositions of $\Vc{\tf}$ by choosing different iso-dissipation forces $\Vc{\tf}'_{iso}$.
In addition, from the symmetry of $\Dissp^{*}$, we have
$\Dissp^{*}(\Vc{\tf})=\Dissp^{*}(\Vc{\tf}_{iso})=\Dissp^{*}(-\Vc{\tf}_{iso})=\Dissp^{*}(-\Vc{\tf})$.
As a consequence of the orthogonality, we obtain 
\begin{align}
\BD[\Vc{\flux}\|\Vc{\flux}_{iso}]=&2 \langle\Vc{\flux},\Vc{\tf}_{A}\rangle \ge 0,\label{eq:Dj_jiso}
\end{align}
where we use $\BD[\Vc{\flux}\|\Vc{\flux}_{iso}]\ge0$ (see Appendix for a proof).
Similarly, by considering the symmetry of $\Dissp^{*}(\Vc{\tf})$, we also have 
\begin{align}
\BD_{\Vc{\frenecy}}[\Vc{\flux}\|-\Vc{\flux}_{iso}]=2 \langle\Vc{\flux},\Vc{\tf}_{S}\rangle\ge 0.    
\end{align}
From these relations, we have a decomposition of $\langle\Vc{\flux},\Vc{\tf} \rangle$ into two nonnegative terms:
\begin{align}
    \langle\Vc{\flux},\Vc{\tf} \rangle&=\langle\Vc{\flux},\Vc{\tf}_{S} \rangle+\langle\Vc{\flux},\Vc{\tf}_{A} \rangle \label{eq:HBdecomp}\\
    &= \frac{\BD[\Vc{\flux}\|-\Vc{\flux}_{iso}]}{2}+\frac{\BD[\Vc{\flux}\|\Vc{\flux}_{iso}]}{2}.
\end{align}
Because $\langle\Vc{\flux},\Vc{\tf}\rangle$ is proportional to the EPR (\eqnref{eq:EPR_def}), this decomposition provides a way to obtain nonnegative decompositions of the EPR.
It should be noted that we have infinitely many decompositions of this type.
In addition, by choosing $\Vc{\flux}_{iso}=\Vc{\flux}$, the decomposition reduces to the fluctuation relation $\EPR= \BD[\Vc{\flux}\|-\Vc{\flux}]$, where we used $\Vc{\flux}=\Vc{\flux}_{iso}$.
This orthogonality has been employed in the MFT \cite{renger2018Entropy,renger2021DiscreteContin.Dyn.Syst.-S,patterson2021ArXiv210314384Math-Ph} to characterize the quasi-potential and the gradient flow aspects of equilibrium and nonequilibrium systems as we see in the next section.
The iso-dissipation hypersurface is known as a level surface of the convex function in Hessian geometry \cite{shima2007} and as a central affine surface in affine differential geometry \cite{nomizu1994}. 
In both cases, it works as a central geometric object.
Moreover, this central affine surface also plays an important role in isobaric thermodynamics, where volume can change in conjunction with reactions \cite{sughiyama2022ArXiv220109417Cond-MatPhysicsphysics}.
The Hilbert orthogonality is also defined in the flux space using $\Dissp(\Vc{\flux})$ similarly to $\Dissp(\Vc{\flux}_{A}+\Vc{\flux}_{S})=\Dissp(\Vc{\flux}_{A}-\Vc{\flux}_{S})$ while we do not use it in this work. 
It should be noted that the orthogonality of $\Vc{\tf}_{A}$ and $\Vc{\tf}_{S}$ generally does not mean the orthogonality of the corresponding $\Vc{\flux}_{A}$ and $\Vc{\flux}_{S}$.

\subsection{Information geometric orthogonality}
The second orthogonality comes from the information geometry \cite{shima2007,amari2016,nielsen2020Entropy}.
For three flux vectors $\Vc{\flux}, \Vc{\flux}', \Vc{\flux}''\in \Jspace$ satisfying $\Vc{\flux}=\Vc{\flux}'+\Vc{\flux}''$, the Bregman divergence $\BD[\Vc{\flux}\|\Vc{0}]$ can be decomposed as
\begin{align}
    \BD[\Vc{\flux}\|\Vc{0}]=\BD[\Vc{\flux}\|\Vc{\flux}'']+\BD[\Vc{\flux}''\|\Vc{0}] +\langle \Vc{\flux},  \Vc{\tf}''\rangle,
\end{align}
where $\Vc{\tf}''$ is the Legendre dual of $\Vc{\flux}''$.
The second orthogonality of $\Vc{\flux}'$ and $\Vc{\flux}''$ is defined by $\langle\Vc{\flux}',\Vc{\tf}'' \rangle=0$.
Then the following generalized Pythagorean theorem (GPT) holds:
\begin{align}
    \BD[\Vc{\flux}\|\Vc{0}]=\BD[\Vc{\flux}\|\Vc{\flux}'']+\BD[\Vc{\flux}''\|\Vc{0}]. \label{eq:GPT1}
\end{align}
For quadratic dissipation functions, the GPT is reduced to the conventional Pythagorean theorem:
\begin{align}
    \|\Vc{\flux}\|^{2}_{{M}}=\|\Vc{\flux}'\|^{2}_{{M}}+\|\Vc{\flux}''\|^{2}_{{M}}.
\end{align}
Thus, \eqnref{eq:GPT1} is a generalization of the usual orthogonality.
We call this orthogonality the information geometric orthogonality.
Similarly, for $\Vc{\tf}=\Vc{\tf}^{\dagger}+\Vc{\tf}^{\ddagger}$, the dual orthogonality between $ \Vc{\tf}^{\dagger},\Vc{\tf}^{\ddagger} \in \Fspace$ is also defined by $\langle\Vc{\flux}^{\dagger},\Vc{\tf}^{\ddagger} \rangle=0$, leading to the dual version of GPT:
\begin{align}
    \BD^{*}[\Vc{\tf}\|\Vc{0}]=\BD^{*}[\Vc{\tf}\|\Vc{\tf}^{\dagger}]+\BD^{*}[\Vc{\tf}^{\dagger}\|\Vc{0}].\label{eq:GPT2}
\end{align}

The Bregman divergence and the information geometric orthogonality are the central geometric tools in information geometry for analyzing manifolds of statistical models and for conducting statistically meaningful projections onto submanifolds \cite{amari2016,nielsen2020Entropy}.
In addition, they have also been recently employed in the thermodynamics of diffusion processes, MJP, and CRN \cite{ito2018Phys.Rev.Lett.,kolchinsky2021Phys.Rev.X,ohga2021ArXiv211211008Cond-Mat,kobayashi2021ArXiv211214910Phys.,sughiyama2022ArXiv220109417Cond-MatPhysicsphysics,sughiyama2022ArXiv220109417Cond-MatPhysicsphysics}. 
However, these works investigate the information and Hessian geometric structures of state space, i.e., the space of the probability vector $\Vc{p}$ or the concentration vector $\Vc{x}$, in which the thermodynamic potential functions such as the Gibbs free energy work as the convex potential function inducing the Hessian structure.
As a result, the Hessian structure of the state space captures the energetic or equilibrium aspects of the systems.
In contrast, the Hessian structure of the flux-force space in this work captures the kinetic and nonequilibrium aspects of the systems.

It should be noted that the Hilbert and information geometric orthogonalities are reduced to usual orthogonality when the dissipation functions are quadratic (see also Appendix ).

\section{Chemical Reaction Dynamics and Entropy Production Decompositions}\label{sec:CRN}
\begin{figure}[t]
\includegraphics[width=0.48\textwidth, bb=0 0 800 500]{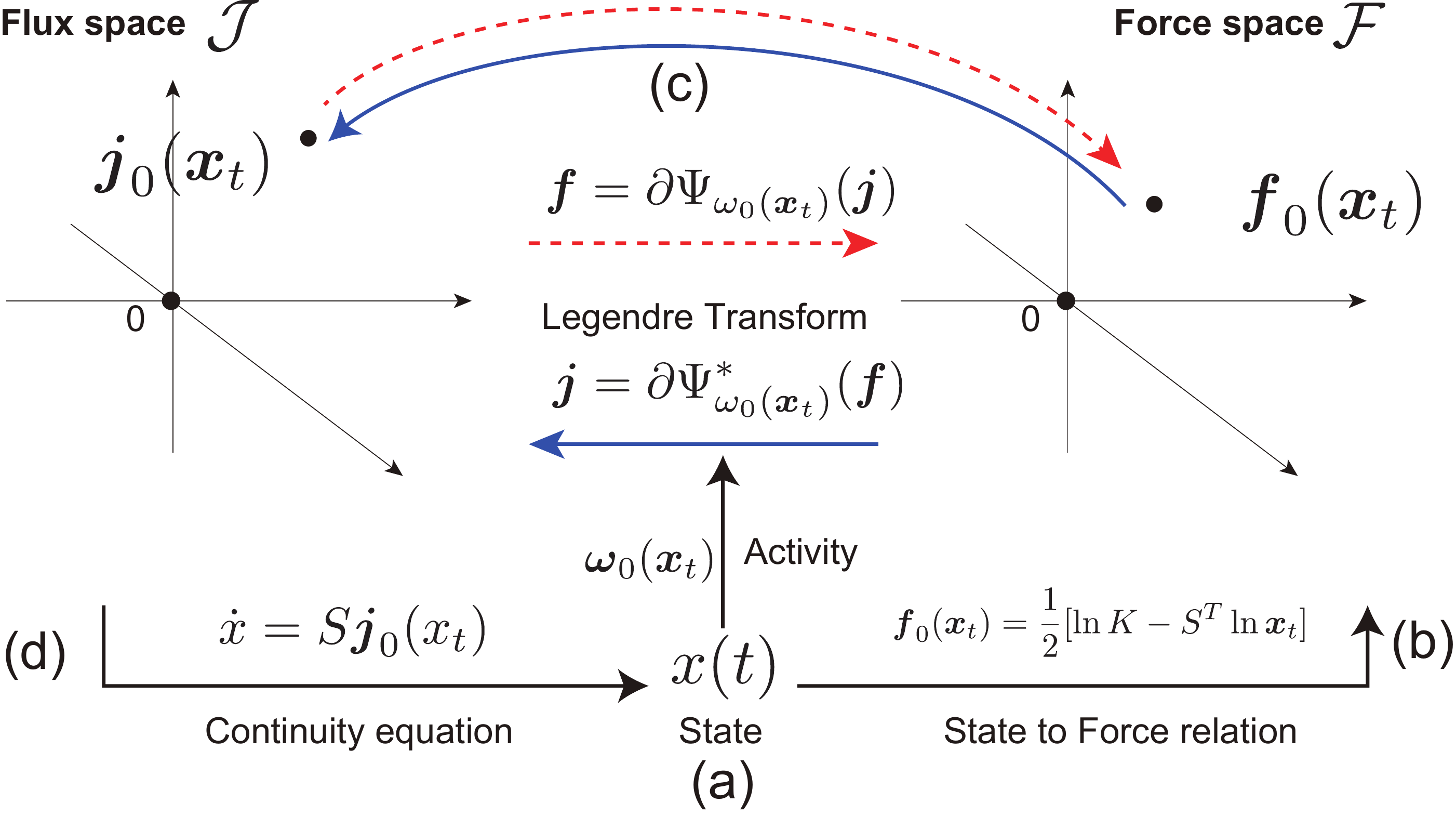}
\caption{Schematic representation of the flux-force relationship for CRN with mass action kinetics and induced CRE (\eqnref{eq:CRN}). Depending on the current state $\Vc{x}(t)$, (a), the force $\Vc{\tf}_{0}(\Vc{x}_{t})$ is induced (b). 
The force is mapped to the corresponding flux $\Vc{\flux}_{0}(\Vc{x}_{t})$ by the Legendre transformation $\partial\Dissp^{*}_{\Vc{\frenecy}_{0}(\Vc{x}_{t})}[\Vc{\tf}_{0}(\Vc{x}_{t})]$ (c). The flux induces the change in $\Vc{x}_{t}$ via the continuity equation (d).}
\label{fig:CRN_Dual}
\end{figure}

In this section, we demonstrate how the two generalizations of orthogonality enable us to extend the different entropy production decompositions to CRN (and MJP).
From the explicit functional forms of the one-way fluxes under mass action kinetics (\eqnref{eq:CRN_rate}), the total flux, force, and activity defined in \eqnref{eq:flux_force} and \eqnref{eq:frenecy} are expressed as
\begin{align}
\begin{aligned}
\Vc{\flux}_{0}(\Vc{x};\Vc{\kcoef}^{\pm})&=(\Vc{\kcoef}^{+}\circ \B_{+}^{\Transpose}-\Vc{\kcoef}^{-}\circ \B_{-}^{\Transpose})\Vc{x}^{\cmMatrix^{\Transpose}}, \\
\Vc{\tf}_{0}(\Vc{x};\Vc{\kcoef}^{\pm})&=\frac{1}{2}\left[\ln\frac{\Vc{\kcoef}^{+}}{\Vc{\kcoef}^{-}}-\stoiMatrix^{\Transpose}\ln\Vc{x}\right], \\
\Vc{\frenecy}_{0}(\Vc{x};\Vc{\kcoef}^{\pm})&=2\sqrt{\Vc{\kcoef}^{+}\circ\Vc{\kcoef}^{-}}\circ \Vc{x}^{[\cmMatrix(\B^{+}+\B^{-})]^{\Transpose}/2}.
\end{aligned}\label{eq:jffMAK}
\end{align}
We use the subscript $(.)_{0}$ to designate the flux, force, and activity given by these particular forms.
With these specific forms of force and activity, the CRN dynamics is described as a generalized flow (\eqnref{eq:CRN} and \fgref{fig:CRN_Dual}).

We further transform the kinetic parameters $\Vc{\kcoef}^{\pm}$ into force parts ($\Vc{K}$) and activity part ($\Vc{\kappa}$) as
\begin{align}
    \Vc{\kcoef}^{+}=\Vc{\kappa}\circ \Vc{K}^{1/2}, \qquad \Vc{\kcoef}^{-}=\Vc{\kappa}\circ \Vc{K}^{-1/2},
\end{align}
where $\Vc{\kappa}\defeq \sqrt{\Vc{\kcoef}^{+}\circ\Vc{\kcoef}^{-}}$ and $\Vc{K}\defeq  \Vc{\kcoef}^{+}/\Vc{\kcoef}^{-}$.
Thus, $(\Vc{\kappa},\Vc{K})$ has the same information as $\Vc{\kcoef}^{\pm}$.
Moreover, we can verify that the force and activity are dependent only on $\Vc{K}$ and $\Vc{\kappa}$, respectively, i.e.,  $\Vc{\tf}_{0}(\Vc{x};\Vc{\kcoef}^{\pm})=\Vc{\tf}_{0}(\Vc{x};\Vc{K})$ and $\Vc{\frenecy}_{0}(\Vc{x};\Vc{\kcoef}^{\pm})=\Vc{\frenecy}_{0}(\Vc{x};\Vc{\kappa})$.

\subsection{Equilibrium dynamics}
First, we consider the case in which the dynamics is equilibrium.
For CNR, the equilibrium states are defined as the set $\Variety_{eq}(\Vc{K},\Vc{\kappa})$ satisfying the detailed balance condition (DBC)
\begin{align}
    \Variety_{eq}(\Vc{K},\Vc{\kappa})\defeq \{\Vc{x}|\Vc{\flux}_{0}(\Vc{x};\Vc{K},\Vc{\kappa})=0\}. \label{eq:DBC}
\end{align}
For a parameter set $(\Vc{K},\Vc{\kappa})$ that admits the existence of equilibrium states, i.e., $\Variety_{eq}(\Vc{K},\Vc{\kappa})\neq \emptyset$, the CRN becomes an equilibrium CRN.
The condition $\Variety_{eq}(\Vc{K},\Vc{\kappa})\neq \emptyset$ is satisfied when the parameter $\Vc{K}$ satisfies the Wegscheider equilibrium (EQ) condition \cite{rao2016Phys.Rev.X,sughiyama2021ArXiv211212403Cond-MatPhysicsphysics,kobayashi2021ArXiv211214910Phys.}: 
\begin{align}
\ln\Vc{K} \in \Img \stoiMatrix^{\Transpose}. \label{eq:Wegscheider}
\end{align}
We denote such $\Vc{K}$ as $\Vc{K}_{eq}$. 
Note that the EQ condition is independent of the activity parameter $\Vc{\kappa}$.
Then, for each initial condition $\Vc{x}_{0}>\Vc{0}$, a unique equilibrium state $\Vc{x}_{eq}$ exists, which is determined by the intersection of $\Variety_{eq}(\Vc{K}_{eq})$ and the stoichiometric compatibility class \cite{craciun2009JournalofSymbolicComputation,kobayashi2021ArXiv211214910Phys.}
\begin{align}
\Polytope(\Vc{x}_{0})\defeq\{\Vc{x}|(\Vc{x}-\Vc{x}_{0})\in \Img[\stoiMatrix]\}
\end{align}
as $\Vc{x}_{eq}=\Variety_{eq}(\Vc{K}_{eq})\cap\Polytope(\Vc{x}_{0})$ \footnote{The uniqueness of the intersection is a consequence of Hessian geometric structure on the concentration-chemical potential space. } \cite{craciun2009JournalofSymbolicComputation,kobayashi2021ArXiv211214910Phys.}.
Moreover, from the DBC, we have $\ln\Vc{K}_{eq}=\stoiMatrix^{\Transpose}\ln \Vc{x}_{eq}$.
Then, the force $\Vc{\tf}_{0}(\Vc{x})$ of the equilibrium dynamics is represented as 
\begin{align}
    \Vc{\tf}_{0}(\Vc{x})&=-\frac{1}{2}\stoiMatrix^{\Transpose}\ln\left(\frac{\Vc{x}}{\Vc{x}_{eq}}\right)=-\frac{1}{2}\stoiMatrix^{\Transpose}\partial_{\Vc{x}}\Gibbs_{eq}(\Vc{x}),
\end{align}
where $\Gibbs_{eq}(\Vc{x})=\KL[\Vc{x}\|\Vc{x}_{eq}]\defeq  \Vc{x}^{\Transpose}\ln\frac{\Vc{x}}{\Vc{x}_{eq}}-\Vc{1}^{\Transpose}(\Vc{x}-\Vc{x}_{eq})$ is the Gibbs free energy of the equilibrium CRN.
The matrix $\stoiMatrix^{\Transpose}$ can be considered as the discrete version of the gradient $\Grad=\stoiMatrix^{\Transpose}$ for the chemical hypergraph because $\stoiMatrix^{\Transpose}$ is the adjoint operator of $\stoiMatrix=-\Div$.
Thus, the equilibrium force is a gradient of $\Gibbs_{eq}(\Vc{x})$:
\begin{align}
    \Vc{\tf}_{0}(\Vc{x})&=-\frac{1}{2}\Grad\left[\partial_{\Vc{x}}\Gibbs_{eq}(\Vc{x})\right]. \label{eq:eq_force}
\end{align}
Additionally, the Wegscheider condition (\eqnref{eq:Wegscheider}) is interpreted as $\ln\Vc{K}_{eq} \in \Img [\Grad]$.
Then the dynamics (\eqnref{eq:CRN}) under the EQ parameter condition becomes a generalized gradient flow (GF) of $\Gibbs_{eq}(\Vc{x})$ \cite{mielke2014PotentialAnal}:
\begin{align}
    \dot{\Vc{x}}=\stoiMatrix\Vc{\flux}_{0}(\Vc{x}) &= \stoiMatrix\partial \Dissp^{*}_{\Vc{x}}\left[-\frac{1}{2}\stoiMatrix^{\Transpose}\partial_{\Vc{x}}\Gibbs_{eq}(\Vc{x})\right]\label{eq:CRNeqGF} \\
    &= -\Div\,\partial \Dissp^{*}_{\Vc{x}}\left[-\frac{1}{2}\Grad\,\left[\partial_{\Vc{x}}\Gibbs_{eq}(\Vc{x})\right]\right],
\end{align}
where $\Dissp^{*}_{\Vc{x}}\defeq\Dissp^{*}_{\Vc{\frenecy}_{0}(\Vc{x})}$.
Because $\Vc{x}(t)$ is the gradient flow of $\Gibbs_{eq}(\Vc{x}(t))$, $\dd\Gibbs_{eq}(\Vc{x}(t))/\dt$ is always non-positive as
\begin{align}
-\frac{\dd \Gibbs_{eq}(\Vc{x})}{\dt}
&=\langle\dot{\Vc{x}},-\partial_{\Vc{x}}\Gibbs_{eq}(\Vc{x}) \rangle=\langle\stoiMatrix\Vc{\flux}_{0}(\Vc{x}), -\partial_{\Vc{x}}\Gibbs_{eq}(\Vc{x})\rangle\notag\\
&=\langle\Vc{\flux}_{0}(\Vc{x}), -\stoiMatrix^{\Transpose}\partial_{\Vc{x}}\Gibbs_{eq}(\Vc{x})\rangle=2\langle\Vc{\flux}_{0}(\Vc{x}),\Vc{\tf}_{0}(\Vc{x})\rangle\notag\\
&=2\left[\Dissp_{\Vc{x}}(\Vc{\flux}_{0}(\Vc{x}))+\Dissp_{\Vc{x}}^{*}(\Vc{\tf}_{0}(\Vc{x}))\right]=\EPR \ge 0,\label{eq:dFeq}
\end{align}
where we use \eqnref{eq:LegendreIdentity} and \eqnref{eq:EPR_def}.
By integration, we have the relation between the change of free energy and the entropy production:
\begin{align}
    \Gibbs_{eq}(\Vc{x}(t))-\Gibbs_{eq}(\Vc{x}(0))=\int_{t'=0}^{t}\EPR_{t'}\dd t'.
\end{align}

\subsection{Complex balanced dynamics, Hilbert orthogonality, and Hatano-Sasa decomposition}
\begin{figure}[t]
\includegraphics[width=0.48\textwidth, bb=0 0 800 450]{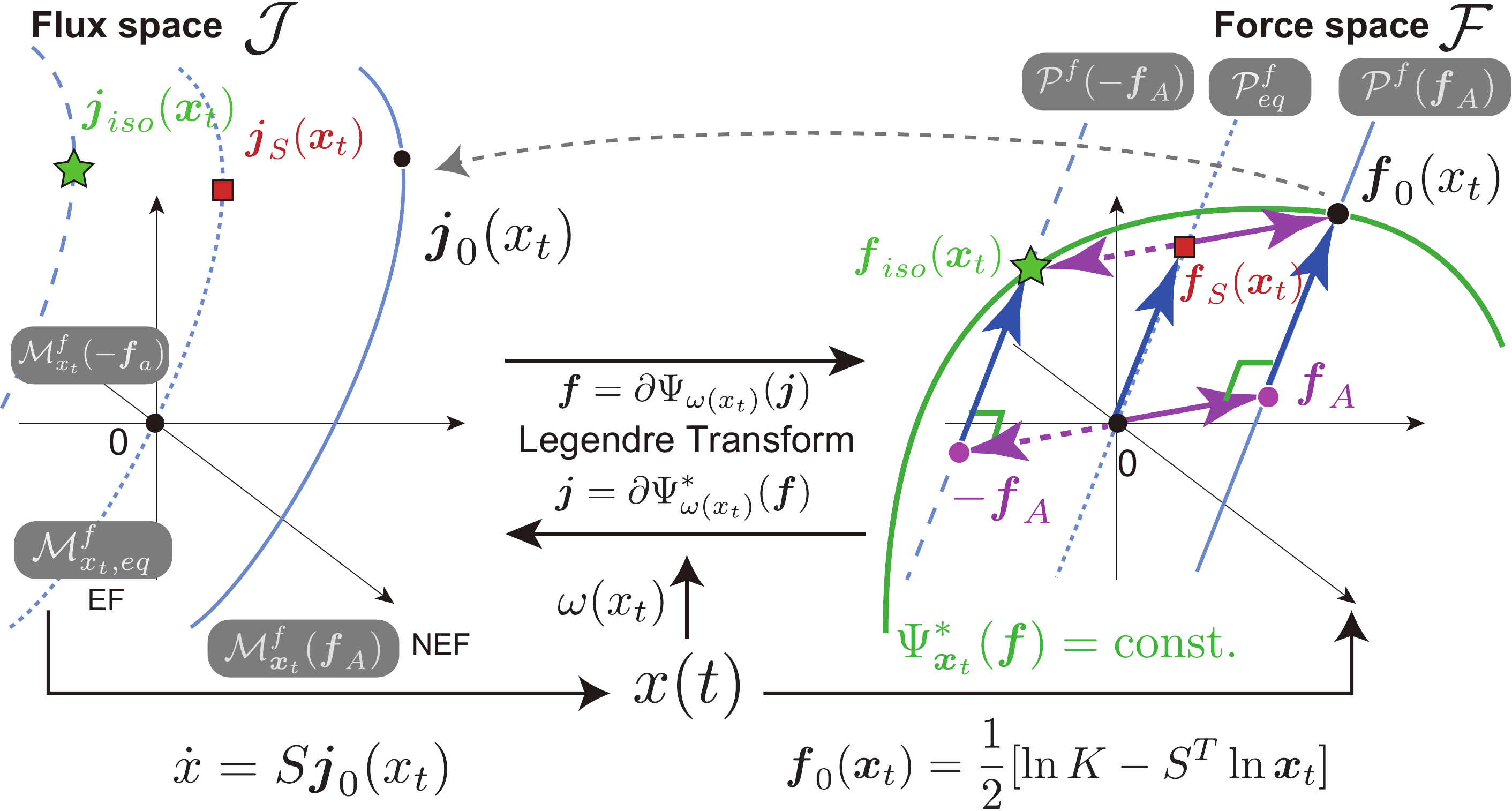}
\caption{Schematic illustration of the Hilbert orthogonality between $\Vc{\tf}_{S}(\Vc{x}_{t})$ (blue arrows) and $\Vc{\tf}_{A}$(magenta arrows). 
The green solid curve represents the iso-dissipation hypersurface:  $\Dissp^{*}_{\Vc{x}}(\Vc{\tf})=\Dissp^{*}_{\Vc{x}}(\Vc{\tf}_{0}(\Vc{x}_{t}))=\mathrm{const.}$. 
The black circle, red square, and green star in $\Fspace$ are $\Vc{\tf}_{0}(\Vc{x}_{t})$, $\Vc{\tf}_{S}(\Vc{x}_{t})$, and $\Vc{\tf}_{iso}(\Vc{x}_{t})$, respectively.
The black circle, red square, and green star in $\Jspace$ are their Legendre transform, i.e., $\Vc{\flux}_{0}(\Vc{x}_{t})$, $\Vc{\flux}_{S}(\Vc{x}_{t})$, and $\Vc{\flux}_{iso}(\Vc{x}_{t})$, respectively. 
The light blue lines in $\Fspace$ are the NEF subspaces, $\Polytope^{f}(\Vc{\tf}_{A})$(solid line) and $\Polytope^{f}(-\Vc{\tf}_{A})$ (dashed line), and the EF subspace $\Polytope^{f}_{eq}$ (dotted line). 
In $\Jspace$, the corresponding NEF manifolds, $\Variety^{f}_{\Vc{x}}(\Vc{\tf}_{A})$ (the light solid blue curve) and $\Variety^{f}_{\Vc{x}}(-\Vc{\tf}_{A})$ (the light dashed blue curve), and the EF manifold $\Variety^{f}_{\Vc{x},eq}$ (the light dotted blue curve) are also depicted.}
\label{fig:Hilbert_Orth}
\end{figure}
Next, we consider the nonequilibrium complex balanced (CB) steady state, and the EPR decomposition induced by the Hilbert orthogonality.
The result here is a reinterpretation of the result in \cite{renger2021DiscreteContin.Dyn.Syst.-S,renger2018Entropy,patterson2021ArXiv210314384Math-Ph} from the viewpoint of Hessian geometry.  Nevertheless, it is worth including because the decomposition is contrasted with the information geometric decomposition in the next subsection.

CB states of CRN are defined as the set satisfying
\begin{align}
    \Variety_{cb}(\Vc{K},\Vc{\kappa})\defeq \{\Vc{x}|\B \Vc{\flux}_{0}(\Vc{x};\Vc{K},\Vc{\kappa})=\Vc{0}\}. \label{eq:CBC}
\end{align}
A parameter set $(\Vc{K}_{cb},\Vc{\kappa}_{cb})$ induces CB dynamics if the CB condition $\Variety_{cb}(\Vc{K}_{cb},\Vc{\kappa}_{cb})\neq\emptyset$ holds.
Then, for each initial condition $\Vc{x}_{0}$, the CB steady state $\Vc{x}_{cb}$ is uniquely determined by the intersection of $\Variety_{cb}(\Vc{K}_{cb},\Vc{\kappa}_{cb})$ and the stoichiometric compatibility class $\Polytope(\Vc{x}_{0})$ as $\Vc{x}_{cb}=\Variety_{cb}(\Vc{K}_{cb},\Vc{\kappa}_{cb})\cap\Polytope(\Vc{x}_{0})$ \cite{craciun2009JournalofSymbolicComputation,kobayashi2021ArXiv211214910Phys.}.

The CB state inherits important properties of the equilibrium state such as the uniqueness, global stability, and the gradient-flow-like aspect \cite{craciun2009JournalofSymbolicComputation,ge2016ChemicalPhysics,rao2016Phys.Rev.X,kobayashi2021ArXiv211214910Phys.}.
By definition, an EQ state is a CB state, but not vice versa.
Additionally, the steady states of MJP always satisfy the CB condition because $\B \Vc{\flux}_{0}(\Vc{x};\Vc{K},\Vc{\kappa})=\Vc{0}$ is nothing but the steady state condition for MJP (\eqnref{eq:MJP_rateEq}).
Thus, MJP are unconditionally complex balanced. 
In contrast, CRN are not always complex balanced depending on the parameter value.

Using $\Vc{x}_{cb}$, the force $\Vc{\tf}_{0}(\Vc{x})$ is represented as 
\begin{align}
    \Vc{\tf}_{0}(\Vc{x})&=-\frac{1}{2}\stoiMatrix^{\Transpose}\partial_{\Vc{x}}\Gibbs_{cb}(\Vc{x})+\frac{1}{2}\ln\frac{\Vc{K}_{cb}}{\tilde{\Vc{K}}_{eq}}, \label{eq:force0}
\end{align}
where $\tilde{\Vc{K}}_{eq}\defeq \stoiMatrix^{\Transpose}\ln \Vc{x}_{cb}$ and $\Gibbs_{cb}(\Vc{x})\defeq \KL[\Vc{x}\|\Vc{x}_{cb}]$.
Let $\Vc{\tf}_{S}(\Vc{x})\defeq-\frac{1}{2}\stoiMatrix^{\Transpose}\partial_{\Vc{x}}\Gibbs_{cb}(\Vc{x})$ and $\Vc{\tf}_{A}\defeq\frac{1}{2}\ln\frac{\Vc{K}_{cb}}{\tilde{\Vc{K}}_{eq}}$.
Then, for all $\Vc{x}\in\X$, the generalized Hilbert orthogonality holds (\fgref{fig:Hilbert_Orth}):
\begin{align}
    \Dissp^{*}_{\Vc{x}}(\Vc{\tf}_{S}(\Vc{x})+\Vc{\tf}_{A})=\Dissp^{*}_{\Vc{x}}(\Vc{\tf}_{S}(\Vc{x})-\Vc{\tf}_{A}),\label{eq:HB_orth_fs_fa}
\end{align}
because of the CB condition $\B\Vc{\flux}_{0}(\Vc{x}_{cb})=\Vc{0}$ \cite{renger2018Entropy,renger2021DiscreteContin.Dyn.Syst.-S,patterson2021ArXiv210314384Math-Ph} (see Appendix  for a proof).
Thus, \eqnref{eq:HBdecomp} admits a nonnegative decomposition of the EPR: 
\begin{align}
\EPR(\Vc{x})=\EPR_{ex}^{GF}(\Vc{x})+\EPR_{hk}^{GF}(\Vc{x}), \label{eq:GFdecomp}   
\end{align}
 where $\EPR_{ex}^{GF}(\Vc{x}) \defeq 2 \langle\Vc{\flux}_{0}(\Vc{x}),\Vc{\tf}_{S}(\Vc{x}) \rangle$ and $\EPR_{hk}^{GF}(\Vc{x})\defeq 2 \langle\Vc{\flux}_{0}(\Vc{x}),\Vc{\tf}_{A} \rangle$.
Owing to this nonnegative decomposition, $\Gibbs_{cb}(\Vc{x})$ works as a Lyapunov function (quasi-potential):
\begin{align}
-\frac{\dd \Gibbs_{cb}(\Vc{x}_{t})}{\dt}
&=\EPR_{ex}^{GF}(\Vc{x}_{t})=\BD[\Vc{\flux}_{0}(\Vc{x}_{t})\|\Vc{\tf}_{iso}(\Vc{x}_{t})] \ge 0,\notag
\end{align}
and $\EPR_{ex}^{GF}(\Vc{x})$ behaves like an excess EPR, i.e., $\lim_{t\to\infty}\EPR_{ex}^{GF}(\Vc{x}_{t})=0$.
Thus, $\Gibbs_{cb}(\Vc{x})$ as a Lyapunov function guarantees the stability of the CB state $\Vc{x}_{cb}$ and the gradient-flow-like property of the CB dynamics.

This decomposition is equivalent to those proposed for CRN in \cite{ge2016ChemicalPhysics,rao2016Phys.Rev.X} as a generalization of the Hatano-Sasa decomposition \cite{hatano2001Phys.Rev.Lett.}.
Therefore, this result is just a reinterpretation of the previous ones.
However, owing to the geometric perspective endowed by Hessian geometry, the non-negativity of $\EPR_{ex}^{GF}(\Vc{x})$ and $\EPR_{hk}^{GF}(\Vc{x})$ becomes transparent, the proof of which in previous works required more technical and less transparent computations\cite{ge2016ChemicalPhysics, rao2016Phys.Rev.X}.

\subsection{Information geometric orthogonality and Maes-Neto\v{c}n\`{y} decomposition}
\begin{figure}[t]
\includegraphics[width=0.48\textwidth, bb=0 0 800 450]{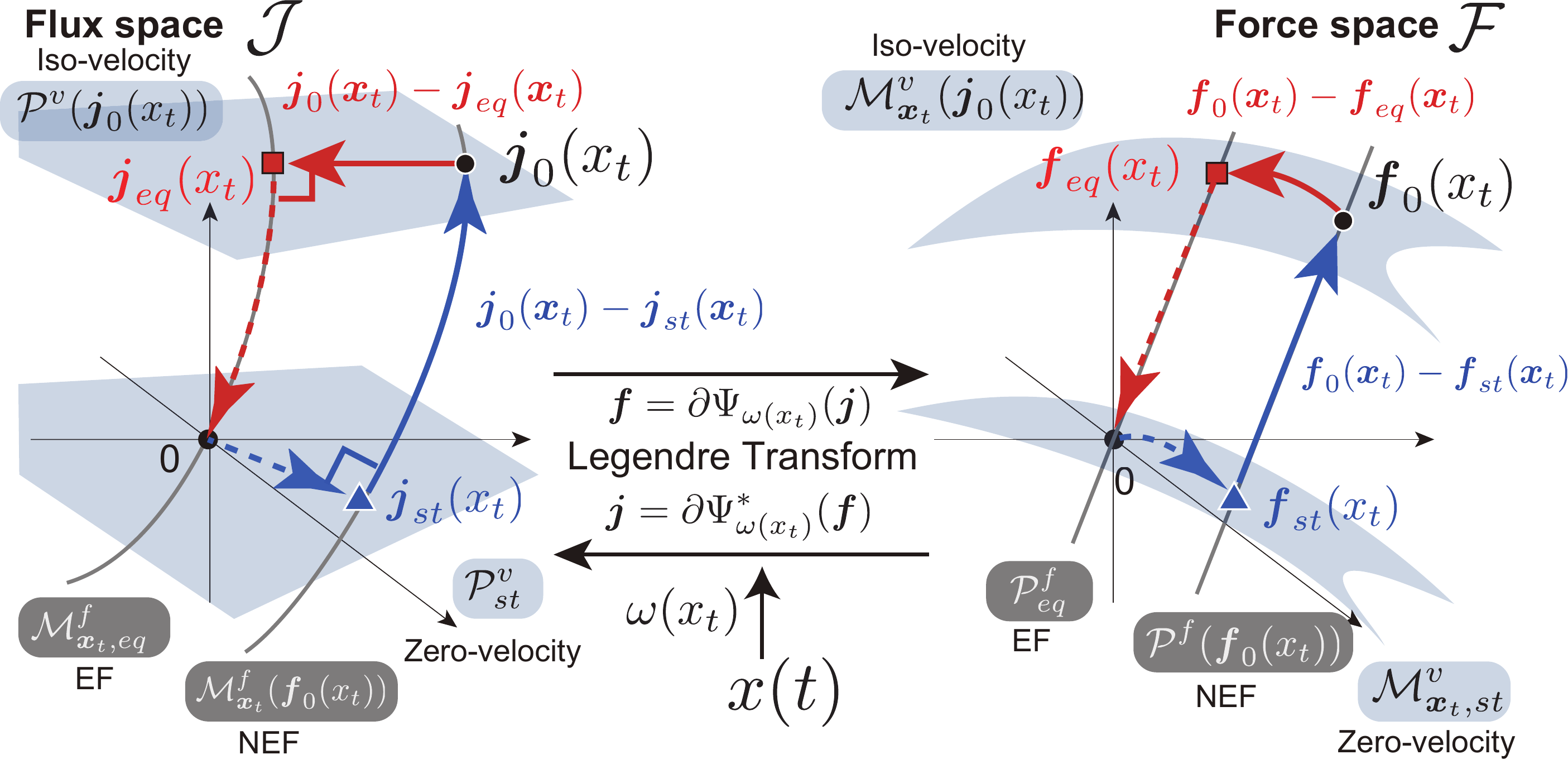}
\caption{Schematic illustration of the information geometric orthogonalities.
(Left) The orthogonality between $\Vc{\flux}_{st}(\Vc{x}_{t})$ and $\Vc{\flux}_{0}(\Vc{x}_{t})-\Vc{\flux}_{st}(\Vc{x}_{t})$ (blue arrows) and the dual orthogonality between
$\Vc{\flux}_{eq}(\Vc{x}_{t})$ and $\Vc{\flux}_{0}(\Vc{x}_{t})-\Vc{\flux}_{eq}(\Vc{x}_{t})$ (red arrows) in $\Jspace$. 
The blue planes are the iso-velocity subspace $\Polytope^{v}(\Vc{\flux}_{0}(\Vc{x}_{t}))$ (the upper plane) and the zero-velocity subspace $\Polytope^{v}_{st}$ (the lower plane).
The grey curves are the NEF manifold $\Variety^{f}_{\Vc{x}}(\Vc{\tf}_{0}(\Vc{x}_{t}))$ and the EF manifold $\Variety^{f}_{\Vc{x},eq}$. 
(Right) The same orthogonalities shown in $\Fspace$ space. 
The orthogonality between $\Vc{\tf}_{st}(\Vc{x}_{t})$ and $\Vc{\tf}_{0}(\Vc{x}_{t})-\Vc{\tf}_{st}(\Vc{x}_{t})$ (blue arrows) and the dual orthogonality between $\Vc{\tf}_{eq}(\Vc{x}_{t})$ and $\Vc{\tf}_{0}(\Vc{x}_{t})-\Vc{\tf}_{eq}(\Vc{x}_{t})$ (red arrows)
 in $\Fspace$. 
The blue surfaces are the iso-velocity manifold $\Variety^{v}_{\Vc{x}_{t}}(\Vc{\flux}_{0}(\Vc{x}_{t}))$ (the upper surface) and the zero-velocity manifold $\Variety^{v}_{\Vc{x}_{t},st}$ (the lower surface). 
The grey lines are the NEF subspaces $\Polytope^{f}(\Vc{\tf}_{0}(\Vc{x}_{t}))$ and the EF subspace $\Polytope^{f}_{\Vc{x},eq}$. 
}
\label{fig:Info_Orth}
\end{figure}

While the gradient flow aspect and associated EPR decomposition by the quasi-potential have been investigated in CRN theory, MFT, and applied mathematics \cite{ge2016ChemicalPhysics,rao2016Phys.Rev.X,renger2021DiscreteContin.Dyn.Syst.-S,craciun2009JournalofSymbolicComputation}, the decomposition does not capture other important thermodynamic aspects such as minimum dissipation or minimum entropy production principles.
In addition, the Hilbert orthogonality for $\Gibbs_{cb}(\Vc{x})$ (\eqnref{eq:HB_orth_fs_fa}) holds only when the CB condition is satisfied.
In CRN, we can have non-complex-balanced steady states depending on the parameter values.

As another type of the EPR decomposition, we have Maes-Neto\v{c}n\`{y} decomposition, which generalizes the Komatsu-Nakagawa-Sasa-Tasaki approach \cite{maes2014JStatPhys,komatsu2008Phys.Rev.Lett.}.
Moreover, its geometric interpretation for overdamped diffusion processes was recently provided in terms of the formal Riemannian geometry of the flux-force space \cite{dechant2022ArXiv210912817Cond-Mat}.
Its extension to MJP and CRN has yet to be achieved because of their nonquadratic dissipation functions.
We show that the information geometric orthogonality is central to this extension.

In the Maes-Neto\v{c}n\`{y} decomposition, for a given $\Vc{x}_{t}$\footnote{$\Vc{x}_{t}$ is not necessarily a steady state here.}, the non-equilibrium flux $\Vc{\flux}_{0}(\Vc{x})$ is decomposed into the quasi-steady flux $\Vc{\flux}_{st}(\Vc{x})$ and the remaining part as $\Vc{\flux}_{0}(\Vc{x})=\Vc{\flux}_{st}(\Vc{x})+(\Vc{\flux}_{0}(\Vc{x})-\Vc{\flux}_{st}(\Vc{x}))$.
The quasi-steady flux $\Vc{\flux}_{st}(\Vc{x})$ is the flux that makes a given state $\Vc{x}$ steady obtained by modulating the conserved force, i.e., being generated by the gradient of potential force in the case of overdamped diffusion.
This means that $\Vc{\flux}_{st}(\Vc{x}) \in \Ker[\Div]$ and $ \Vc{\tf}_{0}(\Vc{x})-\Vc{\tf}_{st}(\Vc{x})\in \Img [\Grad]$ hold.
Furthermore, $\Vc{\flux}_{st}(\Vc{x})$ is also characterized as the minimum EPR flux under the constraint of $\Vc{\tf}_{0}(\Vc{x})-\Vc{\tf}_{st}(\Vc{x})\in \Img [\Grad]$ \cite{maes2014JStatPhys}.
From the viewpoint of vector calculus, 
the decomposition is a special case of the Helmholtz-Hodge decomposition\cite{bhatia2013IEEETrans.Vis.Comput.Graph.}.
With the the Riemannian inner product structure between $\Jspace$ and $\Fspace$ for overdamped diffusion, we can also regard $\Vc{\flux}_{st}(\Vc{x})$ as the orthogonal projection of $\Vc{\flux}_{0}(\Vc{x})$ onto $\Ker[\Div]$ along $\Img [\Grad]$.
Because the Maes-Neto\v{c}n\`{y} decomposition is tightly linked to the orthogonality in the inner product space, it is not trivial how to extend the decomposition to other systems such as CRN (and MJP) with nonquadratic dissipation functions.
We show how information-geometric orthogonality can resolve this problem.

First, we define a linear affine subspace $\Polytope^{v}(\Vc{\flux})$ as 
\begin{align}
\Polytope^{v}(\Vc{\flux})=\left\{\Vc{\flux}'|\Vc{\flux}-\Vc{\flux}' \in \Ker\stoiMatrix\right\} \subset \Jspace.
\end{align}
Because $\Vc{\flux}' \in \Polytope^{v}(\Vc{\flux})$ satisfies $\stoiMatrix\Vc{\flux} = \stoiMatrix\Vc{\flux}'$, $\Polytope^{v}(\Vc{\flux})$ is the set of the fluxes that induces the same instantaneous velocity $\dot{\Vc{x}}$ as $\Vc{\flux}$.
Thus, we call $\Polytope^{v}(\Vc{\flux})$ an iso-velocity subspace in $\Jspace$.
Because $\dot{\Vc{x}}=0$ for the steady state flux, we also denote $\Polytope^{v}_{st}\defeq \Polytope^{v}(\Vc{0})$ as the steady state (zero-velocity) subspace.

As the complementary subspace of $\Polytope^{v}(\Vc{\flux})$, we also define a linear affine subspace $\Polytope^{f}(\Vc{\tf}) \subset\Fspace$ as  
\begin{align}
\Polytope^{f}(\Vc{\tf})\defeq \left\{\Vc{\tf}'|\Vc{\tf}-\Vc{\tf}'\in\Img\stoiMatrix^{\Transpose} \right\}.\label{eq:NEF_subspace}
\end{align}
Because $\stoiMatrix^{\Transpose}$ is the gradient of a CRN, i.e., $\Grad=\stoiMatrix^{\Transpose}$, $\Polytope^{f}(\Vc{\tf})$ is the subspace obtained by shifting $\Img[\Grad]=\Img[\stoiMatrix^{\Transpose}]$ by $\Vc{\tf}$.
We call $\Polytope^{f}(\Vc{\tf})$ an nonequilibrium force (NEF) subspace in $\Fspace$ if $\Vc{\tf}\notin \Img\stoiMatrix^{\Transpose}$ and the equilibrium force (EF) subspace if $\Vc{\tf}\in \Img\stoiMatrix^{\Transpose}$ because, if the equilibrium condition is satisfied, $\Vc{\tf}_{0}(\Vc{x})$ is always in $\Img[\stoiMatrix^{\Transpose}]$ as in \eqnref{eq:eq_force}.
We denote the EF subspace by $\Polytope^{f}_{eq}\defeq \Polytope^{f}(\Vc{0})$.

To obtain the generalized version of $\Vc{\flux}_{st}(\Vc{x})$, we transform $\Polytope^{f}(\Vc{\tf})$ and $\Polytope^{f}_{eq}$ to $\Jspace$ via the Legendre transformation:
\begin{align}
    \Variety^{f}_{\Vc{x}}(\Vc{\tf})&\defeq \partial \Dissp^{*}_{\Vc{x}}\left[\Polytope^{f}(\Vc{\tf})\right]\subset \Jspace,\\
    \Variety^{f}_{\Vc{x},eq}&\defeq \partial \Dissp^{*}_{\Vc{x}}\left[\Polytope^{f}_{eq}\right]\subset \Jspace.
\end{align}
It should be noted that $\Variety^{f}_{\Vc{x}}(\Vc{\tf})$ depends on the current state $\Vc{x}$ of the system even though $\Polytope^{f}(\Vc{\tf})$ does not because the Legendre transformation is dependent on $\Vc{x}$ via the activity $\Vc{\frenecy}(\Vc{x})$.
From the definition, $\Variety^{f}_{\Vc{x}}(\Vc{\tf})$ and $\Variety^{f}_{\Vc{x},eq}$ are the set of nonequilibrium and equilibrium fluxes obtained by modulating the gradient force.

Then the generalized quasi-steady flux $\Vc{\flux}_{st}(\Vc{x})$ is obtained by the intersection of $\Polytope^{v}_{st}$ and $\Variety^{f}_{\Vc{x}}(\Vc{\tf}_{0}(\Vc{x}))$: 
\begin{align}
    \Vc{\flux}_{st}(\Vc{x}) &\defeq\Polytope^{v}_{st}\cap \Variety^{f}_{\Vc{x}}(\Vc{\tf}_{0}(\Vc{x})). \label{eq:def_flux_st}
\end{align}
For the case of overdamped diffusion processes, the Legendre transformation of the corresponding quadratic dissipation function is linear and, therefore, $\Variety^{f}_{\Vc{x}}(\Vc{\tf}_{0}(\Vc{x}))$ becomes a flat subspace.
The existence and uniqueness of $\Vc{\flux}_{st}(\Vc{x})$ follow from the linear algebra.
However, for the generalized $\Vc{\flux}_{st}(\Vc{x})$, the existence and uniqueness of $\Vc{\flux}_{st}(\Vc{x})$ are not guaranteed \textit{a priori}, because $\Variety^{f}_{\Vc{x}}(\Vc{\tf}_{0}(\Vc{x}))$ is a curved manifold as shown in \fgref{fig:Info_Orth} (A).
Nonetheless, $\Vc{\flux}_{st}(\Vc{x})$ in \eqnref{eq:def_flux_st} exists uniquely.
More generally, for any $\Vc{\flux}'$ and $\Vc{\tf}''$, the following intersection is unique and transversal (see Appendix for proof):
\begin{align}
    \Vc{\flux}_{int}(\Vc{\flux}',\Vc{\tf}'')=\Polytope^{v}(\Vc{\flux}') \cap \Variety^{f}_{\Vc{x}}(\Vc{\tf}''). \label{eq:Birch}
\end{align}
This result generalizes the uniqueness of the intersection of complementary subspaces under an inner product structure.
This is one of the notable properties of Hessian geometry.
In information geometry and Hessian geometry, $\Polytope^{v}(\Vc{\flux})$ and $\Variety^{f}_{\Vc{x}}(\Vc{\tf})$ are called dual flat spaces \cite{amari2016,nielsen2020Entropy,shima2007}. 

By virtue of the uniqueness, we obtain a unique decomposition of the flux $\Vc{\flux}_{0}(\Vc{x})$:
\begin{align}
    \Vc{\flux}_{0}(\Vc{x})=\Vc{\flux}_{st}(\Vc{x})+(\Vc{\flux}_{0}(\Vc{x})-\Vc{\flux}_{st}(\Vc{x})). \label{eq:j_st_decomp}
\end{align}
Because $\Vc{\flux}_{st}(\Vc{x}) \in \Ker \stoiMatrix$ and $\Vc{\flux}_{st}(\Vc{x})\in \Variety^{f}_{\Vc{x}}(\Vc{\tf}_{0}(\Vc{x}))$ where $\Variety^{f}_{\Vc{x}}(\Vc{\tf}_{0}(\Vc{x}))=\partial_{\Vc{\flux}}\Dissp^{*}_{\Vc{x}}[\Vc{\tf}_{0}(\Vc{x})+ \Img[\stoiMatrix^{\Transpose}]]$,
$\Vc{\flux}_{st}(\Vc{x})$ is the steady flux that is obtained by modulating the nonequilibrium force $\Vc{\tf}_{0}(\Vc{x})$ by adding the gradient equilibrium force.
We also have $\Vc{\tf}_{0}(\Vc{x})-\Vc{\tf}_{st}(\Vc{x}) \in \Img \stoiMatrix^{\Transpose}$.
Thus, \eqnref{eq:j_st_decomp} is a generalization of Maes-Neto\v{c}n\`{y} decomposition and also of Helmholtz decomposition.
Moreover, from $\langle\Vc{\flux}_{st}(\Vc{x}),\Vc{\tf}_{0}(\Vc{x})-\Vc{\tf}_{st}(\Vc{x})  \rangle =0$, we have the generalized Pythagorean relation: 
\begin{align}
    \BD_{\Vc{x}}^{*}[\Vc{\tf}_{0}(\Vc{x})\|\Vc{0}] &=\BD_{\Vc{x}}^{*}[\Vc{\tf}_{0}(\Vc{x})\|\Vc{\tf}_{st}(\Vc{x})]+\BD_{\Vc{x}}^{*}[\Vc{\tf}_{st}(\Vc{x})\|\Vc{0}],
\end{align}
which is further reduced to a decomposition of the dual dissipation function:
\begin{align}
    \Dissp_{\Vc{x}}^{*}(\Vc{\tf}_{0}(\Vc{x}))&=\BD_{\Vc{x}}^{*}[\Vc{\tf}_{0}(\Vc{x})\|\Vc{\tf}_{st}(\Vc{x})]+\Dissp_{\Vc{x}}^{*}(\Vc{\tf}_{st}(\Vc{x})).\label{eq:Dissp_Decomp_2}
\end{align}
Then, $\Vc{\tf}_{st}$ is characterized variationally as (see Appendix ):  
\begin{align}
    \Vc{\tf}_{st}(\Vc{x}) &= \arg \min_{\Vc{\tf} \in \Polytope^{f}(\Vc{\tf}_{0}(\Vc{x}))} \Dissp_{\Vc{x}}^{*}(\Vc{\tf}).
\end{align}
This variational formula indicates that $\Vc{\tf}_{st}(\Vc{x})$ is the minimum dissipation force that can be obtained by tilting the nonequilibrium force $\Vc{\tf}_{0}(\Vc{x})$ with the equilibrium gradient force.
Thus, $\Vc{\flux}_{st}(\Vc{x})$ is the minimum dissipation (MD) steady state flux.
For quadratic cases where the equivalence of the dissipation functions and the EPR holds (\eqnref{eq:j_st_decomp}), the result reduces to the minimum entropy production principle.

However, for nonquadratic cases, \eqnref{eq:Dissp_Decomp_2} alone does not provide a relevant EPR decomposition because EPR is not determined only by $\Dissp_{\Vc{x}}^{*}(\Vc{\tf}_{0}(\Vc{x}))$.
To resolve the problem, we consider the dual decomposition of $\Vc{\flux}_{0}$ as shown in \fgref{fig:Info_Orth} (A):
\begin{align}
    \Vc{\flux}_{0}(\Vc{x})=\Vc{\flux}_{eq}(\Vc{x})+(\Vc{\flux}(\Vc{x})-\Vc{\flux}_{eq}(\Vc{x})), \label{eq:j_eq_decomp}
\end{align}
where we define $\Vc{\flux}_{eq}(\Vc{x})$ as
\begin{align}
    \Vc{\flux}_{eq}(\Vc{x})\defeq \Polytope^{v}(\Vc{\flux}_{0}(\Vc{x})) \cap \Variety^{f}_{\Vc{x},eq}.
\end{align}
Because $\Vc{\flux}_{eq}(\Vc{x})\in\Polytope^{v}(\Vc{\flux}_{0}(\Vc{x}))$, the flux $\Vc{\flux}_{eq}(\Vc{x})$ induces the same instantaneous velocity $\dot{\Vc{x}}$ as $\Vc{\flux}_{0}(\Vc{x})$. 
Moreover, because $\Vc{\flux}_{eq}(\Vc{x})\in\Variety^{f}_{\Vc{x},eq}$, it is the flux induced by an equilibrium force, i.e., a pure gradient force.
Thus, $\Vc{\flux}_{eq}(\Vc{x})$ is the equilibrium flux that induces the same dynamics as $\Vc{\flux}_{0}(\Vc{x})$. 
It should be noted that $\Vc{\flux}_{0}(\Vc{x})$ is generally induced by a nonequilibrium force.
In addition,  we have $\Vc{\tf}_{eq}(\Vc{x}) \in \Img \stoiMatrix^{\Transpose}$ and 
$\Vc{\flux}(\Vc{x})-\Vc{\flux}_{eq}(\Vc{x}) \in \Ker \stoiMatrix$ from $\Vc{\flux}_{eq}(\Vc{x})\in \Variety^{f}_{\Vc{x},eq}$ and $\Vc{\flux}_{eq}(\Vc{x})\in \Polytope^{v}(\Vc{\flux}_{0}(\Vc{x}))$, respectively.
Thus, \eqnref{eq:j_eq_decomp} is another generalization of Helmholtz-Hodge decomposition of the flux $\Vc{\flux}_{0}(\Vc{x})$.
Moreover, owing to $\langle\Vc{\flux}(\Vc{x})-\Vc{\flux}_{eq}(\Vc{x}), \Vc{\tf}_{eq}(\Vc{x}) \rangle =0$, we have the Pythagorean relation: 
\begin{align}
    \BD_{\Vc{x}}[\Vc{\flux}_{0}(\Vc{x})\| \Vc{0}]&=\BD_{\Vc{x}}[\Vc{\flux}_{0}(\Vc{x})\|\Vc{\flux}_{eq}(\Vc{x})]+\BD_{\Vc{x}}[\Vc{\flux}_{eq}(\Vc{x});\Vc{0}]\label{eq:PGR},
\end{align}
which is reduced to a decomposition of the primal dissipation function:
\begin{align}
    \Dissp_{\Vc{x}}(\Vc{\flux}_{0}(\Vc{x}))&=\BD_{\Vc{x}}[\Vc{\flux}_{0}(\Vc{x})\|\Vc{\flux}_{eq}(\Vc{x})]+\Dissp_{\Vc{x}}(\Vc{\flux}_{eq}(\Vc{x})).\label{eq:Dissp_Decomp_1}
\end{align}
From this, $\Vc{\flux}_{eq}$ can be characterized variationally (see Appendix for proof):  
\begin{align}
    \Vc{\flux}_{eq}(\Vc{x}) &= \arg \min_{\Vc{\flux} \in \Polytope^{v}(\Vc{\flux}_{0}(\Vc{x}))} \Dissp_{\Vc{x}}(\Vc{\flux}).
\end{align}
This variational formula means that, among all fluxes that induce the same velocity as $\Vc{\flux}_{0}(\Vc{x})$, the equilibrium flux $\Vc{\flux}_{eq}(\Vc{x})$ is the flux that minimizes the primal dissipation function. Thus, $\Vc{\flux}_{eq}(\Vc{x})$ is the MD flux.

By combining \eqnref{eq:Dissp_Decomp_1} and \eqnref{eq:Dissp_Decomp_2},
we have an EPR decomposition (see Appendix for proof):
\begin{align}
\EPR(\Vc{x})=\EPR_{hk}^{MD}(\Vc{x})+\EPR_{ex}^{MD}(\Vc{x}) \label{eq:MDdecomp}
\end{align}
where 
 we define housekeeping and excess EPR as 
\begin{align}
    \EPR_{hk}^{MD}(\Vc{x})&\defeq2\left[\BD_{\Vc{x}}[\Vc{\flux}_{0}(\Vc{x})\|\Vc{\flux}_{eq}(\Vc{x})]+\Dissp^{*}_{\Vc{x}}(\Vc{\tf}_{st}(\Vc{x}))\right]\ge0,\notag \\
    \EPR_{ex}^{MD}(\Vc{x})&\defeq2\left[\BD_{\Vc{x}}^{*}[\Vc{\tf}_{0}(\Vc{x})\|\Vc{\tf}_{st}(\Vc{x})]+\Dissp_{\Vc{x}}(\Vc{\flux}_{eq}(\Vc{x}))\right]\ge0.\notag
\end{align}
If the state $\Vc{x}(t)$ converges to a steady state $\Vc{x}_{st}$\footnote{$\Vc{x}_{st}$ is not necessarily a CB state but can be a general steady state}, then 
\begin{align}
\lim_{t\to \infty}\dot{\Sigma}_{hk}(\Vc{x}(t))&=\dot{\Sigma}_{hk}(\Vc{x}_{st})=2 \langle\Vc{\flux}_{0}(\Vc{x}_{st}),\Vc{\tf}_{0}(\Vc{x}_{st})\rangle\\
\lim_{t\to \infty}\dot{\Sigma}_{ex}(\Vc{x}(t))&=\dot{\Sigma}_{ex}(\Vc{x}_{st})=0,
\end{align}
where we used $\Vc{\flux}_{0}(\Vc{x}_{st})=\Vc{\flux}_{st}(\Vc{x}_{st})$ and $\Vc{\flux}_{eq}(\Vc{x}_{st})=\Vc{0}$.
Thus, this EPR decomposition geometrically generalizes the Maes-Neto\v{c}n\`{y} one to MJP and CRN and clarifies its dualistic minimum dissipation principles.
If the state $\Vc{x}(t)$ converges to a CB state $\Vc{x}_{cb}$, then $\lim_{t\to \infty}\dot{\Sigma}_{hk}(\Vc{x}(t))=2\langle\Vc{\flux}_{A},\Vc{\tf}_{A}\rangle$ holds.

\section{Numerical Demonstration}\label{sec:Simulation}
We numerically demonstrate the two decompositions in \eqnref{eq:GFdecomp} and \eqnref{eq:MDdecomp} and the geometric relations among $\Vc{\flux}_{0}(\Vc{x})$, $\Vc{\flux}_{iso}(\Vc{x})$, $\Vc{\flux}_{S}(\Vc{x})$, and $\Vc{\flux}_{A}$ and those among $\Vc{\flux}_{0}(\Vc{x})$, $\Vc{\flux}_{st}(\Vc{x})$, and $\Vc{\flux}_{eq}(\Vc{x})$ by using a CRN in Fig \ref{fig:CRN} (A)\cite{horn1973Proc.R.Soc.Lond.Math.Phys.Sci.,perezmillan2012BullMathBiol}.

\begin{table}[ht]
    \caption{The parameter values used for simulations. }
    \centering
    \begin{tabular}{|r||c|c|c|c|c|c|}
        \hline
        Type & $\kcoef_{1}^{+}$ & $\kcoef_{1}^{-}$ & $\kcoef_{2}^{+}$ & $\kcoef_{2}^{-}$ & $\kcoef_{3}^{+}$ & $\kcoef_{3}^{-}$  \\
        \hline
        \hline
        Equilibrium & $4$ & $1$ & $\frac{3}{\sqrt{2}}$ & $3 \sqrt{2}$& $2\sqrt{2}$ & $4\sqrt{2}$ \\
        \hline
        Complex balanced & $\frac{1}{2}$ & $2$ &  $4$ &  $\frac{47}{4}$ &  $\sqrt{2}$ &  $\frac{15}{2} + 2 \sqrt{2}$ \\
        \hline
        Non-complex balanced &$\frac{1}{2}$ &$2$ & $\frac{1}{17}$  &$\frac{85}{8}$ & $\frac{273}{68}$ & $\frac{137}{68}$  \\
        \hline
        \end{tabular}
    \label{tab:parameter_values}
\end{table}

\begin{figure}[t]
\includegraphics[width=0.48\textwidth, bb=0 0 700 700]{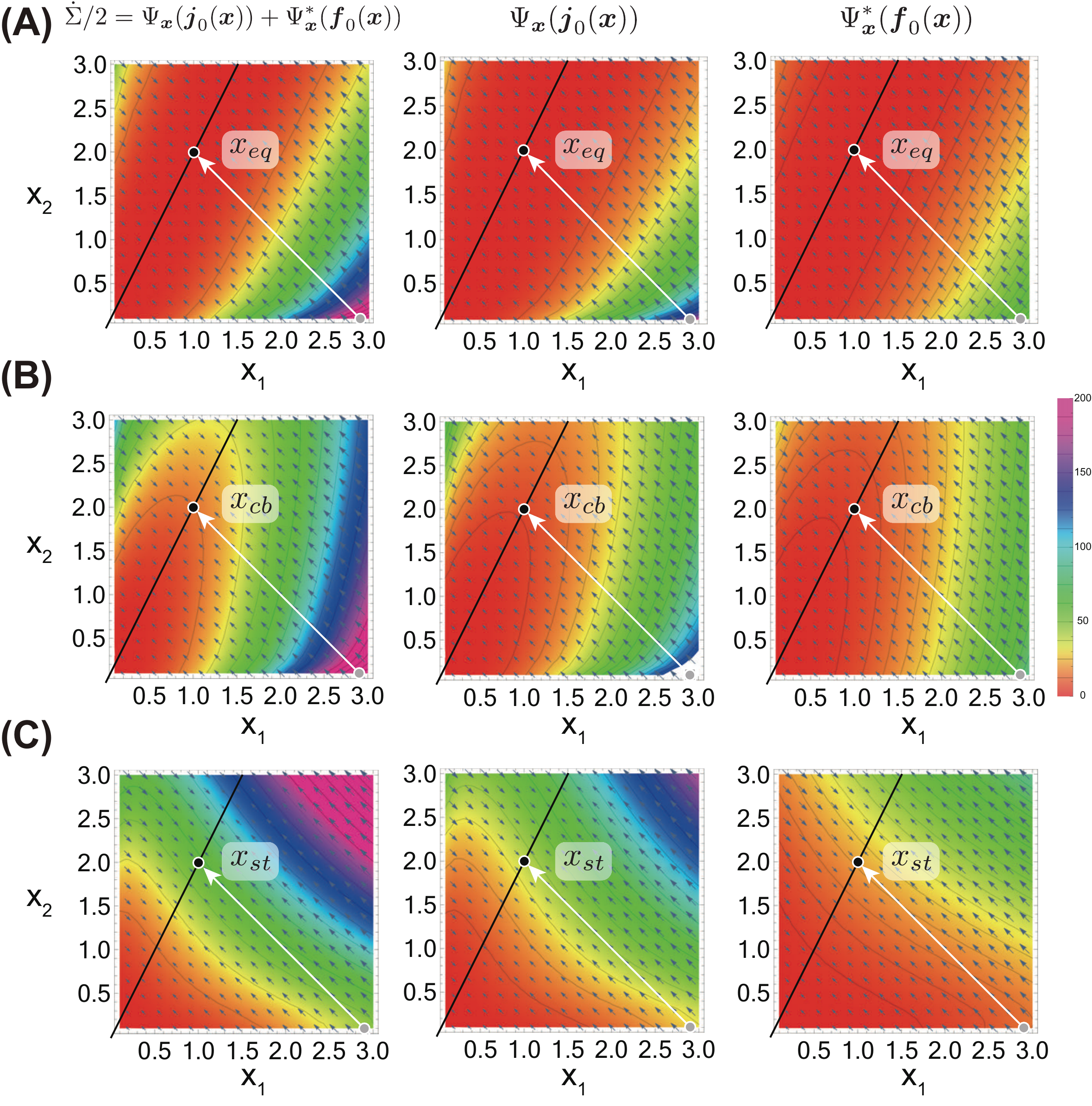}
\caption{Heatmap plots of EPR $\EPR$ (left panels) and the two dissipation functions, $\Dissp_{\Vc{x}}(\Vc{\flux}_{0}(\Vc{x})$ (center panels) and $\Dissp_{\Vc{x}}^{*}(\Vc{\tf}_{0}(\Vc{x})$ (right panels) as functions of $\Vc{x}$ for the equilibrium parameter set (A), the complex balanced parameter set (B), and the non-complex balanced parameter set(C). The parameter sets are listed in Tab. \ref{tab:parameter_values}.  
In each panel, gray and black points are the initial and the steady states, respectively. The white arrow is the trajectory of $\Vc{x}(t)$. The black line in each panel is the set of equilibrium states (A), complex-balanced states (B), and non-complex balanced steady states (C), respectively. }
\label{fig:simulation}
\end{figure}

\begin{figure}[t]
\includegraphics[width=0.48\textwidth, bb=0 0 600 450]{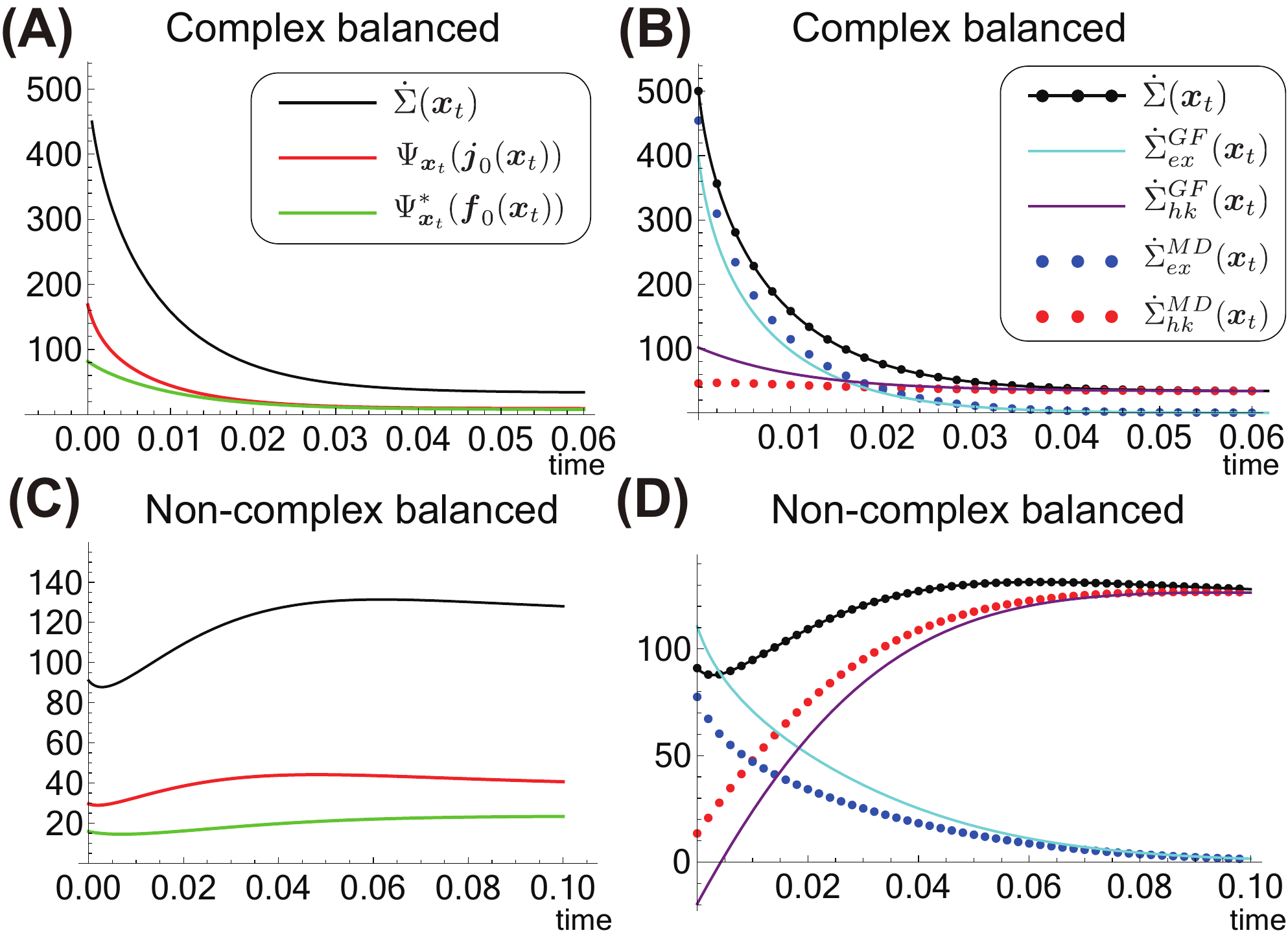}
\caption{The EPR and its decompositions along the trajectory $\Vc{x}_{t}$ for complex balanced (A,B) and non-complex balanced (C,D) parameter sets. 
The decomposition of EPR into two dissipation functions (A, C). 
The decompositions of EPR by Hilbert ($\EPR_{ex}^{GF}(\Vc{x})$, $\EPR_{hk}^{GF}(\Vc{x})$) and information geometric ($\EPR_{ex}^{MD}(\Vc{x})$, $\EPR_{hk}^{MD}(\Vc{x})$) orthogonalities  (B, D).
}
\label{fig:decomposition}
\end{figure}

The CRN depicted in Fig. \ref{fig:CRN} (A) is defined by the following set of chemical reaction equations:
\begin{align}
2\molX_{1} &\xleftrightharpoons[\kcoef^{+}_{1}]{\kcoef^{-}_{1}}2\molX_{2}, 
& 2\molX_{2} &\xleftrightharpoons[\kcoef^{+}_{2}]{\kcoef^{-}_{2}}\molX_{1} + \molX_{2}, 
& \molX_{1} + \molX_{2} &\xleftrightharpoons[\kcoef^{+}_{3}]{\kcoef^{-}_{3}} 2\molX_{1}. 
\end{align}
The corresponding structural quantities are 
\begin{align}
    \stoiMatrix&=
       \begin{pmatrix}
    -2 & +1 & +1\\
    +2 & -1 & -1
    \end{pmatrix},
    &   \IncMatrix&=
    \begin{pmatrix}
    +1 & 0 & -1\\
    -1 & +1 & 0\\
    0 & -1 & +1
    \end{pmatrix},\\
    \cmMatrix&=
    \begin{pmatrix}
    2 & 0 & 1\\
    0 & 2 & 1
    \end{pmatrix}.
\end{align}
One can show that this simple nonlinear CRN has any of the equilibrium, CB, and non-CB steady states depending on the kinetic parameter values\cite{horn1973Proc.R.Soc.Lond.Math.Phys.Sci.,perezmillan2012BullMathBiol}. 
In addition, the flux and force spaces are three-dimensional, and, thereby, the relevant geometric objects and quantities can be visualized computationally.
We used the parameter values in Tab. \ref{tab:parameter_values} for the simulations in Fig. \ref{fig:simulation} and in Fig. \ref{fig:decomposition}.
Because the state space is two-dimensional and the stoichiometric compatibility class is one-dimensional, i.e., $\Polytope(\Vc{x}_{0})\defeq\{\Vc{x}|\Vc{x}-\Vc{x}_{0}\in \Img[\stoiMatrix]\}=\{\Vc{x}|\Vc{x}=\Vc{x}_{0}+\xi (1,-1)^{\Transpose}, \xi \in \Real\}$, the trajectory is restricted in the one-dimensional $\Polytope(\Vc{x}_{0})$ independently of the parameter values.
We also selected the parameter values in Tab. \ref{tab:parameter_values} so that all the sets of equilibrium, CB and non-CB steady states are equivalent to the set,  $\{\Vc{x}|x_{2}=2 x_{1}\}$. 
Thus, the topological properties of the state space are the same for the three sets of parameters.

Figure \ref{fig:simulation} shows the values of the EPR and the associated dissipation functions in the state space.
For the equilibrium parameter (Fig. \ref{fig:simulation} (A)), the EPR and dissipation functions attain their global minimum value of zero on the equilibrium states (black line).
For the CB parameter (Fig. \ref{fig:simulation} (B)), the global minimum is not necessarily attained, which results in the non-zero EPR at the CB steady state $\Vc{x}_{cb}$ (Fig. \ref{fig:decomposition} (A)).
For the non-CB parameter  (Fig. \ref{fig:simulation} (C)), the landscapes of the EPR and dissipation functions become more complicated, and their values along the trajectory $\Vc{x}_{t}$ are not monotonous (Fig. \ref{fig:decomposition} (C)).

For the CB and non-CB cases, the two decompositions, \eqnref{eq:GFdecomp} and \eqnref{eq:MDdecomp}, are computed (Fig. \ref{fig:decomposition} (B) and (D)).
For the CB case, we verify that both decompositions provide the expected behaviors: both housekeeping and excess components stay non-negative; the excess one converges to $0$; the housekeeping one does to a finite non-negative value(Fig. \ref{fig:decomposition} (B)).

For the non-CB case, the expected behaviors are also produced (Fig. \ref{fig:decomposition} (D)) for the information-geometric decomposition, \eqnref{eq:MDdecomp} (a generalization of Maes-Neto\v{c}n\`{y} decomposition. 
On the contrary, the housekeeping component of the Hilbert decomposition, \eqnref{eq:GFdecomp}, becomes negative within a certain time window because the Hilbert orthogonality, \eqnref{eq:HB_orth_fs_fa}, does not hold in this case.
It should be noted that this result does not mean that the Hilbert decomposition is not applicable to the non-CB case. 
The result only indicates that the force decomposition by $\Vc{\tf}_{S}(\Vc{x})\defeq-\frac{1}{2}\stoiMatrix^{\Transpose}\partial_{\Vc{x}}\Gibbs_{cb}(\Vc{x})$ and $\Vc{\tf}_{A}\defeq\frac{1}{2}\ln\frac{\Vc{K}_{cb}}{\Vc{K}_{eq}}$ with the specific form of $\Gibbs_{cb}(\Vc{x})\defeq \KL[\Vc{x}\|\Vc{x}_{cb}]$ no longer satisfies the Hilbert orthogonality. 
We may be able to recover the orthogonality and the associated decomposition\cite{patterson2021ArXiv210314384Math-Ph} by choosing other functional form of $\Gibbs_{cb}(\Vc{x})$.
Finding such $\Gibbs_{cb}(\Vc{x})$ is linked to the computation of quasi-potential, but this problem is still challenging for generic non-CB cases.

\begin{figure}[ht]
\includegraphics[width=0.49\textwidth, bb=0 0 800 400]{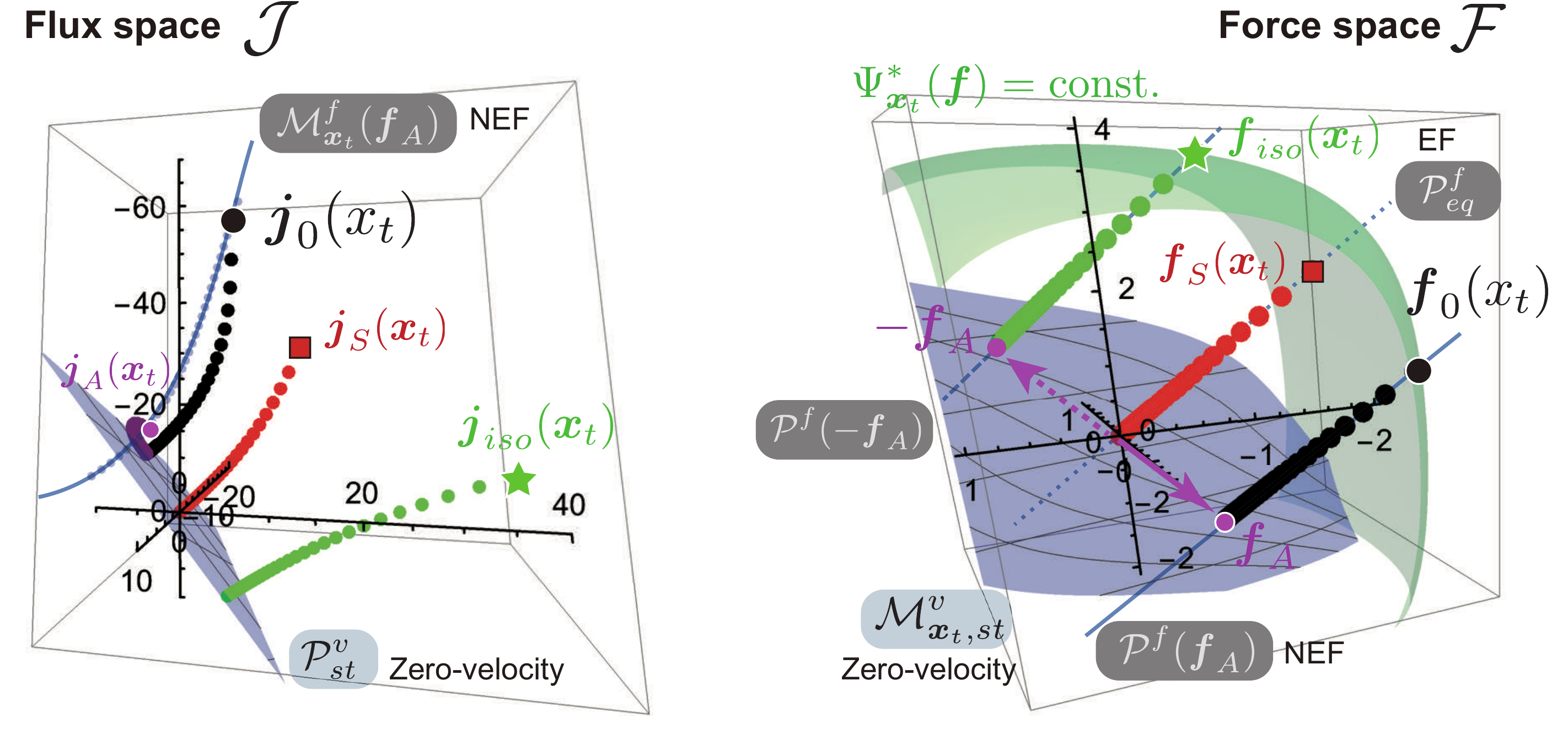}
\caption{Computational visualization of the Hilbert orthogonality between $\Vc{\tf}_{S}(\Vc{x}_{t})$ and $\Vc{\tf}_{A}$ for the CRN in Fig \ref{fig:CRN} (A). 
The green solid surface in $\Fspace$ represents the iso-dissipation hypersurface:  $\Dissp^{*}_{\Vc{x}_{t}}(\Vc{\tf})=\Dissp^{*}_{\Vc{x}_{t}}(\Vc{\tf}_{0}(\Vc{x}_{t}))=\mathrm{const.}$. 
The black circle with a white border, red square with a black border, green star, and magenta circle with a white border in $\Fspace$ are respectively  $\Vc{\tf}_{0}(\Vc{x}_{t})$, $\Vc{\tf}_{S}(\Vc{x}_{t})$, $\Vc{\tf}_{iso}(\Vc{x}_{t})$, and $\Vc{\tf}_{A}$ evaluated at $\Vc{x}_{t}=(5/2,1/2)^{\Transpose}$.
The black circle, red square, green star, and magenta circle with borders in $\Jspace$ are their Legendre transform, i.e., $\Vc{\flux}_{0}(\Vc{x}_{t})$, $\Vc{\flux}_{S}(\Vc{x}_{t})$,  $\Vc{\flux}_{iso}(\Vc{x}_{t})$, and $\Vc{\flux}_{A}(\Vc{x}_{t})$, respectively. 
The black, red, and green circles without a border in $\Fspace$ are the trajectories of $\{\Vc{\tf}_{0}(\Vc{x}_{t})\}$, $\{\Vc{\tf}_{S}(\Vc{x}_{t})\}$, and $\{\Vc{\tf}_{iso}(\Vc{x}_{t})\}$.
The black, red, green, and magenta circles without a border in $\Jspace$ are the trajectories of $\{\Vc{\flux}_{0}(\Vc{x}_{t})\}$, $\{\Vc{\flux}_{S}(\Vc{x}_{t})\}$,  $\{\Vc{\flux}_{iso}(\Vc{x}_{t})\}$, and $\{\Vc{\flux}_{A}(\Vc{x}_{t})\}$, respectively. 
The light blue lines in $\Fspace$ are the NEF subspaces, $\Polytope^{f}(\Vc{\tf}_{A})$(solid line) and $\Polytope^{f}(-\Vc{\tf}_{A})$ (dashed line), and the EF subspace $\Polytope^{f}_{eq}$ (dotted line). 
In $\Jspace$, the corresponding NEF manifolds, $\Variety^{f}_{\Vc{x}}(\Vc{\tf}_{A})$ (the light blue curve with dots), is also depicted.
The blue plane in $\Jspace$ is the zero-velocity subspace $\Polytope^{v}_{st}$, and the blue surface in $\Fspace$ is its Legendre transformation: $\Variety^{v}_{\Vc{x}_{t},st}$.
}
\label{fig:visual_Hilbert}
\end{figure}

\begin{figure}[ht]
\includegraphics[width=0.49\textwidth, bb=0 0 800 400]{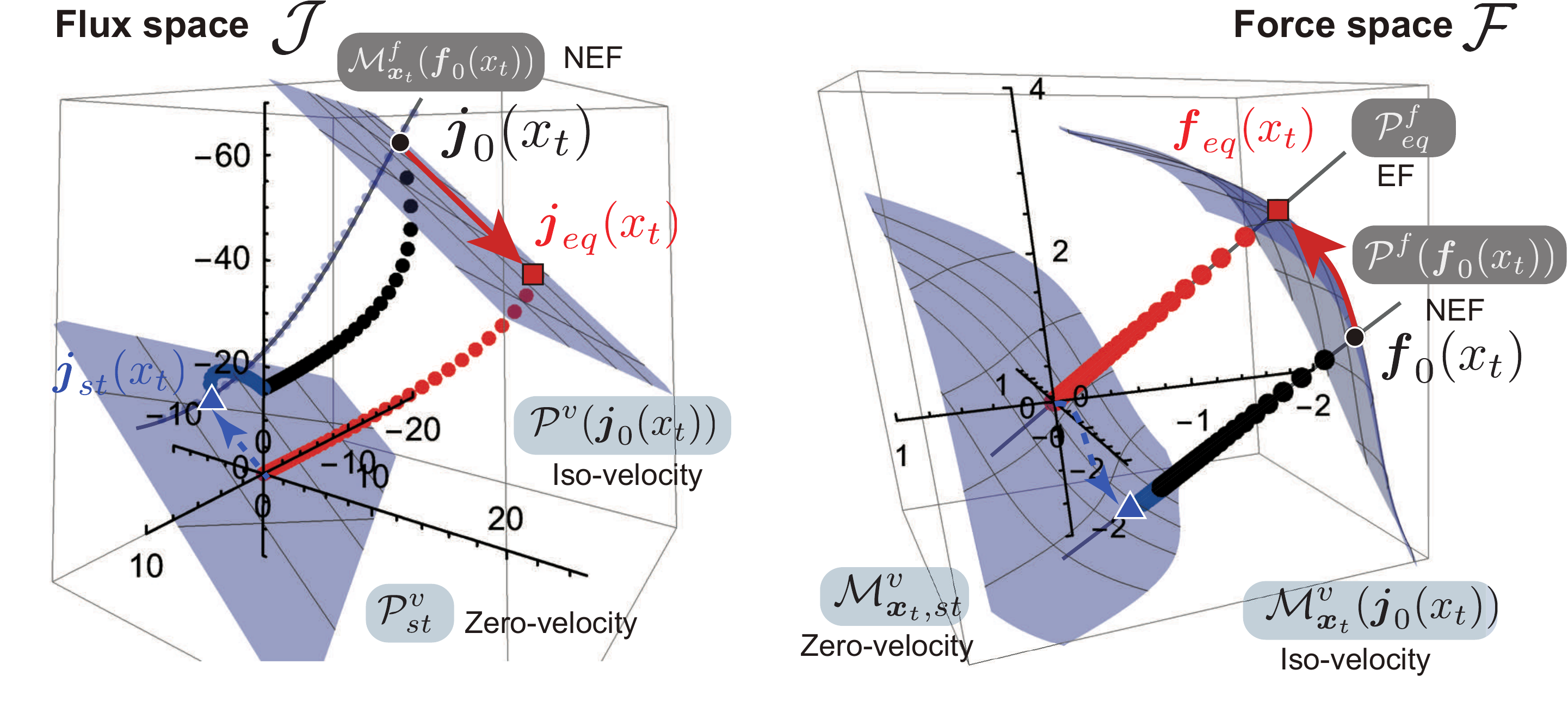}
\caption{Computational visualization of the information geometric orthogonality for the CRN in Fig \ref{fig:CRN} (A). 
(Left) The orthogonality between $\Vc{\flux}_{st}(\Vc{x}_{t})$ (blue dashed arrow) and $\Vc{\flux}_{0}(\Vc{x}_{t})-\Vc{\flux}_{st}(\Vc{x}_{t})$ and the dual orthogonality between
$\Vc{\flux}_{eq}(\Vc{x}_{t})$ and $\Vc{\flux}_{0}(\Vc{x}_{t})-\Vc{\flux}_{eq}(\Vc{x}_{t})$ (red solid arrow) in $\Jspace$. 
The blue planes are the iso-velocity subspace $\Polytope^{v}(\Vc{\flux}_{0}(\Vc{x}_{t}))$ (the upper plane) and the zero-velocity subspace $\Polytope^{v}_{st}$ (the lower plane).
The grey curve with dots is the NEF manifold $\Variety^{f}_{\Vc{x}}(\Vc{\tf}_{0}(\Vc{x}_{t}))$. 
Black, red, and blue circles are the trajectories of $\{\Vc{\flux}_{0}(\Vc{x}_{t})\}$, $\{\Vc{\flux}_{eq}(\Vc{x}_{t})\}$, and $\{\Vc{\flux}_{st}(\Vc{x}_{t})\}$, respectively.
(Right) The same orthogonalities shown in $\Fspace$ space. 
The orthogonality between $\Vc{\tf}_{st}(\Vc{x}_{t})$ (blue curved arrow) and $\Vc{\tf}_{0}(\Vc{x}_{t})-\Vc{\tf}_{st}(\Vc{x}_{t})$ and the dual orthogonality between $\Vc{\tf}_{eq}(\Vc{x}_{t})$ and $\Vc{\tf}_{0}(\Vc{x}_{t})-\Vc{\tf}_{eq}(\Vc{x}_{t})$ (red curved arrow) in $\Fspace$. 
Black, red circles are the trajectories of $\{\Vc{\tf}_{0}(\Vc{x}_{t})\}$, $\{\Vc{\tf}_{eq}(\Vc{x}_{t})\}$, respectively.
The blue surfaces are the iso-velocity manifold $\Variety^{v}_{\Vc{x}_{t}}(\Vc{\flux}_{0}(\Vc{x}_{t}))$ (the upper surface) and the zero-velocity manifold $\Variety^{v}_{\Vc{x}_{t},st}$ (the lower surface). 
The grey lines are the NEF subspaces $\Polytope^{f}(\Vc{\tf}_{0}(\Vc{x}_{t}))$ and the EF subspace $\Polytope^{f}_{\Vc{x},eq}$. }
\label{fig:visual_Info}
\end{figure}

Finally, we computationally visualized the various geometric objects introduced in this work in Figs. \ref{fig:visual_Hilbert} and  \ref{fig:visual_Info} for the CB parameter set in Tab. \ref{tab:parameter_values}.

First, we show the behaviors of the forces, $\Vc{\tf}_{0}(\Vc{x}_{t})$, $\Vc{\tf}_{iso}(\Vc{x}_{t})$, $\Vc{\tf}_{S}(\Vc{x}_{t})$, and $\Vc{\tf}_{A}$, which satisfy the Hilbert orthogonality, and the corresponding fluxes.
In the force space $\Fspace$ (the right panel of Fig. \ref{fig:visual_Hilbert}), we observe that the trajectories of $\Vc{\tf}_{0}(\Vc{x}_{t})$,
$\Vc{\tf}_{iso}(\Vc{x}_{t})$, and $\Vc{\tf}_{S}(\Vc{x}_{t})$ are actually restricted on one dimensional subspaces (lines), $\Polytope^{f}(\Vc{\tf}_{A})$, $\Polytope^{f}(-\Vc{\tf}_{A})$, and $\Polytope^{f}_{eq}=\Polytope^{f}(\Vc{0})$, respectively. 
$\Vc{\tf}_{0}(\Vc{x}_{t})$ and $\Vc{\tf}_{iso}(\Vc{x}_{t})$ are also on the iso-dissipation hypersuface (the green surface) satisfying $\Dissp^{*}_{\Vc{x}_{t}}(\Vc{\tf}_{0}(\Vc{x}_{t}))=\Dissp^{*}_{\Vc{x}_{t}}(\Vc{\tf}_{iso}(\Vc{x}_{t}))$.
In the flux space $\Jspace$ (the left panel of Fig. \ref{fig:visual_Hilbert}), $\Polytope^{f}(\Vc{\tf}_{A})$ is transformed to the one-dimensional curve, $\Variety^{f}_{\Vc{x}_{t}}(\Vc{\tf}_{A})$. 
The trajectories of $\Vc{\flux}_{0}(\Vc{x}_{t})$,
$\Vc{\flux}_{iso}(\Vc{x}_{t})$, and $\Vc{\flux}_{S}(\Vc{x}_{t})$ are also curved in $\Jspace$. 
All the trajectories converge onto the zero-velocity subspace $\Polytope^{v}_{st}$.

Next, we verify the information-geometric orthogonality in Fig. \ref{fig:visual_Info}. 
In the flux space $\Jspace$ (the left panel in Fig. \ref{fig:visual_Info}), we observe that $\Vc{\flux}_{0}(\Vc{x}_{t})$ and the corresponding equilibrium flux $\Vc{\flux}_{eq}(\Vc{x}_{t})$ are on the iso-velocity subspace $\Polytope^{v}(\Vc{\flux}_{0}(\Vc{x}_{t}))$.
The steady state flux  $\Vc{\flux}_{st}(\Vc{x}_{t})$ are on the intersection of the zero-velocity subspace $\Polytope^{v}_{st}$ and the NEF manifold $\Variety^{f}_{\Vc{x}_{t}}(\Vc{\tf}_{0}(\Vc{x}_{t}))$.  
By mapping these objects into $\Fspace$ by the Legendre transform, we observe that $\Vc{\tf}_{0}(\Vc{x}_{t})$ is on the intersection of the NEF subspace $\Polytope^{f}(\Vc{\tf}_{0}(\Vc{x}_{t}))$ and the iso-velocity manifold $\Variety^{v}_{\Vc{x}_{t}}(\Vc{\flux}_{0}(\Vc{x}_{t}))$.
Similarly, $\Vc{\tf}_{eq}(\Vc{x}_{t})$ is on the intersection of the EF subspace $\Polytope^{f}_{eq}$ and the iso-velocity manifold $\Variety^{v}_{\Vc{x}_{t}}(\Vc{\flux}_{0}(\Vc{x}_{t}))$.
By the Legendre transformation, the iso-velocity manifold $\Variety^{v}_{\Vc{x}_{t}}(\Vc{\flux}_{0}(\Vc{x}_{t}))$, and the zero-velocity manifold $\Variety^{v}_{\Vc{x}_{t},st}$ are curved in $\Fspace$.
These graphical representations demonstrate how the nonlinear Legendre transform relates the subspaces and manifolds in $\Jspace$ and $\Fspace$ and also how the generalized notions of orthogonality of Hessian geometry are realized in these spaces.

\section{Summary and Discussion}\label{sec:Discussion}
We have clarified that Hessian geometry is the natural geometric structure of nonequilibrium and nonlinear CRNs as well as MJPs. 
More generally, it can capture the geometry induced by nonquadratic convex dissipation functions.
By employing generalized notions of orthogonality, various aspects of nonequilibrium dynamics are dissected as decompositions of EPRs, which generalize the well-established ones \cite{maes2014JStatPhys,komatsu2008Phys.Rev.Lett.,dechant2022ArXiv210912817Cond-Mat} to CRNs and MJPs.

\subsection{Thermodynamic uncertainty relation and Fisher information}
The nonquadratic property of the dissipation functions of CRN and MJP also appears implicitly in different problems of thermodynamics.
One such example is thermodynamic uncertainty relation (TUR) and its extension to MJP. 
TUR is a relation that bounds the fluctuation of a generalized current in a nonequilibrium steady state by the entropy production\cite{shiraishi2021JStatPhys}:
\begin{align}
    \frac{\mathbb{E}[\mathcal{J}_{d}^{ss}]^{2}}{\mathbb{Var}[\mathcal{J}_{d}^{ss}]}\le \frac{1}{2}\int_{0}^{\tau}\EPR(t') \dd t'\label{eq:TUR}
\end{align}
where $\mathcal{J}_{d}^{ss}$ is a time-integrated generalized current at the steady state, and $\mathbb{E}[\mathcal{J}_{d}^{ss}]$ and $\mathbb{Var}[\mathcal{J}_{d}^{ss}]$ are the mean and variance of $\mathcal{J}_{d}^{ss}$.
Since the proposal of the TUR conjecture and its first proof \cite{gingrich2016Phys.Rev.Lett.,barato2015Phys.Rev.Lett.}, the TUR has been extended to various models and situations, including MJP \cite{shiraishi2021JStatPhys} and CRN \cite{yoshimura2021Phys.Rev.Lett.a}.
However, even though \eqnref{eq:TUR} is tight for overdamped diffusion processes, it is not tight for MJP. 
Instead, for MJP, the fluctuation of the current is bounded tightly by the pseudo entropy production(pEPR) \cite{yoshimura2021Phys.Rev.Lett.a}:
\begin{align}
    \frac{\mathbb{E}[\mathcal{J}_{d}^{ss}]^{2}}{\mathbb{Var}[\mathcal{J}_{d}^{ss}]}\le   \frac{1}{2}\int_{0}^{\tau}\pEPR(t') \dd t'\label{eq:TUR_MJP}
\end{align}
where the pEPR is defined as
\begin{align}
    \pEPR(t) \defeq \frac{(\Vc{\flux}^{+}(\Vc{x}_{t})-\Vc{\flux}^{-}(\Vc{x}_{t}))^{2}}{\Vc{\flux}^{+}(\Vc{x}_{t})+\Vc{\flux}^{+}(\Vc{x}_{t})}.
\end{align}
and $\pEPR(t)\le \EPR(t)$ holds \cite{shiraishi2021JStatPhys}.
We here mention that this gap between the EPR and pEPR is linked to the Fisher information of the dissipation function.
The Fisher information matrix (or metric) for a strictly convex function $\phi^{*}(\Vc{\tf})$ is defined by its Hessian as
\begin{align}
    \FM_{e,e'}^{*}(\Vc{\tf}) \defeq \frac{\partial^{2}\phi^{*}(\Vc{\tf})}{\partial\tf_{e} \partial\tf_{e'}}.
\end{align}
For $\phi^{*}(\Vc{\tf})=\Dissp_{\Vc{\frenecy}}^{*}(\Vc{\tf})$ defined by \eqnref{eq:DissipationFunctions}, we have
\begin{align}
    \FM^{*}(\Vc{\frenecy},\Vc{\tf}) =\diag[\Vc{\frenecy}\circ \cosh(\Vc{\tf})]=(\Vc{\flux}^{+}+\Vc{\flux}^{-}),
\end{align}
where we used \eqnref{eq:flux_force} and \eqnref{eq:frenecy}. For $\phi(\Vc{\flux})=\Dissp_{\Vc{\frenecy}}(\Vc{\flux})$, we have the Fisher information matrix:
\begin{align}
    \FM(\Vc{\frenecy},\Vc{\flux}) =\FM^{*}(\Vc{\frenecy},\Vc{\tf})^{-1}=\frac{1}{\Vc{\flux}^{+}+\Vc{\flux}^{-}}.
\end{align}
Then, the pEPR is represented as
\begin{align}
    \pEPR(\Vc{\flux})=\langle\Vc{\flux}, \FM(\Vc{\frenecy},\Vc{\flux}) \Vc{\flux} \rangle,
\end{align}
which allows us to regard $\pEPR(t)$ as an approximation of the EPR by replacing the actual force $\Vc{\tf}$ associated with $\Vc{\flux}$ with a pseudo-force $\Vc{\tf}^{p}\defeq\FM(\Vc{\frenecy},\Vc{\flux}) \Vc{\flux}$.
If the dissipation function is quadratic, the actual force $\Vc{\tf}$ and the pseudo-force $\Vc{\tf}^{p}$ coincide. 
Thus, the non-tightness of the TUR for MJP is a manifestation of the nonquadratic nature of the dissipation functions.

\subsection{Network thermodynamics}
The information geometric decomposition that we introduced is tightly related to the network thermodynamics \cite{oster1971Nature,hill2005,schnakenberg1976Rev.Mod.Phys.,schnakenberg2012,vanderschaft2013SIAMJ.Appl.Math.a}. 
The network thermodynamics is an attempt to extend the methodology of network theory developed mainly for linear electric circuits to other physical systems \cite{belevitch1962Proc.IRE,chen2012}.
While the linear network theory successfully works for linear electric circuits, which have a linear force and flux relation (Ohm's law) and resulting quadratic dissipation functions, its application to MJP and CRN has encountered difficulties due to the nonlinear relation between force and flux, especially when we evaluate the EPR decomposition for a transient and far from equilibrium state.
The Hessian geometric structure and the generalized decompositions in this work are extensions of those in network thermodynamics. Specifically, the information geometric decomposition is a nonlinear generalization of the cycle-cocycle decomposition \cite{wachtel2015Phys.Rev.E}.
However, network thermodynamics and network theory accommodate a wide variety of methods that are not yet exploited in this work \cite{altaner2012Phys.Rev.E,wachtel2015Phys.Rev.E}. 
For example, the algebraic structure underlying the network is explored by using the integral basis defined by cycles, cords, and spanning trees in the network \cite{schnakenberg1976Rev.Mod.Phys.,polettini2015Math.Technol.Netw.}. 
As an application, the steady EPR is related to cyclic fluxes and affinities in the network.
We may incorporate and exploit the algebraic structure of MJP and CRN more explicitly into our framework.

\subsection{Duality and variational characterizations in thermodynamics}
The notion of duality is the core of thermodynamics. 
In equilibrium thermodynamics, the duality of extensive and intensive variables induced by thermodynamic potential functions characterizes the energetic aspect of physical systems.
The roles and differences of dual pairs of thermodynamic potential functions are well recognized.
In addition, Hessian geometry is central to comprehend the geometric properties of equilibrium thermodynamics \cite{kobayashi2021ArXiv211214910Phys.,sughiyama2021ArXiv211212403Cond-MatPhysicsphysics}.
For the duality between the force and flux explored in this work, 
the dual dissipation functions have been recognized for a while in MFT \cite{bertini2015Rev.Mod.Phys.} and metric gradient flow theory \cite{ambrosio2006}.
Nevertheless, nonquadratic dissipation functions have been investigated only very recently even in MFT \cite{kaiser2018JStatPhys,renger2018Entropy,renger2021DiscreteContin.Dyn.Syst.-S,patterson2021ArXiv210314384Math-Ph} and also in other physics communities \cite{suzuki2012PhysicaA:StatisticalMechanicsanditsApplications}.
As in the case of equilibrium thermodynamics, grasping the roles played by the dual functions should be essential for understanding nonequilibrium and kinetic aspects of thermodynamics. 
Hence, Hessian geometry becomes an indispensable tool for  investigating the geometry induced by the duality and also the variational aspects of nonequilibrium phenomena.
In thermodynamics, there exist continued attempts to characterize nonequilibrium states and relations variationally, which go by the names of minimum entropy production principle \cite{prigogine1968, klein1954Phys.Rev.a,polettini2011Phys.Rev.E}, maximum entropy production principle \cite{zupanovic2004Phys.Rev.E,martyushev2021Phys.-Usp.}, the least dissipation principle \cite{onsager1953Phys.Rev.,gyarmati2013}, and others.
However, all these principles still have limitations in their applicability for nonlinear and far from equilibrium situations \cite{polettini2013Entropy}. 
Hessian geometry and also information geometry, which has an ability to handle nonlinearity induced by convex functions, may contribute to resolving a part of problems in such principles.
Moreover, they can serve as the natural language to integrate the equilibrium (energetic) and nonequilibrium (kinetic) descriptions, and thereby provide us a more universal understanding of thermodynamic systems and extend the applicability of nonequilibrium thermodynamics.

\section{Acknowledgement}
This research is supported by JST (JPMJCR2011,JPMJCR1927) and JSPS (19H05799). We thank Shuhei Horiguchi for valuable suggestions.


%



\begin{thebibliography}{71}%
\makeatletter
\providecommand \@ifxundefined [1]{%
 \@ifx{#1\undefined}
}%
\providecommand \@ifnum [1]{%
 \ifnum #1\expandafter \@firstoftwo
 \else \expandafter \@secondoftwo
 \fi
}%
\providecommand \@ifx [1]{%
 \ifx #1\expandafter \@firstoftwo
 \else \expandafter \@secondoftwo
 \fi
}%
\providecommand \natexlab [1]{#1}%
\providecommand \enquote  [1]{``#1''}%
\providecommand \bibnamefont  [1]{#1}%
\providecommand \bibfnamefont [1]{#1}%
\providecommand \citenamefont [1]{#1}%
\providecommand \href@noop [0]{\@secondoftwo}%
\providecommand \href [0]{\begingroup \@sanitize@url \@href}%
\providecommand \@href[1]{\@@startlink{#1}\@@href}%
\providecommand \@@href[1]{\endgroup#1\@@endlink}%
\providecommand \@sanitize@url [0]{\catcode `\\12\catcode `\$12\catcode
  `\&12\catcode `\#12\catcode `\^12\catcode `\_12\catcode `\%12\relax}%
\providecommand \@@startlink[1]{}%
\providecommand \@@endlink[0]{}%
\providecommand \url  [0]{\begingroup\@sanitize@url \@url }%
\providecommand \@url [1]{\endgroup\@href {#1}{\urlprefix }}%
\providecommand \urlprefix  [0]{URL }%
\providecommand \Eprint [0]{\href }%
\providecommand \doibase [0]{https://doi.org/}%
\providecommand \selectlanguage [0]{\@gobble}%
\providecommand \bibinfo  [0]{\@secondoftwo}%
\providecommand \bibfield  [0]{\@secondoftwo}%
\providecommand \translation [1]{[#1]}%
\providecommand \BibitemOpen [0]{}%
\providecommand \bibitemStop [0]{}%
\providecommand \bibitemNoStop [0]{.\EOS\space}%
\providecommand \EOS [0]{\spacefactor3000\relax}%
\providecommand \BibitemShut  [1]{\csname bibitem#1\endcsname}%
\let\auto@bib@innerbib\@empty
\bibitem [{\citenamefont {Callen}\ and\ \citenamefont
  {Callen}(1985)}]{callen1985}%
  \BibitemOpen
  \bibfield  {author} {\bibinfo {author} {\bibfnamefont {H.~B.}\ \bibnamefont
  {Callen}}\ and\ \bibinfo {author} {\bibfnamefont {H.~B.}\ \bibnamefont
  {Callen}},\ }\href@noop {} {\emph {\bibinfo {title} {Thermodynamics and an
  {{Introduction}} to {{Thermostatistics}}}}}\ (\bibinfo  {publisher}
  {{Wiley}},\ \bibinfo {year} {1985})\BibitemShut {NoStop}%
\bibitem [{\citenamefont {Sughiyama}\ \emph {et~al.}(2021)\citenamefont
  {Sughiyama}, \citenamefont {Loutchko}, \citenamefont {Kamimura},\ and\
  \citenamefont
  {Kobayashi}}]{sughiyama2021ArXiv211212403Cond-MatPhysicsphysics}%
  \BibitemOpen
  \bibfield  {author} {\bibinfo {author} {\bibfnamefont {Y.}~\bibnamefont
  {Sughiyama}}, \bibinfo {author} {\bibfnamefont {D.}~\bibnamefont {Loutchko}},
  \bibinfo {author} {\bibfnamefont {A.}~\bibnamefont {Kamimura}},\ and\
  \bibinfo {author} {\bibfnamefont {T.~J.}\ \bibnamefont {Kobayashi}},\
  }\bibfield  {title} {\bibinfo {title} {A {{Hessian Geometric Structure}} of
  {{Chemical Thermodynamic Systems}} with {{Stoichiometric Constraints}}},\
  }\href@noop {} {\bibfield  {journal} {\bibinfo  {journal} {ArXiv211212403
  Cond-Mat Physicsphysics}\ } (\bibinfo {year} {2021})},\ \Eprint
  {https://arxiv.org/abs/2112.12403} {arXiv:2112.12403 [cond-mat,
  physics:physics]} \BibitemShut {NoStop}%
\bibitem [{\citenamefont {Onsager}(1931{\natexlab{a}})}]{onsager1931Phys.Rev.}%
  \BibitemOpen
  \bibfield  {author} {\bibinfo {author} {\bibfnamefont {L.}~\bibnamefont
  {Onsager}},\ }\bibfield  {title} {\bibinfo {title} {Reciprocal {{Relations}}
  in {{Irreversible Processes}}. {{I}}.},\ }\href
  {https://doi.org/10.1103/PhysRev.37.405} {\bibfield  {journal} {\bibinfo
  {journal} {Phys. Rev.}\ }\textbf {\bibinfo {volume} {37}},\ \bibinfo {pages}
  {405} (\bibinfo {year} {1931}{\natexlab{a}})}\BibitemShut {NoStop}%
\bibitem [{\citenamefont
  {Onsager}(1931{\natexlab{b}})}]{onsager1931Phys.Rev.a}%
  \BibitemOpen
  \bibfield  {author} {\bibinfo {author} {\bibfnamefont {L.}~\bibnamefont
  {Onsager}},\ }\bibfield  {title} {\bibinfo {title} {Reciprocal {{Relations}}
  in {{Irreversible Processes}}. {{II}}.},\ }\href
  {https://doi.org/10.1103/PhysRev.38.2265} {\bibfield  {journal} {\bibinfo
  {journal} {Phys. Rev.}\ }\textbf {\bibinfo {volume} {38}},\ \bibinfo {pages}
  {2265} (\bibinfo {year} {1931}{\natexlab{b}})}\BibitemShut {NoStop}%
\bibitem [{\citenamefont {Onsager}\ and\ \citenamefont
  {Machlup}(1953)}]{onsager1953Phys.Rev.}%
  \BibitemOpen
  \bibfield  {author} {\bibinfo {author} {\bibfnamefont {L.}~\bibnamefont
  {Onsager}}\ and\ \bibinfo {author} {\bibfnamefont {S.}~\bibnamefont
  {Machlup}},\ }\bibfield  {title} {\bibinfo {title} {Fluctuations and
  {{Irreversible Processes}}},\ }\href
  {https://doi.org/10.1103/PhysRev.91.1505} {\bibfield  {journal} {\bibinfo
  {journal} {Phys. Rev.}\ }\textbf {\bibinfo {volume} {91}},\ \bibinfo {pages}
  {1505} (\bibinfo {year} {9月 15, 1953})}\BibitemShut {NoStop}%
\bibitem [{\citenamefont {Machlup}\ and\ \citenamefont
  {Onsager}(1953)}]{machlup1953Phys.Rev.}%
  \BibitemOpen
  \bibfield  {author} {\bibinfo {author} {\bibfnamefont {S.}~\bibnamefont
  {Machlup}}\ and\ \bibinfo {author} {\bibfnamefont {L.}~\bibnamefont
  {Onsager}},\ }\bibfield  {title} {\bibinfo {title} {Fluctuations and
  {{Irreversible Process}}. {{II}}. {{Systems}} with {{Kinetic Energy}}},\
  }\href {https://doi.org/10.1103/PhysRev.91.1512} {\bibfield  {journal}
  {\bibinfo  {journal} {Phys. Rev.}\ }\textbf {\bibinfo {volume} {91}},\
  \bibinfo {pages} {1512} (\bibinfo {year} {9月 15, 1953})}\BibitemShut
  {NoStop}%
\bibitem [{\citenamefont {Bertini}\ \emph {et~al.}(2015)\citenamefont
  {Bertini}, \citenamefont {De~Sole}, \citenamefont {Gabrielli}, \citenamefont
  {{Jona-Lasinio}},\ and\ \citenamefont {Landim}}]{bertini2015Rev.Mod.Phys.}%
  \BibitemOpen
  \bibfield  {author} {\bibinfo {author} {\bibfnamefont {L.}~\bibnamefont
  {Bertini}}, \bibinfo {author} {\bibfnamefont {A.}~\bibnamefont {De~Sole}},
  \bibinfo {author} {\bibfnamefont {D.}~\bibnamefont {Gabrielli}}, \bibinfo
  {author} {\bibfnamefont {G.}~\bibnamefont {{Jona-Lasinio}}},\ and\ \bibinfo
  {author} {\bibfnamefont {C.}~\bibnamefont {Landim}},\ }\bibfield  {title}
  {\bibinfo {title} {Macroscopic fluctuation theory},\ }\href
  {https://doi.org/10.1103/RevModPhys.87.593} {\bibfield  {journal} {\bibinfo
  {journal} {Rev. Mod. Phys.}\ }\textbf {\bibinfo {volume} {87}},\ \bibinfo
  {pages} {593} (\bibinfo {year} {6月 24, 2015})}\BibitemShut {NoStop}%
\bibitem [{\citenamefont {Taniguchi}\ and\ \citenamefont
  {Cohen}(2007)}]{taniguchi2007JStatPhys}%
  \BibitemOpen
  \bibfield  {author} {\bibinfo {author} {\bibfnamefont {T.}~\bibnamefont
  {Taniguchi}}\ and\ \bibinfo {author} {\bibfnamefont {E.~G.~D.}\ \bibnamefont
  {Cohen}},\ }\bibfield  {title} {\bibinfo {title} {Onsager-{{Machlup Theory}}
  for {{Nonequilibrium Steady States}} and {{Fluctuation Theorems}}},\ }\href
  {https://doi.org/10.1007/s10955-006-9252-2} {\bibfield  {journal} {\bibinfo
  {journal} {J Stat Phys}\ }\textbf {\bibinfo {volume} {126}},\ \bibinfo
  {pages} {1} (\bibinfo {year} {2007})}\BibitemShut {NoStop}%
\bibitem [{\citenamefont {Schmiedl}\ and\ \citenamefont
  {Seifert}(2007)}]{schmiedl2007J.Chem.Phys.}%
  \BibitemOpen
  \bibfield  {author} {\bibinfo {author} {\bibfnamefont {T.}~\bibnamefont
  {Schmiedl}}\ and\ \bibinfo {author} {\bibfnamefont {U.}~\bibnamefont
  {Seifert}},\ }\bibfield  {title} {\bibinfo {title} {Stochastic thermodynamics
  of chemical reaction networks},\ }\href {https://doi.org/10.1063/1.2428297}
  {\bibfield  {journal} {\bibinfo  {journal} {J. Chem. Phys.}\ }\textbf
  {\bibinfo {volume} {126}},\ \bibinfo {pages} {044101} (\bibinfo {year}
  {2007})}\BibitemShut {NoStop}%
\bibitem [{\citenamefont {Aurell}\ \emph {et~al.}(2011)\citenamefont {Aurell},
  \citenamefont {{Mej{\'i}a-Monasterio}},\ and\ \citenamefont
  {{Muratore-Ginanneschi}}}]{aurell2011Phys.Rev.Lett.}%
  \BibitemOpen
  \bibfield  {author} {\bibinfo {author} {\bibfnamefont {E.}~\bibnamefont
  {Aurell}}, \bibinfo {author} {\bibfnamefont {C.}~\bibnamefont
  {{Mej{\'i}a-Monasterio}}},\ and\ \bibinfo {author} {\bibfnamefont
  {P.}~\bibnamefont {{Muratore-Ginanneschi}}},\ }\bibfield  {title} {\bibinfo
  {title} {Optimal {{Protocols}} and {{Optimal Transport}} in {{Stochastic
  Thermodynamics}}},\ }\href {https://doi.org/10.1103/PhysRevLett.106.250601}
  {\bibfield  {journal} {\bibinfo  {journal} {Phys. Rev. Lett.}\ }\textbf
  {\bibinfo {volume} {106}},\ \bibinfo {pages} {250601} (\bibinfo {year}
  {2011})}\BibitemShut {NoStop}%
\bibitem [{\citenamefont {Nakazato}\ and\ \citenamefont
  {Ito}(2021)}]{nakazato2021Phys.Rev.Research}%
  \BibitemOpen
  \bibfield  {author} {\bibinfo {author} {\bibfnamefont {M.}~\bibnamefont
  {Nakazato}}\ and\ \bibinfo {author} {\bibfnamefont {S.}~\bibnamefont {Ito}},\
  }\bibfield  {title} {\bibinfo {title} {Geometrical aspects of entropy
  production in stochastic thermodynamics based on {{Wasserstein}} distance},\
  }\href {https://doi.org/10.1103/PhysRevResearch.3.043093} {\bibfield
  {journal} {\bibinfo  {journal} {Phys. Rev. Research}\ }\textbf {\bibinfo
  {volume} {3}},\ \bibinfo {pages} {043093} (\bibinfo {year}
  {2021})}\BibitemShut {NoStop}%
\bibitem [{\citenamefont {Jordan}\ \emph {et~al.}(1997)\citenamefont {Jordan},
  \citenamefont {Kinderlehrer},\ and\ \citenamefont
  {Otto}}]{jordan1997PhysicaD:NonlinearPhenomena}%
  \BibitemOpen
  \bibfield  {author} {\bibinfo {author} {\bibfnamefont {R.}~\bibnamefont
  {Jordan}}, \bibinfo {author} {\bibfnamefont {D.}~\bibnamefont
  {Kinderlehrer}},\ and\ \bibinfo {author} {\bibfnamefont {F.}~\bibnamefont
  {Otto}},\ }\bibfield  {title} {\bibinfo {title} {Free energy and the
  {{Fokker-Planck}} equation},\ }\href
  {https://doi.org/10.1016/S0167-2789(97)00093-6} {\bibfield  {journal}
  {\bibinfo  {journal} {Physica D: Nonlinear Phenomena}\ }\bibinfo {series}
  {16th {{Annual International Conference}} of the {{Center}} for {{Nonlinear
  Studies}}},\ \textbf {\bibinfo {volume} {107}},\ \bibinfo {pages} {265}
  (\bibinfo {year} {1997})}\BibitemShut {NoStop}%
\bibitem [{\citenamefont {Benamou}\ and\ \citenamefont
  {Brenier}(2000)}]{benamou2000Numer.Math.}%
  \BibitemOpen
  \bibfield  {author} {\bibinfo {author} {\bibfnamefont {J.-D.}\ \bibnamefont
  {Benamou}}\ and\ \bibinfo {author} {\bibfnamefont {Y.}~\bibnamefont
  {Brenier}},\ }\bibfield  {title} {\bibinfo {title} {A computational fluid
  mechanics solution to the {{Monge-Kantorovich}} mass transfer problem},\
  }\href {https://doi.org/10.1007/s002110050002} {\bibfield  {journal}
  {\bibinfo  {journal} {Numer. Math.}\ }\textbf {\bibinfo {volume} {84}},\
  \bibinfo {pages} {375} (\bibinfo {year} {2000})}\BibitemShut {NoStop}%
\bibitem [{\citenamefont {Ambrosio}\ \emph {et~al.}(2006)\citenamefont
  {Ambrosio}, \citenamefont {Gigli},\ and\ \citenamefont
  {Savare}}]{ambrosio2006}%
  \BibitemOpen
  \bibfield  {author} {\bibinfo {author} {\bibfnamefont {L.}~\bibnamefont
  {Ambrosio}}, \bibinfo {author} {\bibfnamefont {N.}~\bibnamefont {Gigli}},\
  and\ \bibinfo {author} {\bibfnamefont {G.}~\bibnamefont {Savare}},\
  }\href@noop {} {\emph {\bibinfo {title} {Gradient {{Flows}}: {{In Metric
  Spaces}} and in the {{Space}} of {{Probability Measures}}}}}\ (\bibinfo
  {publisher} {{Springer Science \& Business Media}},\ \bibinfo {year}
  {2006})\BibitemShut {NoStop}%
\bibitem [{\citenamefont {Shima}(2007)}]{shima2007}%
  \BibitemOpen
  \bibfield  {author} {\bibinfo {author} {\bibfnamefont {H.}~\bibnamefont
  {Shima}},\ }\href@noop {} {\emph {\bibinfo {title} {The {{Geometry}} of
  {{Hessian Structures}}}}}\ (\bibinfo  {publisher} {{World Scientific}},\
  \bibinfo {year} {2007})\BibitemShut {NoStop}%
\bibitem [{\citenamefont {Amari}(2016)}]{amari2016}%
  \BibitemOpen
  \bibfield  {author} {\bibinfo {author} {\bibfnamefont {S.-i.}\ \bibnamefont
  {Amari}},\ }\href@noop {} {\emph {\bibinfo {title} {Information {{Geometry}}
  and {{Its Applications}}}}}\ (\bibinfo  {publisher} {{Springer}},\ \bibinfo
  {year} {2016})\BibitemShut {NoStop}%
\bibitem [{\citenamefont {Nielsen}(2020)}]{nielsen2020Entropy}%
  \BibitemOpen
  \bibfield  {author} {\bibinfo {author} {\bibfnamefont {F.}~\bibnamefont
  {Nielsen}},\ }\bibfield  {title} {\bibinfo {title} {An {{Elementary
  Introduction}} to {{Information Geometry}}},\ }\href
  {https://doi.org/10.3390/e22101100} {\bibfield  {journal} {\bibinfo
  {journal} {Entropy}\ }\textbf {\bibinfo {volume} {22}},\ \bibinfo {pages}
  {1100} (\bibinfo {year} {2020})}\BibitemShut {NoStop}%
\bibitem [{\citenamefont {Hill}(2005)}]{hill2005}%
  \BibitemOpen
  \bibfield  {author} {\bibinfo {author} {\bibfnamefont {T.~L.}\ \bibnamefont
  {Hill}},\ }\href@noop {} {\emph {\bibinfo {title} {Free {{Energy
  Transduction}} and {{Biochemical Cycle Kinetics}}}}}\ (\bibinfo  {publisher}
  {{Courier Corporation}},\ \bibinfo {year} {2005})\BibitemShut {NoStop}%
\bibitem [{\citenamefont {Schnakenberg}(1976)}]{schnakenberg1976Rev.Mod.Phys.}%
  \BibitemOpen
  \bibfield  {author} {\bibinfo {author} {\bibfnamefont {J.}~\bibnamefont
  {Schnakenberg}},\ }\bibfield  {title} {\bibinfo {title} {Network theory of
  microscopic and macroscopic behavior of master equation systems},\ }\href
  {https://doi.org/10.1103/RevModPhys.48.571} {\bibfield  {journal} {\bibinfo
  {journal} {Rev. Mod. Phys.}\ }\textbf {\bibinfo {volume} {48}},\ \bibinfo
  {pages} {571} (\bibinfo {year} {1976})}\BibitemShut {NoStop}%
\bibitem [{\citenamefont {Ge}\ and\ \citenamefont
  {Qian}(2016)}]{ge2016ChemicalPhysics}%
  \BibitemOpen
  \bibfield  {author} {\bibinfo {author} {\bibfnamefont {H.}~\bibnamefont
  {Ge}}\ and\ \bibinfo {author} {\bibfnamefont {H.}~\bibnamefont {Qian}},\
  }\bibfield  {title} {\bibinfo {title} {Nonequilibrium thermodynamic formalism
  of nonlinear chemical reaction systems with
  {{Waage}}\textendash{{Guldberg}}'s law of mass action},\ }\href
  {https://doi.org/10.1016/j.chemphys.2016.03.026} {\bibfield  {journal}
  {\bibinfo  {journal} {Chemical Physics}\ }\textbf {\bibinfo {volume} {472}},\
  \bibinfo {pages} {241} (\bibinfo {year} {2016})}\BibitemShut {NoStop}%
\bibitem [{\citenamefont {Rao}\ and\ \citenamefont
  {Esposito}(2016)}]{rao2016Phys.Rev.X}%
  \BibitemOpen
  \bibfield  {author} {\bibinfo {author} {\bibfnamefont {R.}~\bibnamefont
  {Rao}}\ and\ \bibinfo {author} {\bibfnamefont {M.}~\bibnamefont {Esposito}},\
  }\bibfield  {title} {\bibinfo {title} {Nonequilibrium {{Thermodynamics}} of
  {{Chemical Reaction Networks}}: {{Wisdom}} from {{Stochastic
  Thermodynamics}}},\ }\href {https://doi.org/10.1103/PhysRevX.6.041064}
  {\bibfield  {journal} {\bibinfo  {journal} {Phys. Rev. X}\ }\textbf {\bibinfo
  {volume} {6}},\ \bibinfo {pages} {041064} (\bibinfo {year}
  {2016})}\BibitemShut {NoStop}%
\bibitem [{\citenamefont {Beard}\ and\ \citenamefont {Qian}(2008)}]{beard2008}%
  \BibitemOpen
  \bibfield  {author} {\bibinfo {author} {\bibfnamefont {D.~A.}\ \bibnamefont
  {Beard}}\ and\ \bibinfo {author} {\bibfnamefont {H.}~\bibnamefont {Qian}},\
  }\href {https://doi.org/10.1017/CBO9780511803345} {\emph {\bibinfo {title}
  {Chemical {{Biophysics}}: {{Quantitative Analysis}} of {{Cellular
  Systems}}}}},\ Cambridge {{Texts}} in {{Biomedical Engineering}}\ (\bibinfo
  {publisher} {{Cambridge University Press}},\ \bibinfo {address}
  {{Cambridge}},\ \bibinfo {year} {2008})\BibitemShut {NoStop}%
\bibitem [{\citenamefont {Mikhailov}\ and\ \citenamefont
  {Ertl}(2017)}]{mikhailov2017}%
  \BibitemOpen
  \bibfield  {author} {\bibinfo {author} {\bibfnamefont {A.~S.}\ \bibnamefont
  {Mikhailov}}\ and\ \bibinfo {author} {\bibfnamefont {G.}~\bibnamefont
  {Ertl}},\ }\href@noop {} {\emph {\bibinfo {title} {Chemical {{Complexity}}:
  {{Self-Organization Processes}} in {{Molecular Systems}}}}}\ (\bibinfo
  {publisher} {{Springer}},\ \bibinfo {year} {2017})\BibitemShut {NoStop}%
\bibitem [{\citenamefont {Alon}(2019)}]{alon2019}%
  \BibitemOpen
  \bibfield  {author} {\bibinfo {author} {\bibfnamefont {U.}~\bibnamefont
  {Alon}},\ }\href@noop {} {\emph {\bibinfo {title} {An {{Introduction}} to
  {{Systems Biology}}: {{Design Principles}} of {{Biological Circuits}}}}}\
  (\bibinfo  {publisher} {{CRC Press LLC}},\ \bibinfo {year}
  {2019})\BibitemShut {NoStop}%
\bibitem [{\citenamefont {Dechant}\ \emph {et~al.}(2022)\citenamefont
  {Dechant}, \citenamefont {Sasa},\ and\ \citenamefont
  {Ito}}]{dechant2022ArXiv210912817Cond-Mat}%
  \BibitemOpen
  \bibfield  {author} {\bibinfo {author} {\bibfnamefont {A.}~\bibnamefont
  {Dechant}}, \bibinfo {author} {\bibfnamefont {S.-i.}\ \bibnamefont {Sasa}},\
  and\ \bibinfo {author} {\bibfnamefont {S.}~\bibnamefont {Ito}},\ }\bibfield
  {title} {\bibinfo {title} {Geometric decomposition of entropy production in
  out-of-equilibrium systems},\ }\href@noop {} {\bibfield  {journal} {\bibinfo
  {journal} {ArXiv210912817 Cond-Mat}\ } (\bibinfo {year} {2022})},\ \Eprint
  {https://arxiv.org/abs/2109.12817} {arXiv:2109.12817 [cond-mat]} \BibitemShut
  {NoStop}%
\bibitem [{\citenamefont {Grady}\ and\ \citenamefont
  {Polimeni}(2010)}]{grady2010}%
  \BibitemOpen
  \bibfield  {author} {\bibinfo {author} {\bibfnamefont {L.~J.}\ \bibnamefont
  {Grady}}\ and\ \bibinfo {author} {\bibfnamefont {J.~R.}\ \bibnamefont
  {Polimeni}},\ }\href@noop {} {\emph {\bibinfo {title} {Discrete {{Calculus}}:
  {{Applied Analysis}} on {{Graphs}} for {{Computational Science}}}}}\
  (\bibinfo  {publisher} {{Springer Science \& Business Media}},\ \bibinfo
  {year} {2010})\BibitemShut {NoStop}%
\bibitem [{\citenamefont {Gibbs}(1878)}]{gibbs1878Am.J.Sci.}%
  \BibitemOpen
  \bibfield  {author} {\bibinfo {author} {\bibfnamefont {J.~W.}\ \bibnamefont
  {Gibbs}},\ }\bibfield  {title} {\bibinfo {title} {On the equilibrium of
  heterogeneous substances},\ }\href {https://doi.org/10.2475/ajs.s3-16.96.441}
  {\bibfield  {journal} {\bibinfo  {journal} {Am. J. Sci.}\ }\textbf {\bibinfo
  {volume} {s3-16}},\ \bibinfo {pages} {441} (\bibinfo {year}
  {1878})}\BibitemShut {NoStop}%
\bibitem [{\citenamefont {Feinberg}(2019)}]{feinberg2019}%
  \BibitemOpen
  \bibfield  {author} {\bibinfo {author} {\bibfnamefont {M.}~\bibnamefont
  {Feinberg}},\ }\href@noop {} {\emph {\bibinfo {title} {Foundations of
  {{Chemical Reaction Network Theory}}}}}\ (\bibinfo  {publisher}
  {{Springer}},\ \bibinfo {year} {2019})\BibitemShut {NoStop}%
\bibitem [{\citenamefont {Klamt}\ \emph {et~al.}(2009)\citenamefont {Klamt},
  \citenamefont {Haus},\ and\ \citenamefont
  {Theis}}]{klamt2009PLOSComputationalBiology}%
  \BibitemOpen
  \bibfield  {author} {\bibinfo {author} {\bibfnamefont {S.}~\bibnamefont
  {Klamt}}, \bibinfo {author} {\bibfnamefont {U.-U.}\ \bibnamefont {Haus}},\
  and\ \bibinfo {author} {\bibfnamefont {F.}~\bibnamefont {Theis}},\ }\bibfield
   {title} {\bibinfo {title} {Hypergraphs and {{Cellular Networks}}},\ }\href
  {https://doi.org/10.1371/journal.pcbi.1000385} {\bibfield  {journal}
  {\bibinfo  {journal} {PLOS Computational Biology}\ }\textbf {\bibinfo
  {volume} {5}},\ \bibinfo {pages} {e1000385} (\bibinfo {year}
  {2009})}\BibitemShut {NoStop}%
\bibitem [{\citenamefont {Maes}(2017{\natexlab{a}})}]{maes2017Phys.Rev.Lett.}%
  \BibitemOpen
  \bibfield  {author} {\bibinfo {author} {\bibfnamefont {C.}~\bibnamefont
  {Maes}},\ }\bibfield  {title} {\bibinfo {title} {Frenetic {{Bounds}} on the
  {{Entropy Production}}},\ }\href
  {https://doi.org/10.1103/PhysRevLett.119.160601} {\bibfield  {journal}
  {\bibinfo  {journal} {Phys. Rev. Lett.}\ }\textbf {\bibinfo {volume} {119}},\
  \bibinfo {pages} {160601} (\bibinfo {year} {2017}{\natexlab{a}})}\BibitemShut
  {NoStop}%
\bibitem [{\citenamefont {Kaiser}\ \emph {et~al.}(2018)\citenamefont {Kaiser},
  \citenamefont {Jack},\ and\ \citenamefont {Zimmer}}]{kaiser2018JStatPhys}%
  \BibitemOpen
  \bibfield  {author} {\bibinfo {author} {\bibfnamefont {M.}~\bibnamefont
  {Kaiser}}, \bibinfo {author} {\bibfnamefont {R.~L.}\ \bibnamefont {Jack}},\
  and\ \bibinfo {author} {\bibfnamefont {J.}~\bibnamefont {Zimmer}},\
  }\bibfield  {title} {\bibinfo {title} {Canonical {{Structure}} and
  {{Orthogonality}} of {{Forces}} and {{Currents}} in {{Irreversible Markov
  Chains}}},\ }\href {https://doi.org/10.1007/s10955-018-1986-0} {\bibfield
  {journal} {\bibinfo  {journal} {J Stat Phys}\ }\textbf {\bibinfo {volume}
  {170}},\ \bibinfo {pages} {1019} (\bibinfo {year} {2018})}\BibitemShut
  {NoStop}%
\bibitem [{\citenamefont {Renger}(2018)}]{renger2018Entropy}%
  \BibitemOpen
  \bibfield  {author} {\bibinfo {author} {\bibfnamefont {D.~R.~M.}\
  \bibnamefont {Renger}},\ }\bibfield  {title} {\bibinfo {title} {Gradient and
  {{GENERIC Systems}} in the {{Space}} of {{Fluxes}}, {{Applied}} to {{Reacting
  Particle Systems}}},\ }\href {https://doi.org/10.3390/e20080596} {\bibfield
  {journal} {\bibinfo  {journal} {Entropy}\ }\textbf {\bibinfo {volume} {20}},\
  \bibinfo {pages} {596} (\bibinfo {year} {2018})}\BibitemShut {NoStop}%
\bibitem [{\citenamefont {Renger}\ and\ \citenamefont
  {Zimmer}(2021)}]{renger2021DiscreteContin.Dyn.Syst.-S}%
  \BibitemOpen
  \bibfield  {author} {\bibinfo {author} {\bibfnamefont {D.~R.~M.}\
  \bibnamefont {Renger}}\ and\ \bibinfo {author} {\bibfnamefont
  {J.}~\bibnamefont {Zimmer}},\ }\bibfield  {title} {\bibinfo {title}
  {Orthogonality of fluxes in general nonlinear reaction networks},\ }\href
  {https://doi.org/10.3934/dcdss.2020346} {\bibfield  {journal} {\bibinfo
  {journal} {Discrete Contin. Dyn. Syst. - S}\ }\textbf {\bibinfo {volume}
  {14}},\ \bibinfo {pages} {205} (\bibinfo {year} {2021})}\BibitemShut
  {NoStop}%
\bibitem [{\citenamefont {Patterson}\ \emph {et~al.}(2021)\citenamefont
  {Patterson}, \citenamefont {Renger},\ and\ \citenamefont
  {Sharma}}]{patterson2021ArXiv210314384Math-Ph}%
  \BibitemOpen
  \bibfield  {author} {\bibinfo {author} {\bibfnamefont {R.~I.~A.}\
  \bibnamefont {Patterson}}, \bibinfo {author} {\bibfnamefont {D.~R.~M.}\
  \bibnamefont {Renger}},\ and\ \bibinfo {author} {\bibfnamefont
  {U.}~\bibnamefont {Sharma}},\ }\bibfield  {title} {\bibinfo {title}
  {Variational structures beyond gradient flows: A macroscopic
  fluctuation-theory perspective},\ }\href@noop {} {\bibfield  {journal}
  {\bibinfo  {journal} {ArXiv210314384 Math-Ph}\ } (\bibinfo {year} {2021})},\
  \Eprint {https://arxiv.org/abs/2103.14384} {arXiv:2103.14384 [math-ph]}
  \BibitemShut {NoStop}%
\bibitem [{\citenamefont {Maes}(2021)}]{maes2021SciPostPhys.Lect.Notes}%
  \BibitemOpen
  \bibfield  {author} {\bibinfo {author} {\bibfnamefont {C.}~\bibnamefont
  {Maes}},\ }\bibfield  {title} {\bibinfo {title} {Local detailed balance},\
  }\href {https://doi.org/10.21468/SciPostPhysLectNotes.32} {\bibfield
  {journal} {\bibinfo  {journal} {SciPost Phys. Lect. Notes}\ ,\ \bibinfo
  {pages} {032}} (\bibinfo {year} {2021})}\BibitemShut {NoStop}%
\bibitem [{\citenamefont {Maes}(2017{\natexlab{b}})}]{maes2017}%
  \BibitemOpen
  \bibfield  {author} {\bibinfo {author} {\bibfnamefont {C.}~\bibnamefont
  {Maes}},\ }\href@noop {} {\emph {\bibinfo {title} {Non-{{Dissipative
  Effects}} in {{Nonequilibrium Systems}}}}}\ (\bibinfo  {publisher}
  {{Springer}},\ \bibinfo {year} {2017})\BibitemShut {NoStop}%
\bibitem [{\citenamefont {Kobayashi}\ \emph {et~al.}(2021)\citenamefont
  {Kobayashi}, \citenamefont {Loutchko}, \citenamefont {Kamimura},\ and\
  \citenamefont {Sughiyama}}]{kobayashi2021ArXiv211214910Phys.}%
  \BibitemOpen
  \bibfield  {author} {\bibinfo {author} {\bibfnamefont {T.~J.}\ \bibnamefont
  {Kobayashi}}, \bibinfo {author} {\bibfnamefont {D.}~\bibnamefont {Loutchko}},
  \bibinfo {author} {\bibfnamefont {A.}~\bibnamefont {Kamimura}},\ and\
  \bibinfo {author} {\bibfnamefont {Y.}~\bibnamefont {Sughiyama}},\ }\bibfield
  {title} {\bibinfo {title} {Kinetic {{Derivation}} of the {{Hessian Geometric
  Structure}} in {{Chemical Reaction Systems}}},\ }\href@noop {} {\bibfield
  {journal} {\bibinfo  {journal} {ArXiv211214910 Phys.}\ } (\bibinfo {year}
  {2021})},\ \Eprint {https://arxiv.org/abs/2112.14910} {arXiv:2112.14910
  [physics]} \BibitemShut {NoStop}%
\bibitem [{\citenamefont {Sughiyama}\ \emph {et~al.}(2022)\citenamefont
  {Sughiyama}, \citenamefont {Kamimura}, \citenamefont {Loutchko},\ and\
  \citenamefont
  {Kobayashi}}]{sughiyama2022ArXiv220109417Cond-MatPhysicsphysics}%
  \BibitemOpen
  \bibfield  {author} {\bibinfo {author} {\bibfnamefont {Y.}~\bibnamefont
  {Sughiyama}}, \bibinfo {author} {\bibfnamefont {A.}~\bibnamefont {Kamimura}},
  \bibinfo {author} {\bibfnamefont {D.}~\bibnamefont {Loutchko}},\ and\
  \bibinfo {author} {\bibfnamefont {T.~J.}\ \bibnamefont {Kobayashi}},\
  }\bibfield  {title} {\bibinfo {title} {The {{Chemical Thermodynamics}} for
  {{Growing Systems}}},\ }\href@noop {} {\bibfield  {journal} {\bibinfo
  {journal} {ArXiv220109417 Cond-Mat Physicsphysics}\ } (\bibinfo {year}
  {2022})},\ \Eprint {https://arxiv.org/abs/2201.09417} {arXiv:2201.09417
  [cond-mat, physics:physics]} \BibitemShut {NoStop}%
\bibitem [{\citenamefont
  {Renger}(2021)}]{renger2021ArXiv211112164Cond-MatPhysicsmath-Ph}%
  \BibitemOpen
  \bibfield  {author} {\bibinfo {author} {\bibfnamefont {D.~R.~M.}\
  \bibnamefont {Renger}},\ }\bibfield  {title} {\bibinfo {title} {Anisothermal
  {{Chemical Reactions}}: {{Onsager-Machlup}} and {{Macroscopic Fluctuation
  Theory}}},\ }\href@noop {} {\bibfield  {journal} {\bibinfo  {journal}
  {ArXiv211112164 Cond-Mat Physicsmath-Ph}\ } (\bibinfo {year} {2021})},\
  \Eprint {https://arxiv.org/abs/2111.12164} {arXiv:2111.12164 [cond-mat,
  physics:math-ph]} \BibitemShut {NoStop}%
\bibitem [{\citenamefont {Nomizu}(1994)}]{nomizu1994}%
  \BibitemOpen
  \bibfield  {author} {\bibinfo {author} {\bibfnamefont {K.}~\bibnamefont
  {Nomizu}},\ }\href@noop {} {\emph {\bibinfo {title} {Affine {{Differential
  Geometry}}: {{Geometry}} of {{Affine Immersions}}}}}\ (\bibinfo  {publisher}
  {{Cambridge University Press}},\ \bibinfo {year} {1994})\BibitemShut
  {NoStop}%
\bibitem [{\citenamefont {Ito}(2018)}]{ito2018Phys.Rev.Lett.}%
  \BibitemOpen
  \bibfield  {author} {\bibinfo {author} {\bibfnamefont {S.}~\bibnamefont
  {Ito}},\ }\bibfield  {title} {\bibinfo {title} {Stochastic {{Thermodynamic
  Interpretation}} of {{Information Geometry}}},\ }\href
  {https://doi.org/10.1103/PhysRevLett.121.030605} {\bibfield  {journal}
  {\bibinfo  {journal} {Phys. Rev. Lett.}\ }\textbf {\bibinfo {volume} {121}},\
  \bibinfo {pages} {030605} (\bibinfo {year} {2018})}\BibitemShut {NoStop}%
\bibitem [{\citenamefont {Kolchinsky}\ and\ \citenamefont
  {Wolpert}(2021)}]{kolchinsky2021Phys.Rev.X}%
  \BibitemOpen
  \bibfield  {author} {\bibinfo {author} {\bibfnamefont {A.}~\bibnamefont
  {Kolchinsky}}\ and\ \bibinfo {author} {\bibfnamefont {D.~H.}\ \bibnamefont
  {Wolpert}},\ }\bibfield  {title} {\bibinfo {title} {Work, {{Entropy
  Production}}, and {{Thermodynamics}} of {{Information}} under {{Protocol
  Constraints}}},\ }\href {https://doi.org/10.1103/PhysRevX.11.041024}
  {\bibfield  {journal} {\bibinfo  {journal} {Phys. Rev. X}\ }\textbf {\bibinfo
  {volume} {11}},\ \bibinfo {pages} {041024} (\bibinfo {year}
  {2021})}\BibitemShut {NoStop}%
\bibitem [{\citenamefont {Ohga}\ and\ \citenamefont
  {Ito}(2021)}]{ohga2021ArXiv211211008Cond-Mat}%
  \BibitemOpen
  \bibfield  {author} {\bibinfo {author} {\bibfnamefont {N.}~\bibnamefont
  {Ohga}}\ and\ \bibinfo {author} {\bibfnamefont {S.}~\bibnamefont {Ito}},\
  }\bibfield  {title} {\bibinfo {title} {Information-geometric {{Legendre}}
  duality in stochastic thermodynamics},\ }\href@noop {} {\bibfield  {journal}
  {\bibinfo  {journal} {ArXiv211211008 Cond-Mat}\ } (\bibinfo {year} {2021})},\
  \Eprint {https://arxiv.org/abs/2112.11008} {arXiv:2112.11008 [cond-mat]}
  \BibitemShut {NoStop}%
\bibitem [{\citenamefont {Craciun}\ \emph {et~al.}(2009)\citenamefont
  {Craciun}, \citenamefont {Dickenstein}, \citenamefont {Shiu},\ and\
  \citenamefont {Sturmfels}}]{craciun2009JournalofSymbolicComputation}%
  \BibitemOpen
  \bibfield  {author} {\bibinfo {author} {\bibfnamefont {G.}~\bibnamefont
  {Craciun}}, \bibinfo {author} {\bibfnamefont {A.}~\bibnamefont
  {Dickenstein}}, \bibinfo {author} {\bibfnamefont {A.}~\bibnamefont {Shiu}},\
  and\ \bibinfo {author} {\bibfnamefont {B.}~\bibnamefont {Sturmfels}},\
  }\bibfield  {title} {\bibinfo {title} {Toric dynamical systems},\ }\href
  {https://doi.org/10.1016/j.jsc.2008.08.006} {\bibfield  {journal} {\bibinfo
  {journal} {Journal of Symbolic Computation}\ }\bibinfo {series} {In
  {{Memoriam Karin Gatermann}}},\ \textbf {\bibinfo {volume} {44}},\ \bibinfo
  {pages} {1551} (\bibinfo {year} {2009})}\BibitemShut {NoStop}%
\bibitem [{\citenamefont {Mielke}\ \emph {et~al.}(2014)\citenamefont {Mielke},
  \citenamefont {Peletier},\ and\ \citenamefont
  {Renger}}]{mielke2014PotentialAnal}%
  \BibitemOpen
  \bibfield  {author} {\bibinfo {author} {\bibfnamefont {A.}~\bibnamefont
  {Mielke}}, \bibinfo {author} {\bibfnamefont {M.~A.}\ \bibnamefont
  {Peletier}},\ and\ \bibinfo {author} {\bibfnamefont {D.~R.~M.}\ \bibnamefont
  {Renger}},\ }\bibfield  {title} {\bibinfo {title} {On the {{Relation}}
  between {{Gradient Flows}} and the {{Large-Deviation Principle}}, with
  {{Applications}} to {{Markov Chains}} and {{Diffusion}}},\ }\href
  {https://doi.org/10.1007/s11118-014-9418-5} {\bibfield  {journal} {\bibinfo
  {journal} {Potential Anal}\ }\textbf {\bibinfo {volume} {41}},\ \bibinfo
  {pages} {1293} (\bibinfo {year} {2014})}\BibitemShut {NoStop}%
\bibitem [{\citenamefont {Hatano}\ and\ \citenamefont
  {Sasa}(2001)}]{hatano2001Phys.Rev.Lett.}%
  \BibitemOpen
  \bibfield  {author} {\bibinfo {author} {\bibfnamefont {T.}~\bibnamefont
  {Hatano}}\ and\ \bibinfo {author} {\bibfnamefont {S.-i.}\ \bibnamefont
  {Sasa}},\ }\bibfield  {title} {\bibinfo {title} {Steady-{{State
  Thermodynamics}} of {{Langevin Systems}}},\ }\href
  {https://doi.org/10.1103/PhysRevLett.86.3463} {\bibfield  {journal} {\bibinfo
   {journal} {Phys. Rev. Lett.}\ }\textbf {\bibinfo {volume} {86}},\ \bibinfo
  {pages} {3463} (\bibinfo {year} {4月 16, 2001})}\BibitemShut {NoStop}%
\bibitem [{\citenamefont {Maes}\ and\ \citenamefont {Neto{\v
  c}n{\'y}}(2014)}]{maes2014JStatPhys}%
  \BibitemOpen
  \bibfield  {author} {\bibinfo {author} {\bibfnamefont {C.}~\bibnamefont
  {Maes}}\ and\ \bibinfo {author} {\bibfnamefont {K.}~\bibnamefont {Neto{\v
  c}n{\'y}}},\ }\bibfield  {title} {\bibinfo {title} {A {{Nonequilibrium
  Extension}} of the {{Clausius Heat Theorem}}},\ }\href
  {https://doi.org/10.1007/s10955-013-0822-9} {\bibfield  {journal} {\bibinfo
  {journal} {J Stat Phys}\ }\textbf {\bibinfo {volume} {154}},\ \bibinfo
  {pages} {188} (\bibinfo {year} {2014})}\BibitemShut {NoStop}%
\bibitem [{\citenamefont {Komatsu}\ \emph {et~al.}(2008)\citenamefont
  {Komatsu}, \citenamefont {Nakagawa}, \citenamefont {Sasa},\ and\
  \citenamefont {Tasaki}}]{komatsu2008Phys.Rev.Lett.}%
  \BibitemOpen
  \bibfield  {author} {\bibinfo {author} {\bibfnamefont {T.~S.}\ \bibnamefont
  {Komatsu}}, \bibinfo {author} {\bibfnamefont {N.}~\bibnamefont {Nakagawa}},
  \bibinfo {author} {\bibfnamefont {S.-i.}\ \bibnamefont {Sasa}},\ and\
  \bibinfo {author} {\bibfnamefont {H.}~\bibnamefont {Tasaki}},\ }\bibfield
  {title} {\bibinfo {title} {Steady-{{State Thermodynamics}} for {{Heat
  Conduction}}: {{Microscopic Derivation}}},\ }\href
  {https://doi.org/10.1103/PhysRevLett.100.230602} {\bibfield  {journal}
  {\bibinfo  {journal} {Phys. Rev. Lett.}\ }\textbf {\bibinfo {volume} {100}},\
  \bibinfo {pages} {230602} (\bibinfo {year} {6月 13, 2008})}\BibitemShut
  {NoStop}%
\bibitem [{\citenamefont {Bhatia}\ \emph {et~al.}(2013)\citenamefont {Bhatia},
  \citenamefont {Norgard}, \citenamefont {Pascucci},\ and\ \citenamefont
  {Bremer}}]{bhatia2013IEEETrans.Vis.Comput.Graph.}%
  \BibitemOpen
  \bibfield  {author} {\bibinfo {author} {\bibfnamefont {H.}~\bibnamefont
  {Bhatia}}, \bibinfo {author} {\bibfnamefont {G.}~\bibnamefont {Norgard}},
  \bibinfo {author} {\bibfnamefont {V.}~\bibnamefont {Pascucci}},\ and\
  \bibinfo {author} {\bibfnamefont {P.-T.}\ \bibnamefont {Bremer}},\ }\bibfield
   {title} {\bibinfo {title} {The {{Helmholtz-Hodge
  Decomposition}}\textemdash{{A Survey}}},\ }\href
  {https://doi.org/10.1109/TVCG.2012.316} {\bibfield  {journal} {\bibinfo
  {journal} {IEEE Trans. Vis. Comput. Graph.}\ }\textbf {\bibinfo {volume}
  {19}},\ \bibinfo {pages} {1386} (\bibinfo {year} {2013})}\BibitemShut
  {NoStop}%
\bibitem [{\citenamefont {Horn}\ and\ \citenamefont
  {Rowlinson}(1973)}]{horn1973Proc.R.Soc.Lond.Math.Phys.Sci.}%
  \BibitemOpen
  \bibfield  {author} {\bibinfo {author} {\bibfnamefont {F.}~\bibnamefont
  {Horn}}\ and\ \bibinfo {author} {\bibfnamefont {J.~S.}\ \bibnamefont
  {Rowlinson}},\ }\bibfield  {title} {\bibinfo {title} {Stability and complex
  balancing in mass-action systems with three short complexes},\ }\href
  {https://doi.org/10.1098/rspa.1973.0095} {\bibfield  {journal} {\bibinfo
  {journal} {Proc. R. Soc. Lond. Math. Phys. Sci.}\ }\textbf {\bibinfo {volume}
  {334}},\ \bibinfo {pages} {331} (\bibinfo {year} {1973})}\BibitemShut
  {NoStop}%
\bibitem [{\citenamefont {P{\'e}rez~Mill{\'a}n}\ \emph
  {et~al.}(2012)\citenamefont {P{\'e}rez~Mill{\'a}n}, \citenamefont
  {Dickenstein}, \citenamefont {Shiu},\ and\ \citenamefont
  {Conradi}}]{perezmillan2012BullMathBiol}%
  \BibitemOpen
  \bibfield  {author} {\bibinfo {author} {\bibfnamefont {M.}~\bibnamefont
  {P{\'e}rez~Mill{\'a}n}}, \bibinfo {author} {\bibfnamefont {A.}~\bibnamefont
  {Dickenstein}}, \bibinfo {author} {\bibfnamefont {A.}~\bibnamefont {Shiu}},\
  and\ \bibinfo {author} {\bibfnamefont {C.}~\bibnamefont {Conradi}},\
  }\bibfield  {title} {\bibinfo {title} {Chemical {{Reaction Systems}} with
  {{Toric Steady States}}},\ }\href {https://doi.org/10.1007/s11538-011-9685-x}
  {\bibfield  {journal} {\bibinfo  {journal} {Bull Math Biol}\ }\textbf
  {\bibinfo {volume} {74}},\ \bibinfo {pages} {1027} (\bibinfo {year}
  {2012})}\BibitemShut {NoStop}%
\bibitem [{\citenamefont {Shiraishi}(2021)}]{shiraishi2021JStatPhys}%
  \BibitemOpen
  \bibfield  {author} {\bibinfo {author} {\bibfnamefont {N.}~\bibnamefont
  {Shiraishi}},\ }\bibfield  {title} {\bibinfo {title} {Optimal {{Thermodynamic
  Uncertainty Relation}} in {{Markov Jump Processes}}},\ }\href
  {https://doi.org/10.1007/s10955-021-02829-8} {\bibfield  {journal} {\bibinfo
  {journal} {J Stat Phys}\ }\textbf {\bibinfo {volume} {185}},\ \bibinfo
  {pages} {19} (\bibinfo {year} {2021})}\BibitemShut {NoStop}%
\bibitem [{\citenamefont {Gingrich}\ \emph {et~al.}(2016)\citenamefont
  {Gingrich}, \citenamefont {Horowitz}, \citenamefont {Perunov},\ and\
  \citenamefont {England}}]{gingrich2016Phys.Rev.Lett.}%
  \BibitemOpen
  \bibfield  {author} {\bibinfo {author} {\bibfnamefont {T.~R.}\ \bibnamefont
  {Gingrich}}, \bibinfo {author} {\bibfnamefont {J.~M.}\ \bibnamefont
  {Horowitz}}, \bibinfo {author} {\bibfnamefont {N.}~\bibnamefont {Perunov}},\
  and\ \bibinfo {author} {\bibfnamefont {J.~L.}\ \bibnamefont {England}},\
  }\bibfield  {title} {\bibinfo {title} {Dissipation {{Bounds All Steady-State
  Current Fluctuations}}},\ }\href
  {https://doi.org/10.1103/PhysRevLett.116.120601} {\bibfield  {journal}
  {\bibinfo  {journal} {Phys. Rev. Lett.}\ }\textbf {\bibinfo {volume} {116}},\
  \bibinfo {pages} {120601} (\bibinfo {year} {2016})}\BibitemShut {NoStop}%
\bibitem [{\citenamefont {Barato}\ and\ \citenamefont
  {Seifert}(2015)}]{barato2015Phys.Rev.Lett.}%
  \BibitemOpen
  \bibfield  {author} {\bibinfo {author} {\bibfnamefont {A.~C.}\ \bibnamefont
  {Barato}}\ and\ \bibinfo {author} {\bibfnamefont {U.}~\bibnamefont
  {Seifert}},\ }\bibfield  {title} {\bibinfo {title} {Thermodynamic
  {{Uncertainty Relation}} for {{Biomolecular Processes}}},\ }\href
  {https://doi.org/10.1103/PhysRevLett.114.158101} {\bibfield  {journal}
  {\bibinfo  {journal} {Phys. Rev. Lett.}\ }\textbf {\bibinfo {volume} {114}},\
  \bibinfo {pages} {158101} (\bibinfo {year} {2015})}\BibitemShut {NoStop}%
\bibitem [{\citenamefont {Yoshimura}\ and\ \citenamefont
  {Ito}(2021)}]{yoshimura2021Phys.Rev.Lett.a}%
  \BibitemOpen
  \bibfield  {author} {\bibinfo {author} {\bibfnamefont {K.}~\bibnamefont
  {Yoshimura}}\ and\ \bibinfo {author} {\bibfnamefont {S.}~\bibnamefont
  {Ito}},\ }\bibfield  {title} {\bibinfo {title} {Thermodynamic {{Uncertainty
  Relation}} and {{Thermodynamic Speed Limit}} in {{Deterministic Chemical
  Reaction Networks}}},\ }\href
  {https://doi.org/10.1103/PhysRevLett.127.160601} {\bibfield  {journal}
  {\bibinfo  {journal} {Phys. Rev. Lett.}\ }\textbf {\bibinfo {volume} {127}},\
  \bibinfo {pages} {160601} (\bibinfo {year} {2021})}\BibitemShut {NoStop}%
\bibitem [{\citenamefont {Oster}\ \emph {et~al.}(1971)\citenamefont {Oster},
  \citenamefont {Perelson},\ and\ \citenamefont
  {Katchalsky}}]{oster1971Nature}%
  \BibitemOpen
  \bibfield  {author} {\bibinfo {author} {\bibfnamefont {G.}~\bibnamefont
  {Oster}}, \bibinfo {author} {\bibfnamefont {A.}~\bibnamefont {Perelson}},\
  and\ \bibinfo {author} {\bibfnamefont {A.}~\bibnamefont {Katchalsky}},\
  }\bibfield  {title} {\bibinfo {title} {Network {{Thermodynamics}}},\ }\href
  {https://doi.org/10.1038/234393a0} {\bibfield  {journal} {\bibinfo  {journal}
  {Nature}\ }\textbf {\bibinfo {volume} {234}},\ \bibinfo {pages} {393}
  (\bibinfo {year} {1971})}\BibitemShut {NoStop}%
\bibitem [{\citenamefont {Schnakenberg}(2012)}]{schnakenberg2012}%
  \BibitemOpen
  \bibfield  {author} {\bibinfo {author} {\bibfnamefont {J.}~\bibnamefont
  {Schnakenberg}},\ }\href@noop {} {\emph {\bibinfo {title} {Thermodynamic
  {{Network Analysis}} of {{Biological Systems}}}}}\ (\bibinfo  {publisher}
  {{Springer Science \& Business Media}},\ \bibinfo {year} {2012})\BibitemShut
  {NoStop}%
\bibitem [{\citenamefont {{van der Schaft}}\ \emph {et~al.}(2013)\citenamefont
  {{van der Schaft}}, \citenamefont {Rao},\ and\ \citenamefont
  {Jayawardhana}}]{vanderschaft2013SIAMJ.Appl.Math.a}%
  \BibitemOpen
  \bibfield  {author} {\bibinfo {author} {\bibfnamefont {A.}~\bibnamefont {{van
  der Schaft}}}, \bibinfo {author} {\bibfnamefont {S.}~\bibnamefont {Rao}},\
  and\ \bibinfo {author} {\bibfnamefont {B.}~\bibnamefont {Jayawardhana}},\
  }\bibfield  {title} {\bibinfo {title} {On the {{Mathematical Structure}} of
  {{Balanced Chemical Reaction Networks Governed}} by {{Mass Action
  Kinetics}}},\ }\href {https://doi.org/10.1137/11085431X} {\bibfield
  {journal} {\bibinfo  {journal} {SIAM J. Appl. Math.}\ }\textbf {\bibinfo
  {volume} {73}},\ \bibinfo {pages} {953} (\bibinfo {year} {2013})}\BibitemShut
  {NoStop}%
\bibitem [{\citenamefont {Belevitch}(1962)}]{belevitch1962Proc.IRE}%
  \BibitemOpen
  \bibfield  {author} {\bibinfo {author} {\bibfnamefont {V.}~\bibnamefont
  {Belevitch}},\ }\bibfield  {title} {\bibinfo {title} {Summary of the
  {{History}} of {{Circuit Theory}}},\ }\href
  {https://doi.org/10.1109/JRPROC.1962.288301} {\bibfield  {journal} {\bibinfo
  {journal} {Proc. IRE}\ }\textbf {\bibinfo {volume} {50}},\ \bibinfo {pages}
  {848} (\bibinfo {year} {1962})}\BibitemShut {NoStop}%
\bibitem [{\citenamefont {Chen}(2012)}]{chen2012}%
  \BibitemOpen
  \bibfield  {author} {\bibinfo {author} {\bibfnamefont {W.-K.}\ \bibnamefont
  {Chen}},\ }\href@noop {} {\emph {\bibinfo {title} {Applied {{Graph
  Theory}}}}}\ (\bibinfo  {publisher} {{Elsevier}},\ \bibinfo {year}
  {2012})\BibitemShut {NoStop}%
\bibitem [{\citenamefont {Wachtel}\ \emph {et~al.}(2015)\citenamefont
  {Wachtel}, \citenamefont {Vollmer},\ and\ \citenamefont
  {Altaner}}]{wachtel2015Phys.Rev.E}%
  \BibitemOpen
  \bibfield  {author} {\bibinfo {author} {\bibfnamefont {A.}~\bibnamefont
  {Wachtel}}, \bibinfo {author} {\bibfnamefont {J.}~\bibnamefont {Vollmer}},\
  and\ \bibinfo {author} {\bibfnamefont {B.}~\bibnamefont {Altaner}},\
  }\bibfield  {title} {\bibinfo {title} {Fluctuating currents in stochastic
  thermodynamics. {{I}}. {{Gauge}} invariance of asymptotic statistics},\
  }\href {https://doi.org/10.1103/PhysRevE.92.042132} {\bibfield  {journal}
  {\bibinfo  {journal} {Phys. Rev. E}\ }\textbf {\bibinfo {volume} {92}},\
  \bibinfo {pages} {042132} (\bibinfo {year} {2015})}\BibitemShut {NoStop}%
\bibitem [{\citenamefont {Altaner}\ \emph {et~al.}(2012)\citenamefont
  {Altaner}, \citenamefont {Grosskinsky}, \citenamefont {Herminghaus},
  \citenamefont {Katth{\"a}n}, \citenamefont {Timme},\ and\ \citenamefont
  {Vollmer}}]{altaner2012Phys.Rev.E}%
  \BibitemOpen
  \bibfield  {author} {\bibinfo {author} {\bibfnamefont {B.}~\bibnamefont
  {Altaner}}, \bibinfo {author} {\bibfnamefont {S.}~\bibnamefont
  {Grosskinsky}}, \bibinfo {author} {\bibfnamefont {S.}~\bibnamefont
  {Herminghaus}}, \bibinfo {author} {\bibfnamefont {L.}~\bibnamefont
  {Katth{\"a}n}}, \bibinfo {author} {\bibfnamefont {M.}~\bibnamefont {Timme}},\
  and\ \bibinfo {author} {\bibfnamefont {J.}~\bibnamefont {Vollmer}},\
  }\bibfield  {title} {\bibinfo {title} {Network representations of
  nonequilibrium steady states: {{Cycle}} decompositions, symmetries, and
  dominant paths},\ }\href {https://doi.org/10.1103/PhysRevE.85.041133}
  {\bibfield  {journal} {\bibinfo  {journal} {Phys. Rev. E}\ }\textbf {\bibinfo
  {volume} {85}},\ \bibinfo {pages} {041133} (\bibinfo {year}
  {2012})}\BibitemShut {NoStop}%
\bibitem [{\citenamefont {Polettini}(2015)}]{polettini2015Math.Technol.Netw.}%
  \BibitemOpen
  \bibfield  {author} {\bibinfo {author} {\bibfnamefont {M.}~\bibnamefont
  {Polettini}},\ }\bibfield  {title} {\bibinfo {title} {System/{{Environment
  Duality}} of {{Nonequilibrium Network Observables}}},\ }in\ \href
  {https://doi.org/10.1007/978-3-319-16619-3_13} {\emph {\bibinfo {booktitle}
  {Mathematical {{Technology}} of {{Networks}}}}},\ \bibinfo {series and
  number} {Springer {{Proceedings}} in {{Mathematics}} \& {{Statistics}}},\
  \bibinfo {editor} {edited by\ \bibinfo {editor} {\bibfnamefont
  {D.}~\bibnamefont {Mugnolo}}}\ (\bibinfo  {publisher} {{Springer
  International Publishing}},\ \bibinfo {address} {{Cham}},\ \bibinfo {year}
  {2015})\ pp.\ \bibinfo {pages} {191--205}\BibitemShut {NoStop}%
\bibitem [{\citenamefont
  {Suzuki}(2012)}]{suzuki2012PhysicaA:StatisticalMechanicsanditsApplications}%
  \BibitemOpen
  \bibfield  {author} {\bibinfo {author} {\bibfnamefont {M.}~\bibnamefont
  {Suzuki}},\ }\bibfield  {title} {\bibinfo {title} {Irreversibility and
  entropy production in transport phenomena, {{II}}: {{Statistical}}\textendash
  mechanical theory on steady states including thermal disturbance and energy
  supply},\ }\href {https://doi.org/10.1016/j.physa.2011.09.033} {\bibfield
  {journal} {\bibinfo  {journal} {Physica A: Statistical Mechanics and its
  Applications}\ }\textbf {\bibinfo {volume} {391}},\ \bibinfo {pages} {1074}
  (\bibinfo {year} {2012})}\BibitemShut {NoStop}%
\bibitem [{\citenamefont {Prigogine}(1968)}]{prigogine1968}%
  \BibitemOpen
  \bibfield  {author} {\bibinfo {author} {\bibfnamefont {I.}~\bibnamefont
  {Prigogine}},\ }\href@noop {} {\emph {\bibinfo {title} {Introduction to
  {{Thermodynamics}} of {{Irreversible Processes}}}}}\ (\bibinfo  {publisher}
  {{Wiley}},\ \bibinfo {year} {1968})\BibitemShut {NoStop}%
\bibitem [{\citenamefont {Klein}\ and\ \citenamefont
  {Meijer}(1954)}]{klein1954Phys.Rev.a}%
  \BibitemOpen
  \bibfield  {author} {\bibinfo {author} {\bibfnamefont {M.~J.}\ \bibnamefont
  {Klein}}\ and\ \bibinfo {author} {\bibfnamefont {P.~H.~E.}\ \bibnamefont
  {Meijer}},\ }\bibfield  {title} {\bibinfo {title} {Principle of {{Minimum
  Entropy Production}}},\ }\href {https://doi.org/10.1103/PhysRev.96.250}
  {\bibfield  {journal} {\bibinfo  {journal} {Phys. Rev.}\ }\textbf {\bibinfo
  {volume} {96}},\ \bibinfo {pages} {250} (\bibinfo {year} {1954})}\BibitemShut
  {NoStop}%
\bibitem [{\citenamefont {Polettini}(2011)}]{polettini2011Phys.Rev.E}%
  \BibitemOpen
  \bibfield  {author} {\bibinfo {author} {\bibfnamefont {M.}~\bibnamefont
  {Polettini}},\ }\bibfield  {title} {\bibinfo {title} {Macroscopic constraints
  for the minimum entropy production principle},\ }\href
  {https://doi.org/10.1103/PhysRevE.84.051117} {\bibfield  {journal} {\bibinfo
  {journal} {Phys. Rev. E}\ }\textbf {\bibinfo {volume} {84}},\ \bibinfo
  {pages} {051117} (\bibinfo {year} {2011})}\BibitemShut {NoStop}%
\bibitem [{\citenamefont {{\v Z}upanovi{\'c}}\ \emph
  {et~al.}(2004)\citenamefont {{\v Z}upanovi{\'c}}, \citenamefont
  {Jureti{\'c}},\ and\ \citenamefont {Botri{\'c}}}]{zupanovic2004Phys.Rev.E}%
  \BibitemOpen
  \bibfield  {author} {\bibinfo {author} {\bibfnamefont {P.}~\bibnamefont {{\v
  Z}upanovi{\'c}}}, \bibinfo {author} {\bibfnamefont {D.}~\bibnamefont
  {Jureti{\'c}}},\ and\ \bibinfo {author} {\bibfnamefont {S.}~\bibnamefont
  {Botri{\'c}}},\ }\bibfield  {title} {\bibinfo {title} {Kirchhoff's loop law
  and the maximum entropy production principle},\ }\href
  {https://doi.org/10.1103/PhysRevE.70.056108} {\bibfield  {journal} {\bibinfo
  {journal} {Phys. Rev. E}\ }\textbf {\bibinfo {volume} {70}},\ \bibinfo
  {pages} {056108} (\bibinfo {year} {2004})}\BibitemShut {NoStop}%
\bibitem [{\citenamefont {Martyushev}(2021)}]{martyushev2021Phys.-Usp.}%
  \BibitemOpen
  \bibfield  {author} {\bibinfo {author} {\bibfnamefont {L.~M.}\ \bibnamefont
  {Martyushev}},\ }\bibfield  {title} {\bibinfo {title} {Maximum entropy
  production principle: History and current status},\ }\href
  {https://doi.org/10.3367/UFNe.2020.08.038819} {\bibfield  {journal} {\bibinfo
   {journal} {Phys.-Usp.}\ }\textbf {\bibinfo {volume} {64}},\ \bibinfo {pages}
  {558} (\bibinfo {year} {2021})}\BibitemShut {NoStop}%
\bibitem [{\citenamefont {Gyarmati}(2013)}]{gyarmati2013}%
  \BibitemOpen
  \bibfield  {author} {\bibinfo {author} {\bibfnamefont {I.}~\bibnamefont
  {Gyarmati}},\ }\href@noop {} {\emph {\bibinfo {title} {Non-Equilibrium
  {{Thermodynamics}}: {{Field Theory}} and {{Variational Principles}}}}}\
  (\bibinfo  {publisher} {{Springer Science \& Business Media}},\ \bibinfo
  {year} {2013})\BibitemShut {NoStop}%
\bibitem [{\citenamefont {Polettini}(2013)}]{polettini2013Entropy}%
  \BibitemOpen
  \bibfield  {author} {\bibinfo {author} {\bibfnamefont {M.}~\bibnamefont
  {Polettini}},\ }\bibfield  {title} {\bibinfo {title} {Fact-{{Checking
  Ziegler}}'s {{Maximum Entropy Production Principle}} beyond the {{Linear
  Regime}} and towards {{Steady States}}},\ }\href
  {https://doi.org/10.3390/e15072570} {\bibfield  {journal} {\bibinfo
  {journal} {Entropy}\ }\textbf {\bibinfo {volume} {15}},\ \bibinfo {pages}
  {2570} (\bibinfo {year} {2013})}\BibitemShut {NoStop}%
\end{thebibliography}%

\end{document}